\documentclass[twocolumn]{aastex631}
\usepackage{subfigure}
\usepackage{float}
\usepackage{graphicx}
\usepackage{multirow}

\begin{document}

\title{JAGB 2.0: Improved Constraints on the J-region Asymptotic Giant Branch-based Hubble Constant from an Expanded Sample of JWST Observations}

\author[0000-0002-8623-1082]{Siyang Li}
\affiliation{Department of Physics and Astronomy, Johns Hopkins University, Baltimore, MD 21218, USA}

\author[0000-0002-6124-1196]{Adam G.~Riess}
\affiliation{Space Telescope Science Institute, 3700 San Martin Drive, Baltimore, MD 21218, USA}
\affiliation{Department of Physics and Astronomy, Johns Hopkins University, Baltimore, MD 21218, USA}

\author[0000-0002-4934-5849]{Daniel Scolnic}
\affiliation{Department of Physics, Duke University, Durham, NC 27708, USA}

\author{Stefano Casertano}
\affiliation{Space Telescope Science Institute, 3700 San Martin Drive, Baltimore, MD 21218, USA}

\author[0000-0002-5259-2314]{Gagandeep S.~Anand}
\affiliation{Space Telescope Science Institute, 3700 San Martin Drive, Baltimore, MD 21218, USA}

\begin{abstract}
The J-region Asymptotic Giant Branch (JAGB) is an overdensity of stars in the near-infrared, attributed to carbon-rich AGB stars, and recently used as a standard candle for measuring extragalactic distances and the Hubble constant. Using \emph{JWST} in Cycle 2, we extend JAGB measurements to 6 hosts of 9 Type~Ia supernovae (SNe~Ia) (NGC~2525, NGC~3147, NGC~3370, NGC~3447, NGC~5468, and NGC~5861), with two at $D\sim40$ Mpc, all calibrated by the maser host NGC~4258. We investigate the effects of incompleteness and are unable to recover a JAGB measurement in NGC~3147. We compile all JWST JAGB observations in SNe~Ia hosts, 15 galaxies hosting 18 SNe~Ia, from the SH0ES and CCHP programs and employ all literature measures. We find no significant mean difference between these distances and those from HST Cepheids, $-0.03\pm0.02$(stat)$\pm$0.05(sys)~mag. We find a difference of 0.11$\pm$0.022~mag between JAGB mode measurements in the CCHP analyses of two fields in NGC~4258, a feature also seen in two SH0ES fields (see field-to-field variations in \citealt{Li_2024ApJ...966...20L}), indicating significant variation of NGC~4258 JAGB measurements which produce a large absolute calibration uncertainty. Variations are also seen in the shape of the JAGB LF across galaxies so that different measures produce different values of the Hubble constant. We look for but do not (yet) find a standardizing relation between JAGB LF skew or color dependence and the apparent variation. Using the middle result of all JAGB measures to calibrate SNe~Ia yields a Hubble constant of $H_0$~=~73.3~$\pm$~1.4~(stat)$\pm$2.0~(sys)~km/s/Mpc with the systematic dominated by apparent differences across NGC 4258 calibrating fields or their measures.
\end{abstract}


\section{Introduction} \label{sec:intro}

The Hubble constant, $H_0$, parameterizes the current expansion rate of our universe.  Broadly speaking, there are two main approaches to measuring $H_0$: cosmologically from an early relic such as the cosmic microwave background fit to a cosmological model such as $\Lambda$CDM (for instance, from \citealt{Planck_2020A&A...641A...6P}), and locally, with the observed redshift $-$ distance relation for distant objects in the Hubble flow ($z>0.01$ e.g. to Type Ia supernovae most extensively by the SH0ES Team, \citealt{Riess_2022ApJ...934L...7R, Murakami_2023JCAP...11..046M, Breuval_2024ApJ...973...30B}). The ``Hubble Tension" refers to a $>$ 5$\sigma$ discrepancy between $H_0$ measured using these two approaches (see reviews by \citealp{Di_valentino_2021CQGra..38o3001D, Verde_2023arXiv231113305V}) and may suggest a flaw in the standard $\Lambda$CDM model used for the cosmological measurements. An alternative solution could involve a degree of unrecognized systematics in the measurement themselves, however, after roughly a decade of scrutiny there has not yet been compelling evidence for measurement systematics providing a satisfactory solution to the Hubble Tension.\footnote{For an extensive compilation of such studies, see \url{https://djbrout.github.io/SH0ESrefs.html}.}

The strongest constraints on $H_0$ from local measurements use an extragalactic distance ladder constructed with Cepheid variables in the hosts of 42 Type~Ia supernovae. Cepheids are favored for this role because
they offer 1) the largest number of independent sources of geometric calibration, 2) observability with the same telescope between rungs, and 3) the greatest reach to the small sample of local SN Ia.
For instance, the Supernovae, $H_0$, for the Equation of State (SH0ES) team used a three-rung distance ladder based on Cepheid variables and Type~Ia supernovae to measure $H_0$ = 73.17 $\pm$ 0.86 km s$^{-1}$ Mpc$^{-1}$ \citep{Riess_2022ApJ...934L...7R, Murakami_2023JCAP...11..046M, Breuval_2024ApJ...973...30B}. One of the most common approaches to test the systematics of this distance ladder is to replace rungs of the distance ladder with alternative standard candles.

The J-region Asymptotic Giant Branch (JAGB) refers to an overdensity of stars in a near-infrared color-magnitude diagram (CMD) whose composition is dominated by carbon-rich asymptotic giant branch stars. The JAGB has recently been used as a distance indicator to cross-check systematics of standard candles such as Cepheid variables. The concept for using carbon stars as distance indicators was pioneered in the 1980s \citep{Richer_1981ApJ...243..744R, Richer_1984ApJ...287..138R, Richer_1985ApJ...298..240R, Pritchet_1987ApJ...323...79P, Cook_1986ApJ...305..634C} and has recently experienced a revival \citep{Battinelli_2005AA...442..159B, Ripoche_2020MNRAS.495.2858R, Madore_2020ApJ...899...66M, Freedman_2020ApJ...899...67F,  Parada_2021MNRAS.501..933P, Parada_2023MNRAS.522..195P, Zgirski_2021ApJ...916...19Z, 
Lee_MW_2021ApJ...923..157L,
Lee_WLM_2021ApJ...907..112L, Madore_MW_2022ApJ...938..125M,
Lee_M33_2022ApJ...933..201L, Lee_M31_2023_ApJ...956...15L, Lee_JWST_2024ApJ...961..132L}. 

The foundational development of the JAGB method through these works has also recently led to measurements of distances to SN Ia hosts using JWST \citep{Li_2024ApJ...966...20L, Lee_JWST_JAGB_H0_2024arXiv240803474L, Freedman_JWST_H0_2024arXiv240806153F}. While two groups of JAGB measurements yield distances which are consistent with those from HST Cepheids \citep{Riess_JWST_H0_2024arXiv240811770R}, the samples of SNe Ia they calibrate are small and thus can yield fluctuations in $H_0$ as seen from the Cepheid sub-samples as well.  For instance, \cite{Li_2024ApJ...966...20L} found $H_0 = $ 74.7 $\pm$ 2.1 (stat) $\pm$ 2.3 (sys) km/s/Mpc using five galaxies from JWST Cycle 1, program GO-1685 (PI: A. Riess, \citealt{Riess_2021jwst.prop.1685R}). Meanwhile, the Carnegie-Chicago Hubble Program (CCHP) team measured JAGB in seven SN Ia hosts from \emph{JWST} program GO-1995 (PI: W. Freedman, \citealt{Freedman_2021jwst.prop.1995F}) and found $H_0 = $ 68.0 $\pm$ 1.9 (stat) $\pm$ 1.9 (sys) km/s/Mpc \citep{Lee_JWST_JAGB_H0_2024arXiv240803474L}. \footnote{We note that HST Cepheids yield $H_0 = $ 70 km/s/Mpc from the same CCHP JAGB sample, both calibrated to NGC 4258 \citep{Riess_JWST_H0_2024arXiv240811770R}. A further reduction by 2.0 km/s/Mpc in the CCHP analysis can be traced use of the Carnegie Supernovae Project \citep{Uddin_CSP_2024ApJ...970...72U} sample rather than the larger Pantheon+ \citep{Scolnic_Pantheon_2022ApJ...938..113S}, reducing $H_0$ by 0.7 km/s/Mpc, and use of the JAGB mode rather than considering other measurement measures by another 1.1 km/s/Mpc (see full discussion in \citealt{Riess_JWST_H0_2024arXiv240811770R}).} Although these JAGB-based studies of $H_0$ are limited in sample size, resulting in large statistical uncertainties in both JAGB and SNe~Ia measurements, the JAGB method has the potential to become more competitive in $H_0$ precision with more \emph{JWST} observations and can still regardless provide a powerful means of crosschecking systematics of Cepheid- and TRGB-based distances. The JAGB method involves using some form of the mean, median, mode, or a model fit \citep{Madore_2020ApJ...899...66M, Freedman_2020ApJ...899...67F, Lee_M31_2023_ApJ...956...15L, Ripoche_2020MNRAS.495.2858R, Zgirski_2021ApJ...916...19Z} of the near-infrared luminosity function of stars in the `J'-region, as categorized by \cite{Weinberg_2001ApJ...548..712W}, as the standard candle reference magnitude. 

In this paper, we extend recent JAGB measurements in two ways:
1) augmenting the JWST Cycle 1 data with Cycle 2 data of 5 hosts of 8 SN Ia; NGC 2525, NGC 3370, NGC 5861, NGC 3447, and NGC 3147 from \emph{JWST} Cycle 2 program GO-2875 \citep{Riess_2023jwst.prop.2875R} and 
2) extending the range of JAGB measurements by a magnitude past $\mu_0 =$ 32~mag, which has been the furthest JAGB distance currently analyzed in the literature, using \emph{JWST} observations of NGC 5468 from \emph{JWST} Cycle 1 program GO-1685 \citep{Riess_2021jwst.prop.1685R} and NGC 3147 from Cycle 2. We find that at greater distances incomplete sampling of the JAGB LF becomes a significant issue without deeper images than that obtained for Cepheid follow-up.

In Section \ref{sec:extending_JAGB}, we present analysis and distance measurements for the new Cycle 2 sample (plus one, NGC 5468 previously unanalyzed from Cycle 1).  In Section \ref{sec:distances}, we consistently combine these results with previously published cycle 1 programs from \cite{Lee_JWST_JAGB_H0_2024arXiv240803474L, Freedman_JWST_H0_2024arXiv240806153F} and \cite{Li_2024ApJ...966...20L}. We use these to measure $H_0$ in Section \ref{sec:Hubble_constant}. We conclude by discussing our results in Section \ref{sec:discussion}. In the Appendix, we describe the incompleteness and signal-to-noise limitations that preclude using observations of NGC 3147 from \emph{JWST} Cycle 2 program GO-2875 \citep[PI: A. Riess;][]{Riess_2023jwst.prop.2875R} for a robust JAGB measurement as well as compare JAGB distances to NGC 5643 in the \emph{F115W} and \emph{F150W} filters. 

\section{New JAGB Observations} \label{sec:extending_JAGB}

\emph{JWST} Cycle 1 and 2 programs GO-1685, 2875 included observations of five SNe~Ia host galaxies: NGC 2525, NGC 3147, NGC 3370, NGC 5468, and NGC 5861, that lie beyond $\mu_0 = 32$~mag, or about 25 Mpc \citep{Riess_2022ApJ...934L...7R}, further than any currently published JAGB distance. Past measurements of $H_0$ that use the JAGB have been limited in statistical leverage due to the number of available \emph{JWST} observations \citep{Li_2024ApJ...966...20L, Lee_JWST_JAGB_H0_2024arXiv240803474L, Freedman_JWST_H0_2024arXiv240806153F}. With three targeted JAGB programs now completed, we can combine the measurements used in the literature with those presented here to reduce statistical uncertainties associated with a $H_0$ measurement. In this section, we present an analysis of these five galaxies from \emph{JWST} Cycle 1 and 2 programs GO-1685, 2875 as well consistently combine the galaxies observed in \cite{Lee_JWST_JAGB_H0_2024arXiv240803474L} and \cite{Freedman_JWST_H0_2024arXiv240806153F}.

\begin{figure*}[ht!]
  \centering
  \includegraphics[angle=0, width=0.9\linewidth]{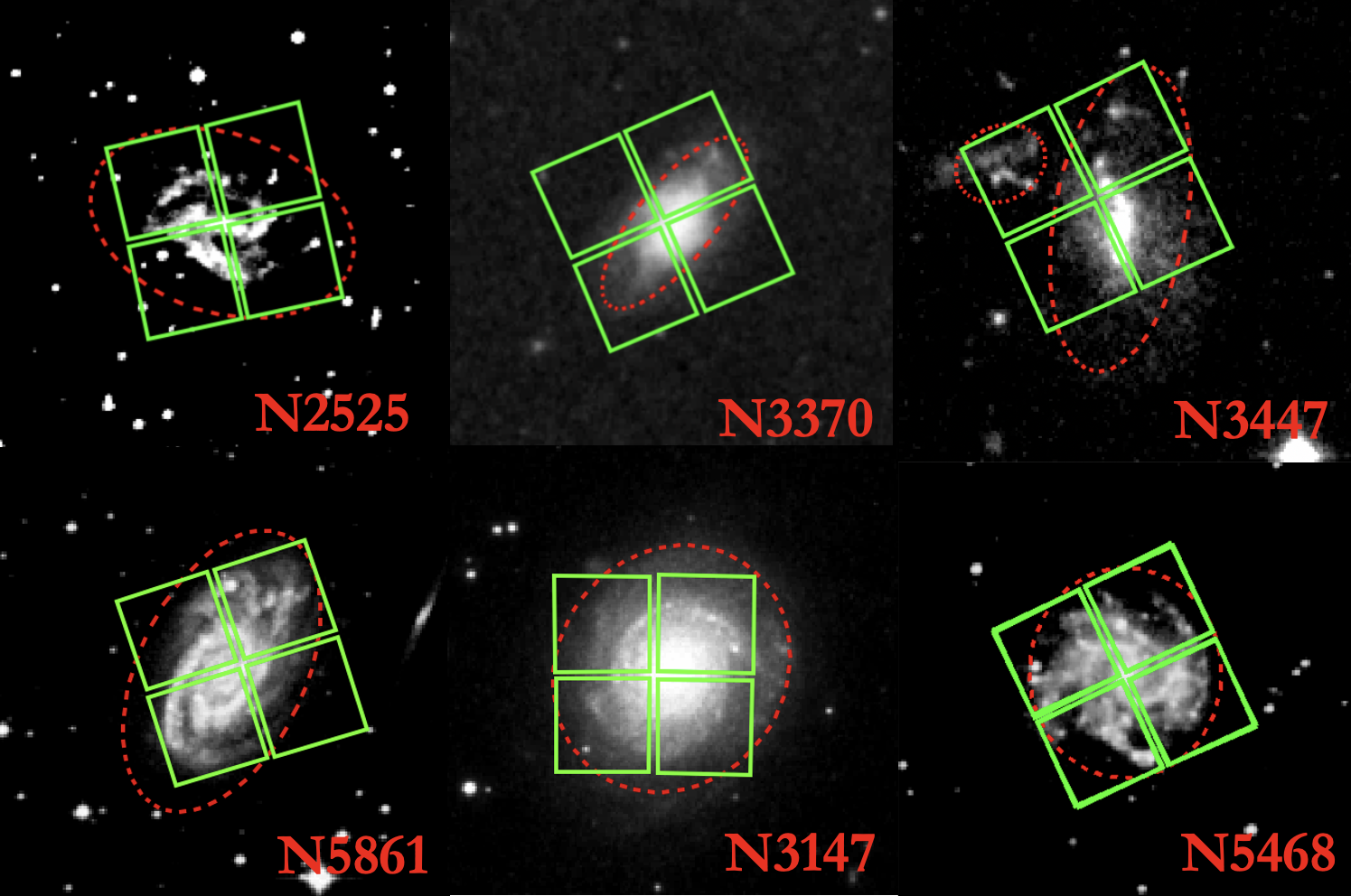}
  \caption{Footprints of the observations for NGC 2525, NGC 3370, NGC 3447, NGC 5861, NGC 3147, and NGC 5468 used for this study. \emph{JWST} NIRCam modules corresponds to green squares, and JAGB reference magnitudes are measured using stars in the regions outside the dashed red ellipses. Observations for NGC 5468 were taken in two epochs, with overlapping footprints for NGC 5468. Observations of NGC 4258, which are used to anchor the distances measured here, can be found in \cite{Li_2024ApJ...966...20L}. Galaxy images were retrieved from the ESO Digital Sky Survey, taken with blue photometric plates.}
  \label{fig:footprints}
\end{figure*}
\subsection{Data Selection} \label{sec:data_selection}

\begin{deluxetable*}{ccccccccc}
\caption{Summary of JAGB New Observations in JWST Cycles 1 and 2}
\label{tab:Observations_Table}
\tablehead{\colhead{Galaxy} & \colhead{Program} & \colhead{Observation} & \colhead{Observation Date} & \colhead{Filters} & \colhead{Exposure Time [s]} & \colhead{Context File Number} & \colhead{E(B-V)}}
\startdata
NGC$\,$4258 & GO-1685 & 5 & 2023-05-02 & \emph{F090W}/\emph{F277W} & 257.7 x 4 & 1215 & 0.014 \\
NGC$\,$4258 & GO-1685 & 5 & 2023-05-02 & \emph{F150W}/\emph{F277W} & 365.1 x 4 & " & " \\
NGC$\,$4258 & GO-1685 & 6 & 2023-05-17 & \emph{F090W}/\emph{F277W} & 257.7 x 4 & " & "\\
NGC$\,$4258 & GO-1685 & 6 & 2023-05-17 & \emph{F150W}/\emph{F277W} & 365.1 x 4 & " & "\\
NGC$\,$5468 & GO-1685 & 7 & 2023-06-28 & \emph{F090W}/\emph{F277W} & 306.0 x 4 & " & 0.021\\
NGC$\,$5468 & GO-1685 & 7 & 2023-06-28 & \emph{F150W}/\emph{F277W} & 708.6 x 4 & " & " \\
NGC$\,$5468 & GO-1685 & 8 & 2023-07-14 & \emph{F090W}/\emph{F277W} & 389.2 x 4 & " & " \\
NGC$\,$5468 & GO-1685 & 8 & 2023-07-14 & \emph{F150W}/\emph{F277W} & 657.6 x 4 & " & " \\
NGC$\,$2525 & GO-2875 & 1 & 2023-04-23 & \emph{F090W}/\emph{F277W} & 472.4 x 3 & 1225 & 0.050 \\
NGC$\,$2525 & GO-2875 & 1 & 2023-04-23 & \emph{F150W}/\emph{F277W} & 526.1 x 3 & " & "\\
NGC$\,$3147 & GO-2875 & 3 & 2024-02-23 & \emph{F090W}/\emph{F277W} & 526.1 x 4 & 1210 & 0.021 \\
NGC$\,$3147 & GO-2875 & 3 & 2024-02-23 & \emph{F150W}/\emph{F277W} & 622.7 x 4 & "  & " \\
NGC$\,$3370 & GO-2875 & 2 & 2023-06-28 & \emph{F090W}/\emph{F277W} & 472.4 x 4 & 1234 & 0.027\\
NGC$\,$3370 & GO-2875 & 2 & 2023-06-28 & \emph{F150W}/\emph{F277W} & 526.1 x 4 & "  & " \\
NGC$\,$3447 & GO-2875 & 12 & 2024-05-23 & \emph{F090W}/\emph{F277W} & 472.4 x 4 & " & 0.026\\
NGC$\,$3447 & GO-2875 & 12 & 2023-05-23 & \emph{F150W}/\emph{F277W} & 622.7 x 4 & "  & " \\
NGC$\,$5861 & GO-2875 & 14 & 2024-07-30 & \emph{F090W}/\emph{F277W} & 472.4 x 4 & " & 0.093\\
NGC$\,$5861 & GO-2875 & 14 & 2024-07-30 & \emph{F150W}/\emph{F277W} & 622.7 x 4 & " & " \\
\enddata
\tablecomments{Summary table for the \emph{JWST} NIRCAM observations of NGC 4258, NGC 5468, NGC 2525, NGC 3147, NGC 3370, NGC 3447, and NGC 5861 used in this study. Columns from left to right are: galaxy name, program number, observation number, observation date, filters, effective exposure time per dither, context (\texttt{pmap}) file number, and E(B-V) from \cite{Schlafly_2011ApJ...737..103S}.}
\end{deluxetable*}

We retrieved \emph{JWST} \emph{F150W} and \emph{F277W} images of NGC 2525, NGC 3147, NGC 3370, NGC 3447, NGC 5468, and NGC 5861 from the Barbara A. Mikulski Archive for Space Telescopes (MAST)\footnote{\url{https://mast.stsci.edu/portal/Mashup/Clients/Mast/Portal.html}}. These images were taken as part of \emph{JWST} Cycle 1 program GO-1685 \citep{Riess_2021jwst.prop.1685R} and \emph{JWST} Cycle 2 program GO-2875 \citep{Riess_2023jwst.prop.2875R}. While both of these programs were primarily designed to observe Cepheid variables, they also contain observations of the outer disk regions of the target galaxies that have carbon stars usable for the JAGB method. We provide a list of these observations in Table \ref{tab:Observations_Table} and show the footprints for these observations in Fig. \ref{fig:footprints}. For N2525, we exclude the fourth dither, which we found to be blurred from the Fine Guidance Sensor wandering due to the guide star landing on a region of bad read out pixels. We apply foreground extinction corrections using \emph{E(B-V)} values from \cite{Schlafly_2011ApJ...737..103S} and adopt the \cite{Fitzpatrick_1999PASP..111...63F} $R_v$ = 3.1 reddening law with $A_{\lambda}/A_V$ = 0.6021 and 0.2527 for \emph{F150W} and \emph{F277W}, respectively \citep{Riess_2024ApJ...962L..17R, Anand_2024ApJ...966...89A, Li_2024ApJ...966...20L}.


\begin{deluxetable*}{cccccccc}
\caption{Ellipses used to Define the Outer Disk}
\scriptsize
\label{tab:spatial_cuts}
\tablehead{
\colhead{Galaxy} & 
\colhead{RA} & 
\colhead{Dec} & 
\colhead{PA} & 
\colhead{Minor/Major Axes Ratio} & 
\colhead{SMA} & 
\multicolumn{2}{c}{Mean Crowd Corr.} \\ \cline{7-8}
& & & & & & 
\colhead{(\emph{F150W})} & 
\colhead{(\emph{F277W})} \\
}
\startdata
NGC 4258 & 12h18m57.5s & +47d18m14.3s & NA & NA & NA & 0.01 & 0.01 \\
NGC 2525 & 08h05m38.0s & -11d25m37.3s & 161.6 & 0.661 & 90 & 0.04 & 0.05 \\
NGC 3147 & 10h16m53.7s & +73d24m02.7s & 145.0 & 0.891 & 85 & NA & NA \\
NGC 3370 & 10h47m04.1s & +17d16m25.0s & 141.0 & 0.367 & 70 & 0.02 & 0.02  \\
NGC 3447 & 10h53m24.0s & +16d46m26.8s & 170 & 0.429 & 100 & 0.03 & 0.01 \\
NGC 3447 & 10h53m29.4s & +16d46m19.5s & 115 & 0.8 & 30 & " & " \\
NGC 5468 & 14h06m34.9s & -05d27m11.2s & 145.0 & 0.912 & 70 & 0.03 & 0.03 \\
NGC 5861 & 15h09m16.1s & -11d19m18.0s & 155.0 & 0.550 & 100 & 0.03 & 0.03 \\
\enddata
\tablecomments{Summary table for elliptical cuts used to define the outer disk and mean crowding corrections, where applicable, for NGC 4258, NGC 2525, NGC 3147, NGC 3370, NGC 3447, NGC 5468, and NGC 5861. Two ellipses are used to isolate the region used for the JAGB measurement in NGC 3447 due to its two distinct stellar groups arising from its merger history. Columns from left to right are: Galaxy name, RA, Dec, position angle (PA), elliptical minor to major axis ratio, elliptical semi-major axis, mean crowding correction (input - output) in \emph{F150W} at 22.3~mag for NGC 4258 and 25~mag for NGC 2525, NGC 3370, and NGC 5861, mean crowding correction (input - output) in \emph{F277W} at 21.05~mag for NGC 4258 and 23.75~mag (i.e. corresponding to stars at a color of $F150W - F277W$=1.25~mag for NGC 2525, NGC 3370, and NGC 5861. We do not apply an elliptical cut for NGC 4258 due to the placement of fields which already lie in the outer disk and do not apply crowding corrections for NGC 3147 because we do not use it for a JAGB measurement. }
\end{deluxetable*}

\subsection{Photometry} \label{sec:photometry}

We measure PSF photometry using the December 2022 version of the DOLPHOT\footnote{\url{http://americano.dolphinsim.com/dolphot/}} software package \citep{Dolphin_2000PASP..112.1383D, Dolphin_2016ascl.soft08013D} together with the February 2024 version of the \emph{JWST}/NIRCam module \citep{Weisz_2024ApJS..271...47W} of that same package. We perform photometry directly on the stage 2 *cal.fits images and use stage 3 \emph{F150W} *i2d.fits files as reference frames.  We use the Vega$-$Vega system to remain consistent with earlier versions of DOLPHOT; this zero-point has a Sirius-Vega $-$ Vega-Vega = $\sim$0.04~mag offset in \emph{F090W} from the Vega$-$Sirius system that is currently default in DOLPHOT NIRCam module, noting that \cite{Li_2024ApJ...966...20L} use the Vega-Vega zeropoint and some other \emph{JWST} JAGB studies, such as \cite{Lee_JWST_JAGB_H0_2024arXiv240803474L}, use the Sirius-Vega zeropoint.
Past studies have found issues with source matching when both short and long wavelength photometry is performed simultaneously using the DOLPHOT NIRCam module \citep{Riess_2024ApJ...962L..17R, Anand_2024ApJ...966...89A}. To circumvent this, we perform the warmstart procedure as outlined in the DOLPHOT manual: First, we perform a round of photometry using only short wavelength (\emph{F150W}) images, then use the source list from this first round to perform a second round of photometry with both short and long wavelength images (\emph{F150W} and \emph{F277W}). We use ``-etctime" option, allowing use of the \texttt{TMEASURE} header keyword for S/N calculations. We adopt the same DOLPHOT quality cuts as used in \cite{Li_2024ApJ...966...20L}, which are similar to those used in \cite{Weisz_2024ApJS..271...47W}: \texttt{SN} $>$ 3 (in \emph{F150W}), \texttt{crowd} $<$ 0.5, \texttt{sharp$^2$} $<$ 0.01, \texttt{flag} $\le$ 2, and \texttt{type} $\le$ 2, where the \texttt{crowd}, \texttt{sharp}, \texttt{flag}, and \texttt{type} cuts are applied separately in each band. For NGC 3370, we find that the lower exposure times causes the \texttt{crowd} cut to artificially create a sharp horizontal cutoff in the CMD at $\sim$28.2~mag. We lower the \texttt{crowd} cut in \emph{F277W} slightly to 0.4 for NGC 3370 only, noting that we explicitly correct for crowding, as described in Section \ref{sec:crowding_corrections}.

NGC 5468 from \emph{JWST} program GO-1685 was observed in two epochs with the intention of using phase information to estimate the mean magnitudes of Cepheid variables \citep{Riess_2024ApJ...962L..17R}. Because we do not need to know the phase information for our measurements, we perform photometery on the two epochs simultaneously using the combined stage 3 *i2d.fits image from MAST. We decided to perform photometry on the two images simultaneously in increase the signal-to-noise of the photometry to minimize the effects of incompleteness, which will be described in the Appendix. The other host galaxies, NGC 2525, NGC 3147, NGC 3370, NGC 3447, and NGC 5861 were all imaged and reduced with a single epoch. We apply spatial cuts using the ellipses centered on each galaxy (such that only stars outside the ellipses are retained), shown in Fig. \ref{fig:footprints} and listed in Table \ref{tab:spatial_cuts}, to isolate the outer disk for measurements. 

\subsection{Crowding Corrections} \label{sec:crowding_corrections}

The moderate density of stars in the outer disk of a galaxy where the JAGB is typically measured can cause a bright-ward bias in the photometric measurement of a given star. This bias is primarily due to contaminating flux from nearby stars in the field of view and can be corrected for in a \emph{statistical} manner by using artificial star tests. This correction procedure involves placing artificially generated, `fake' stars with known magnitudes on an image and measuring the magnitude of that star using the same photometric process used for the primary photometry \citep{Dolphin_2000PASP..112.1383D, Dolphin_2002MNRAS.332...91D, Dolphin_2016ascl.soft08013D}. The crowding bias is estimated statistically by repeating this procedure for many stars and finding the mean magnitude of the differences between the input and recovered magnitudes across these repetitions. We note that while artificial star tests can correct for the bias that arises from crowding, it does not necessarily reduce the scatter in magnitudes from this effect (that is, it will not reduce the width of the J-region luminosity function).

We ran artificial star tests to estimate the crowding biases for our JAGB measurements using the DOLPHOT software and a procedure similar to that described in \cite{Li_2024ApJ...966...20L}. \cite{Li_2024ApJ...966...20L} performed photometric bias corrections by dividing each \emph{JWST} NIRCam chip into a 10x10 grid and placing 200 artificial stars in each grid cell at three different magnitudes spaced 1~mag apart. Then, they calculated the mean difference between the input and recovered magnitudes for each magnitude and applied a linear least-squares fit to the mean differences as a function of magnitude. These interpolations were used as the crowding bias estimates and to correct the magnitudes of the real stars inside each grid cell.

We improve the procedure described in \cite{Li_2024ApJ...966...20L} by creating an exclusion zone using an ellipse with a semi-major axis (SMA) 20$-$30$\%$ smaller less than that used to define the outer disk used for the JAGB measurements, with the exception of NGC 4258 for which we do not apply a spatial cut due to the placement of the fields which fully lie on the outer disk. This exclusion zone is solely used to increase the efficiency of the artificial star tests and prevent artificial stars from being placed in inner regions of the disk that would be cut out later by the outer disk spatial cuts listed in Table \ref{tab:spatial_cuts}. We use a smaller ellipse here than that used for the spatial cuts used to define the outer disk for JAGB measurements to ensure that grid cells that are cut off by the elliptical cut used to define the outer disk are adequately populated by artificial stars. 

We perform two rounds of artificial star tests, first with \emph{F150W} only, then \emph{F150W} and \emph{F277W} using the warmstart photometry. We use only the \emph{F150W} results from the first, \emph{F150W}-only round of artificial stars and \emph{F277W} from the second round using the warmstart photometry. As in \cite{Li_2024ApJ...966...20L}, we parallelize our artificial star tests using GNU parallel \citep{Tange_2011a} and run the tests on the Advanced Research Computing at Hopkins (ARCH) Rockfish cluster. We use at least 200,000 stars per module across the three input magnitudes spaced 1~mag apart, centered on the approximate JAGB mean before crowding corrections. The mean crowding corrections for NGC 4258, NGC 2525, NGC 3370, NGC 3447, and NGC 5861 are listed in Table \ref{tab:spatial_cuts}.

\begin{deluxetable*}{cccccc}
\caption{Measurement Variants}
\tablehead{\colhead{\textbf{Variant}} & \colhead{} & \colhead{}  & \colhead{\textbf{Range Considered}} & \colhead{} & \colhead{}}
\startdata
\multicolumn{1}{c|}{\textbf{Method}}  & \multicolumn{1}{c|}{Median}  & \multicolumn{1}{c|}{Clipped Mean} & \multicolumn{1}{c|}{Mode} & \multicolumn{1}{c|}{Gaussian+Quadratic Model Fit} &\\
\tableline
\multicolumn{1}{c|}{\textbf{Smoothing}}  & \multicolumn{1}{c|}{}  & \multicolumn{1}{c|}{} & \multicolumn{1}{c|}{0.25, 0.35} & \multicolumn{1}{c|}{} & \\
\tableline
\textbf{Color Range} & Narrow  & Baseline & Wide & Blue 5\% &  Red 5\%  \\
\textbf{(All 5 used for} & Both sides 5\% narrower && Both sides 5\% wider & Blue side 5\% redder &Red side 5\% bluer \\
\textbf{each method)} & $1.05-1.425$~mag & $1-1.5$~mag & $0.95-1.575$~mag & $1.05-1.5$~mag & $1-1.425$~mag \\
\enddata
\label{tab:Measurement_Variants}
\tablecomments{Summary table for the 25 measurement variants used in this analysis.}
\end{deluxetable*}
\begin{figure*}[h!]
  \centering
  \begin{tabular}{cc}
    \includegraphics[width=0.4\linewidth]{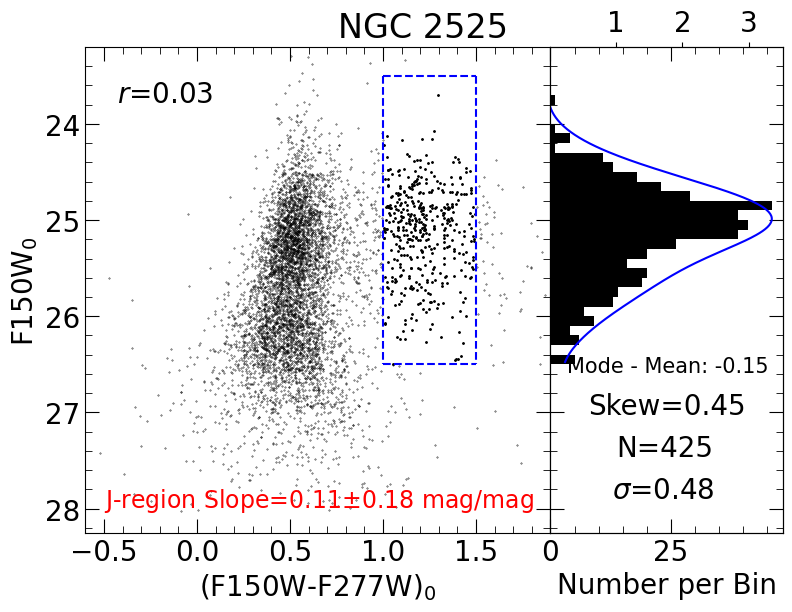} &
    \includegraphics[width=0.4\linewidth]{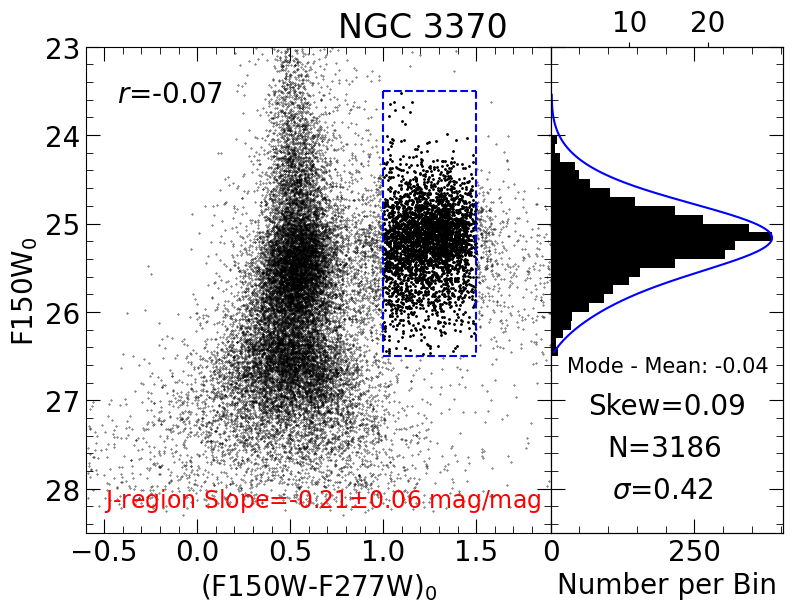} \\
    \includegraphics[width=0.4\linewidth]{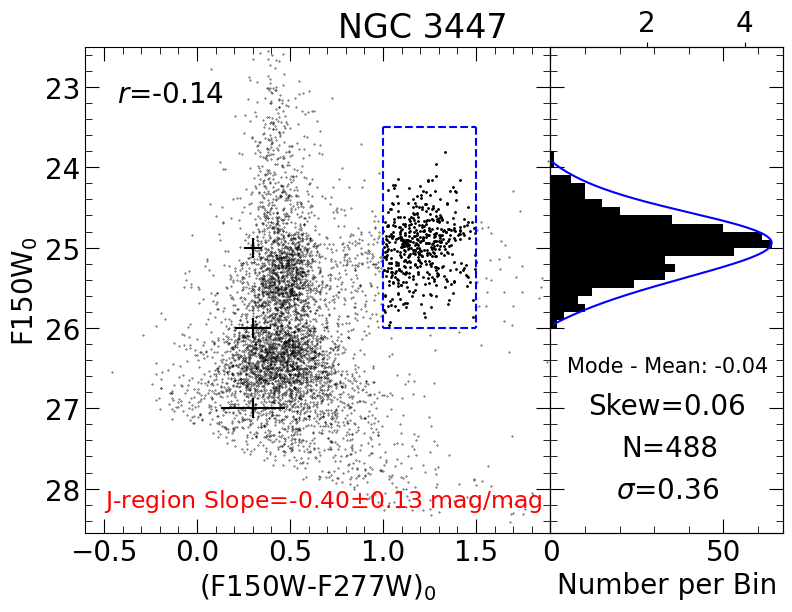} &
    \includegraphics[width=0.4\linewidth]{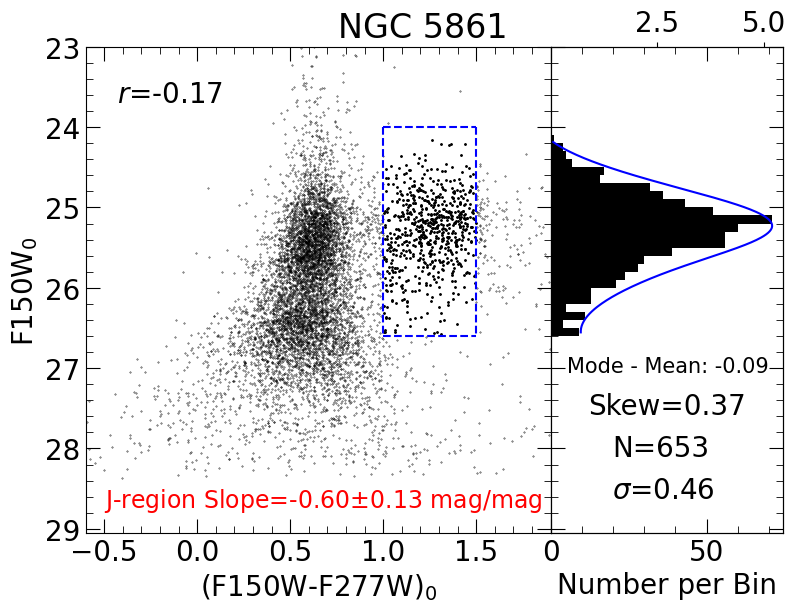} \\
    \includegraphics[width=0.4\linewidth]{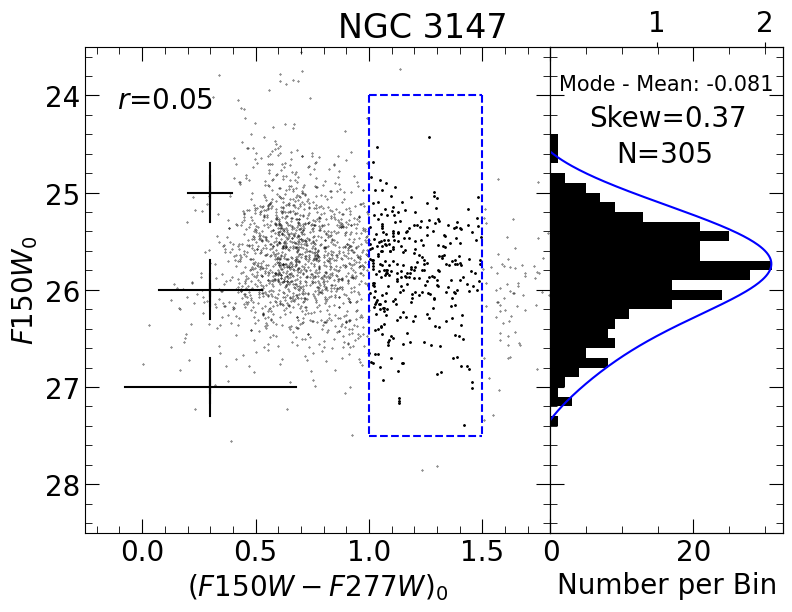} &
    \includegraphics[width=0.4\linewidth]{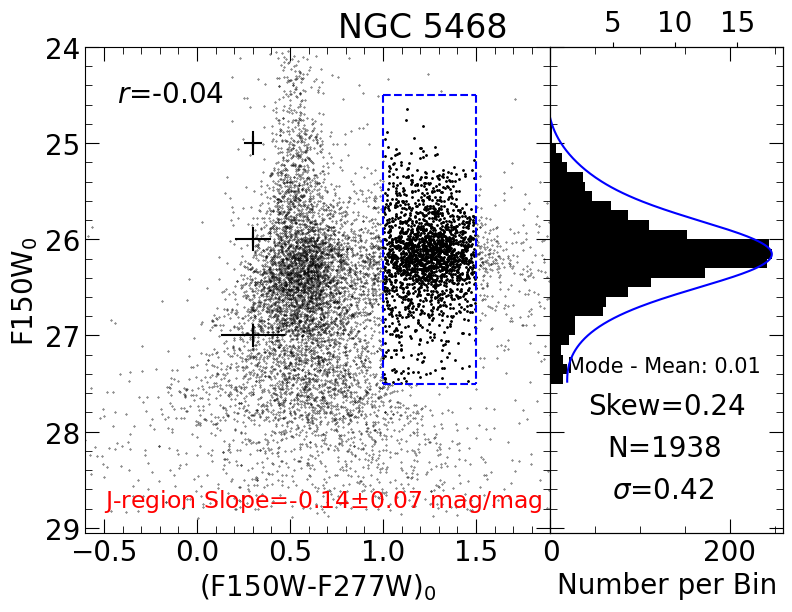} \\
  \end{tabular}
  \caption{Color magnitude diagrams and luminosity functions for NGC 2525, NGC 3370, NGC 5861, NGC 3147, NGC 3447, and NGC 5468. The left panels of each plot show the \emph{F150W} vs. \emph{F150W $-$ F277W} color magnitude diagrams after foreground extinction and crowding corrections. The vertical blue dashed lines use the color cuts of 1.0 $<$ \emph{F150W $-$ F277W} $<$ 1.5~mag and magnitude cuts $\sim$5 $\sigma$ from the anticipated JAGB, selected to fully encapsulate the J-region. Points inside the dashed blue box are enlarged for emphasis. The value in the upper left corner of each plot shows the Pearson correlation coefficient for the stars inside the dashed blue box. The right panel shows the luminosity function for stars inside the dashed blue box, with a GLOESS smoothed luminosity function (s=0.25~mag) shown in blue. We also list the difference between the mode (s=0.25~mag) and mean reference magnitudes, Fisher-Pearson coefficient of skewness using the Python \texttt{scipy.stats.skew}, the number (N) of stars in the dashed-blue box, and the standard deviation of the luminosity function. For NGC 3147 and NGC 5458, both of which lie at $D \sim$40 Mpc \citep{Riess_2022ApJ...934L...7R}, we show typically magnitude and color uncertainties for stars at a color of 1.25~mag at 25, 26, and 27~mag with the black vertical and horizontal lines. We do not measure a JAGB for NGC 3147 due to excessive incompleteness and so do not provide a J-region slope measurement. The CMD for NGC 4258 can be found in \cite{Li_2024ApJ...966...20L}.}
  \label{fig:CMDs}
\end{figure*}

\begin{figure*}[ht!]
  \centering
  \fbox{ 
    \begin{tabular}{ccc}
      \includegraphics[height=0.303\textwidth]{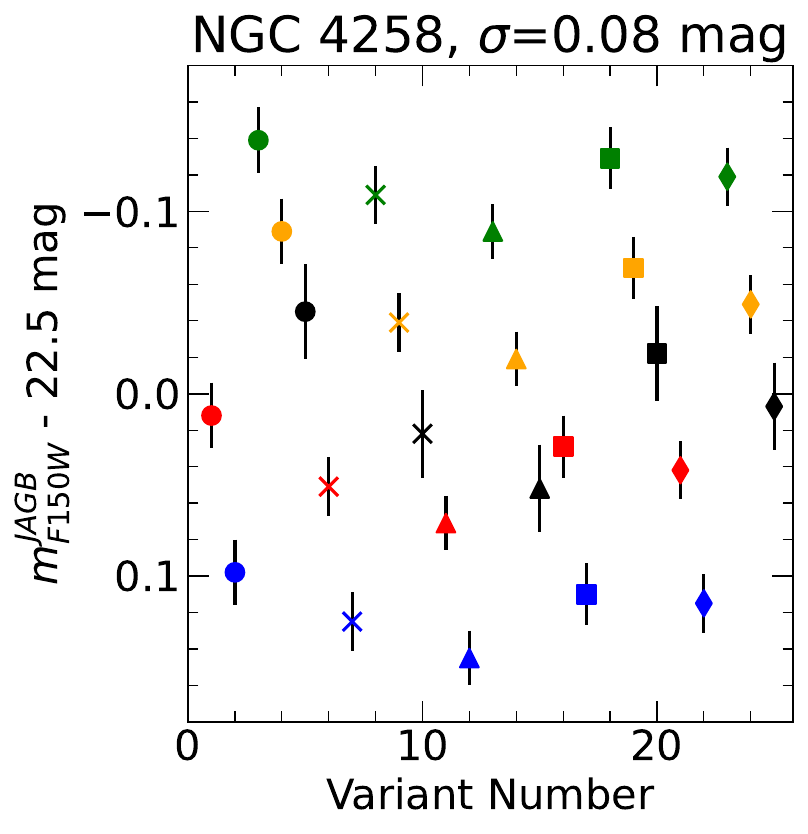} &
      \includegraphics[width=0.3\linewidth]{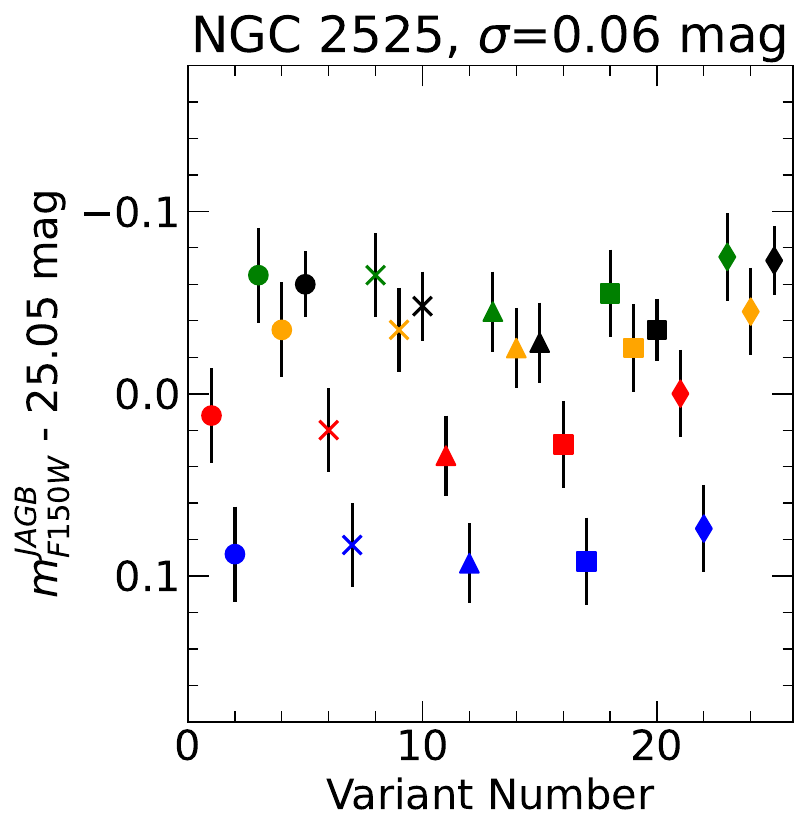} &
      \includegraphics[width=0.3\linewidth]{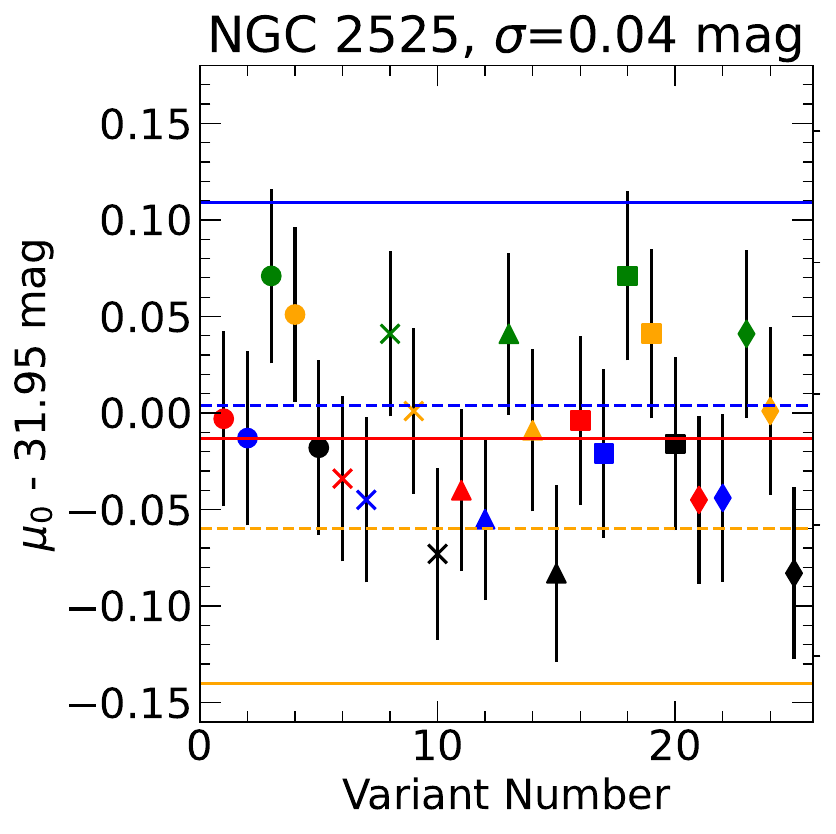} \\
      \multicolumn{3}{c}{\includegraphics[width=0.45\linewidth]{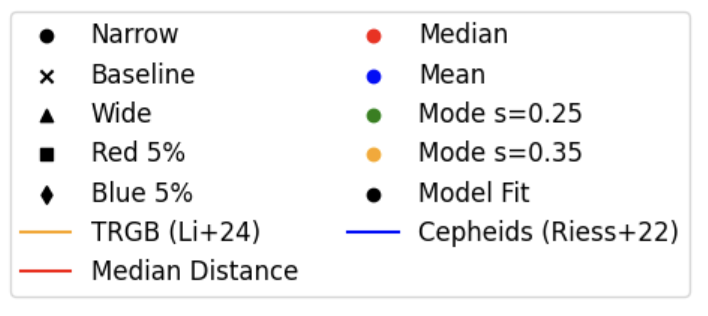}}
    \end{tabular}
  } 
  \caption{The left two subplots show the JAGB reference magnitudes as a function of the 25 measurement variants, as listed in Table \ref{tab:Measurement_Variants}, for NGC 4258 and NGC 2525. The right-most subplot shows distances to NGC 2525 across the same variants. The Cepheid \citep[fit variant 10; ][]{Riess_2022ApJ...934L...7R, Riess_JWST_H0_2024arXiv240811770R} and TRGB \citep{Li_TRGB_vs_Ceph_2024arXiv240800065L} distances are shown in blue and orange, respectively. The middle of all variants is show in with the red line. We also list the standard deviation of all variants in the plot titles, noting partial cancellation of variations and asymmetry along the distance ladder. Uncertainties for the Cepheid and TRGB distances are shown with the blue and orange dashed lines, respectively.}
  \label{fig:Variants_N2525}
\end{figure*}

\begin{figure*}[htbp]
  \centering
  \begin{tabular}{ccc}
    \includegraphics[width=0.3\linewidth]{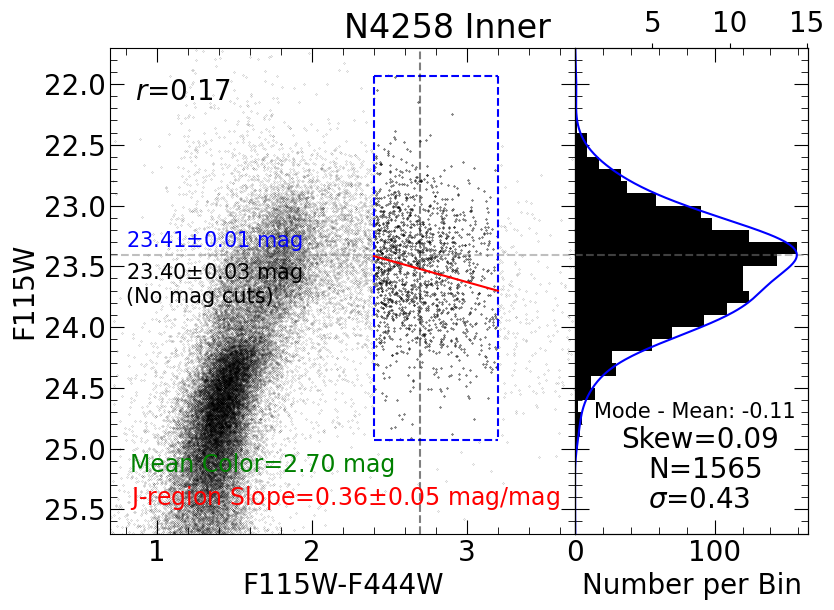} &
    \includegraphics[width=0.3\linewidth]{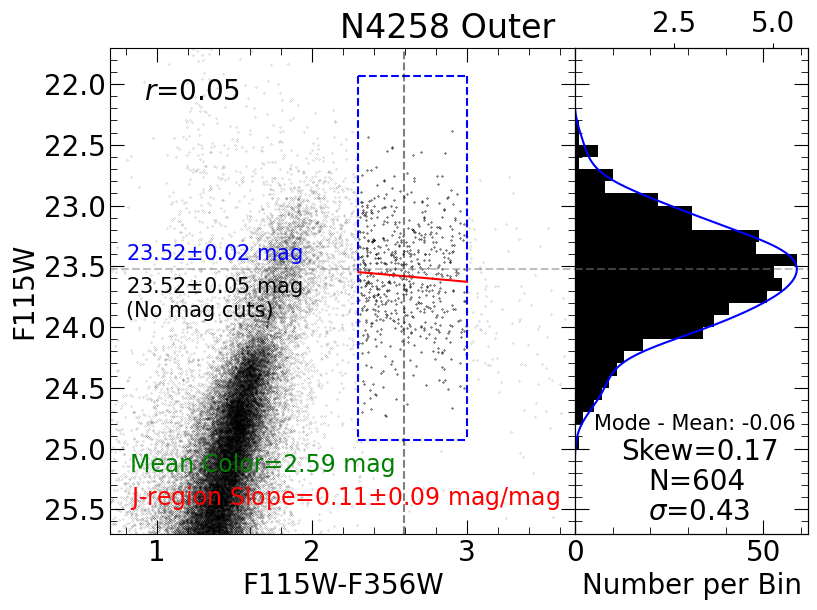} &
    \includegraphics[width=0.3\linewidth]{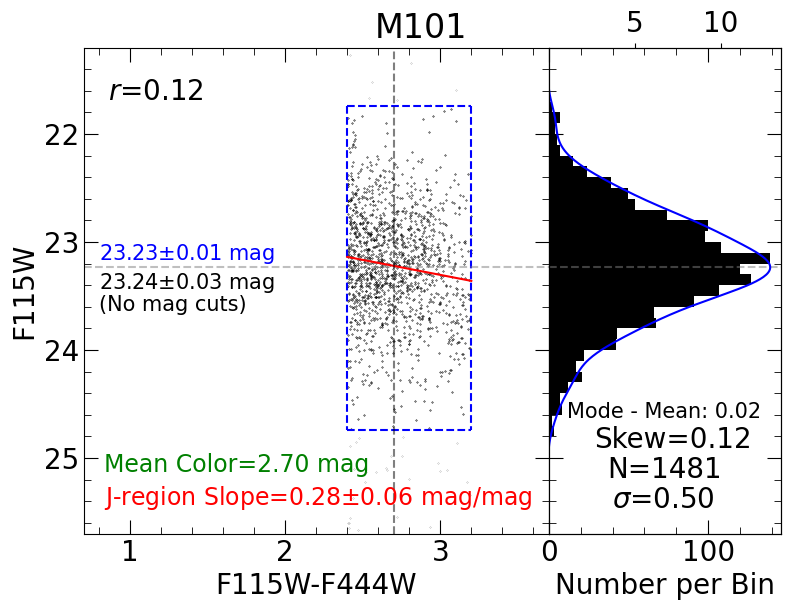} \\
    \includegraphics[width=0.3\linewidth]{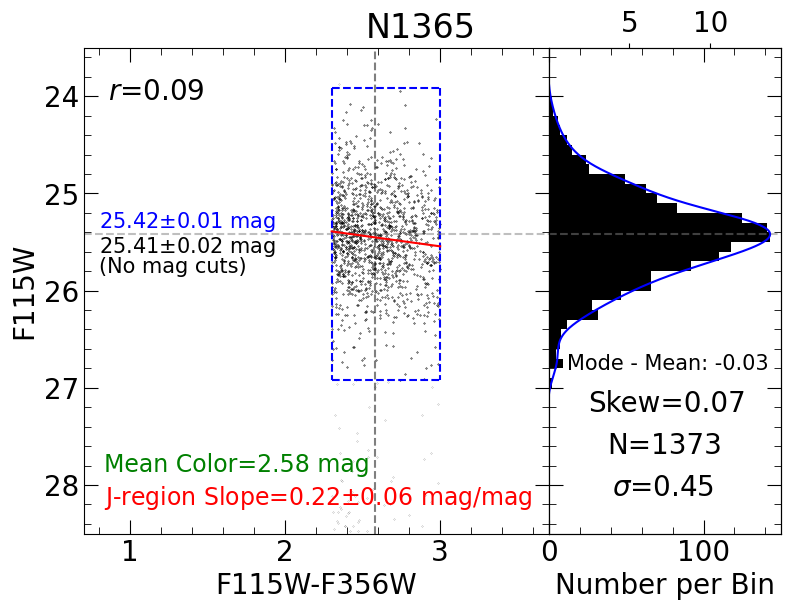} &
    \includegraphics[width=0.3\linewidth]{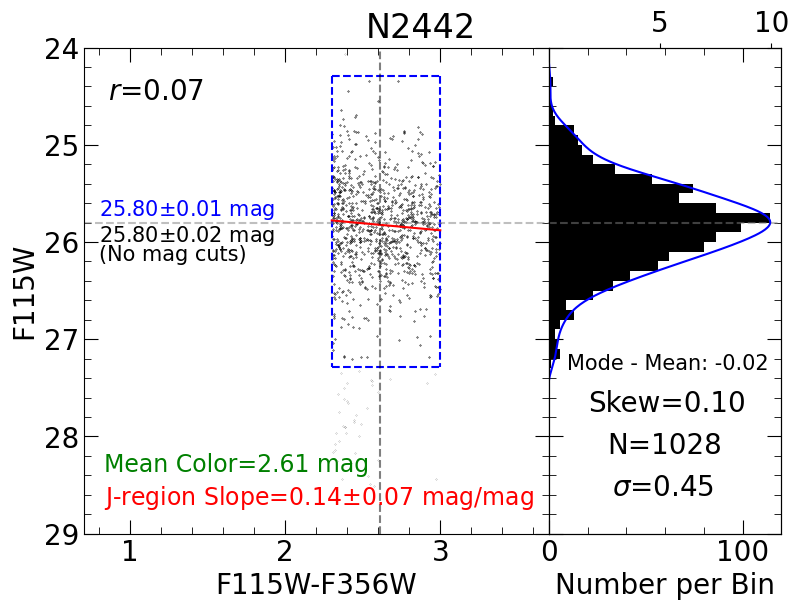} &
    \includegraphics[width=0.3\linewidth]{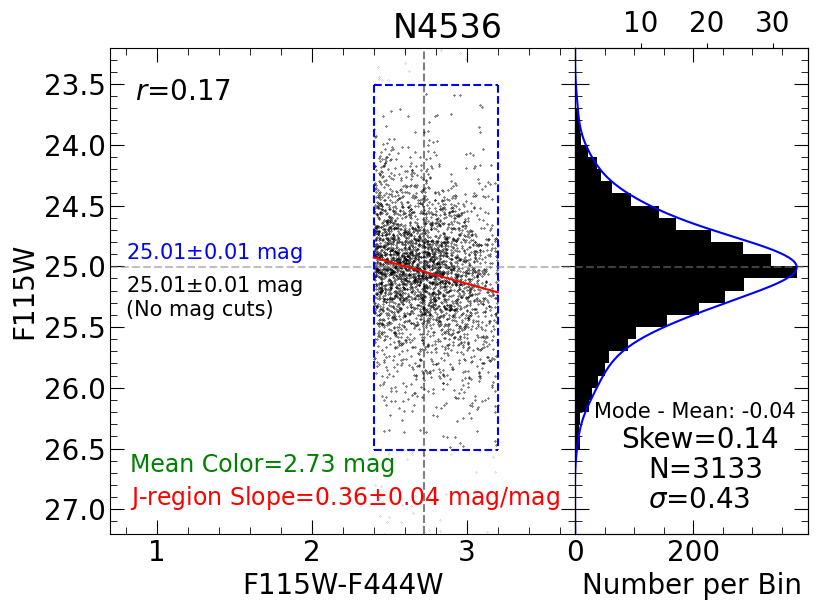} \\
    \includegraphics[width=0.3\linewidth]{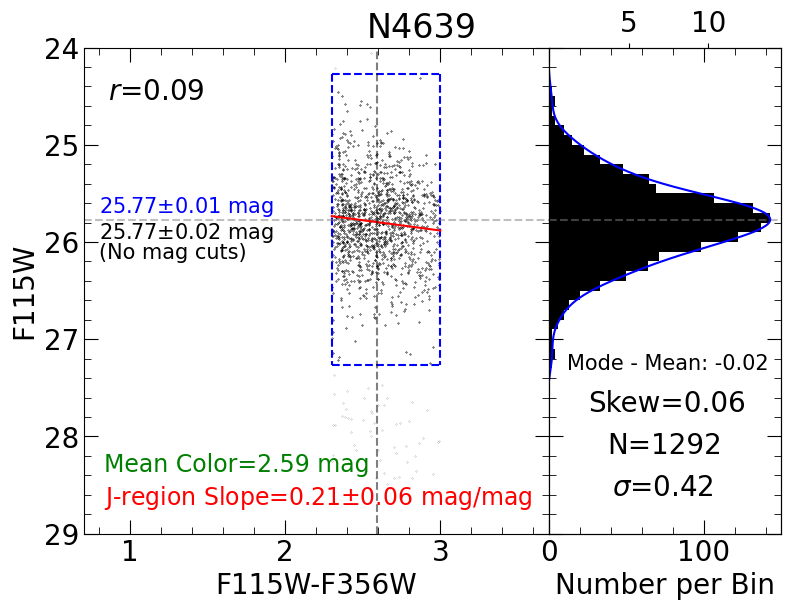} &
    \includegraphics[width=0.3\linewidth]{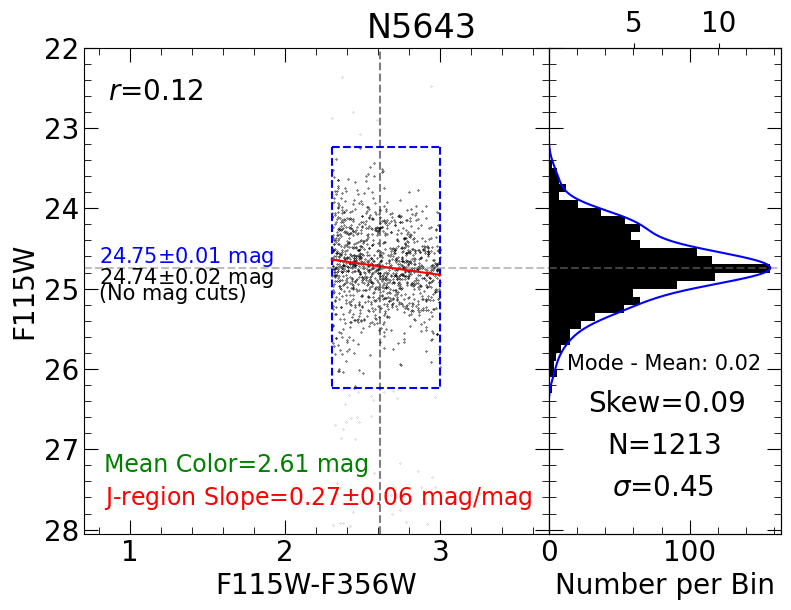} &
    \includegraphics[width=0.3\linewidth]{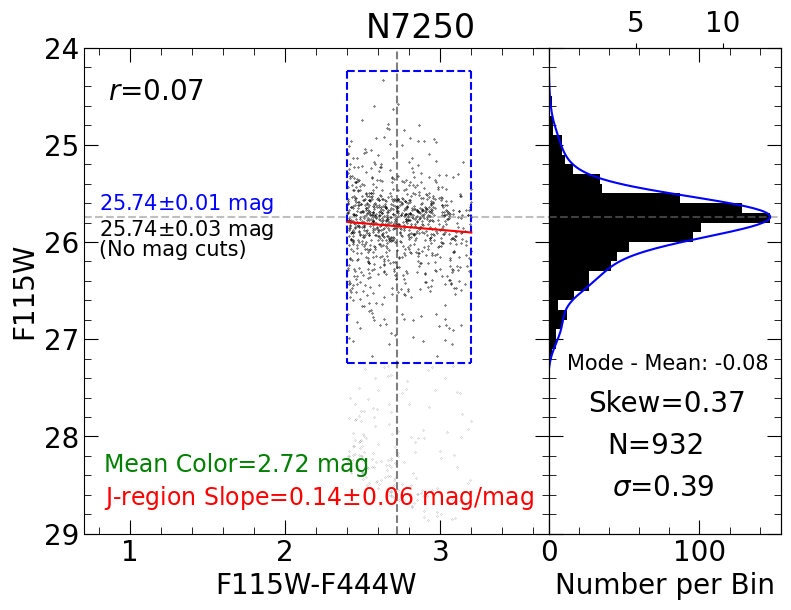}
  \end{tabular}
 {\footnotesize  \caption{Color magnitude diagrams and luminosity functions for NGC 4258 (Inner and Outer fields), M101, NGC 1365, NGC 2442, NGC 4536, NGC 4639, NGC 5643, and NGC 7250 from \cite{Lee_JWST_JAGB_H0_2024arXiv240803474L} beyond the JAGB measurement minimum radii. The left panels of each plot show the color magnitude diagrams using data as retrieved from \cite{Lee_JWST_JAGB_H0_2024arXiv240803474L} and after applying their spatial cuts and spiral arm cuts, where applicable. The blue dashed lines show color and magnitude cuts (color cuts are from \cite{Lee_JWST_JAGB_H0_2024arXiv240803474L}. The value in the upper left corner of each plot shows the Pearson correlation coefficient for the stars inside the dashed blue box. The right panel shows the luminosity function for stars inside the dashed blue box, with a GLOESS smoothed luminosity function (s=0.25~mag) shown in blue. We also list the difference between the mode (s=0.25~mag) and mean reference magnitudes, Fisher-Pearson coefficient of skewness using the Python \texttt{scipy.stats.skew}, the number (N) of stars in the dashed-blue box, and the standard deviation of the luminosity function. The blue text and horizontal dashed line shows the JAGB measured using stars inside the dashed blue lines, while the black text underneath shows the JAGB magnitude measured without magnitude cuts as done in \cite{Lee_JWST_JAGB_H0_2024arXiv240803474L}. We show and list the J-region slope in red, and the mean J-region color in green and the vertical dashed line.}}
  \label{fig:CMDs_J_band_slope}
\end{figure*}

\subsection{Measurement Techniques and Variants} \label{subsec:measurement_variants}

We plot the CMDs and J-region luminosity functions for the new GO-1685 \& GO-2875 samples in Fig. \ref{fig:CMDs}. For these galaxies, which have not yet been analyzed in the literature, we can immediately see well defined J-region `clumps' in \emph{F150W} vs. $F150W - F277W$ space. These regions are separated by the dashed blue lines that use color cuts of 1.0~mag $<$ $F150W - F277W$ $<$ 1.5~mag \citep{Li_2024ApJ...966...20L} and magnitude cuts with widths of $\pm$ $\sim$5$-$6 $\sigma$ of the luminosity function, selected to adequately encapsulate the J-region while avoiding truncating the edges of the luminosity function. We find standard deviations of the luminosity functions to be $\sim$0.4$-$0.5~mag.  Tighter magnitude cuts of $\pm$0.75~mag have been used in the literature and naturally produce a smaller standard deviation closer to 0.3~mag \citep[see cuts used to calculate standard deviations in][]{Lee_JWST_JAGB_H0_2024arXiv240803474L, Freedman_JWST_H0_2024arXiv240806153F} as it cuts off a significant number of stars off the two edges of the luminosity function. Truncating the luminosity function this way may be useful depending on the goal, but it represents a $1.5-2 \sigma$ cut which will necessarily underestimate the underlying luminosity function dispersion by removing the tails of the distribution and can underestimate the statistical uncertainty on the mean, where applicable.

\begin{figure*}[htbp]
  \centering
    \fbox{ 
    \includegraphics[width=0.7\linewidth]{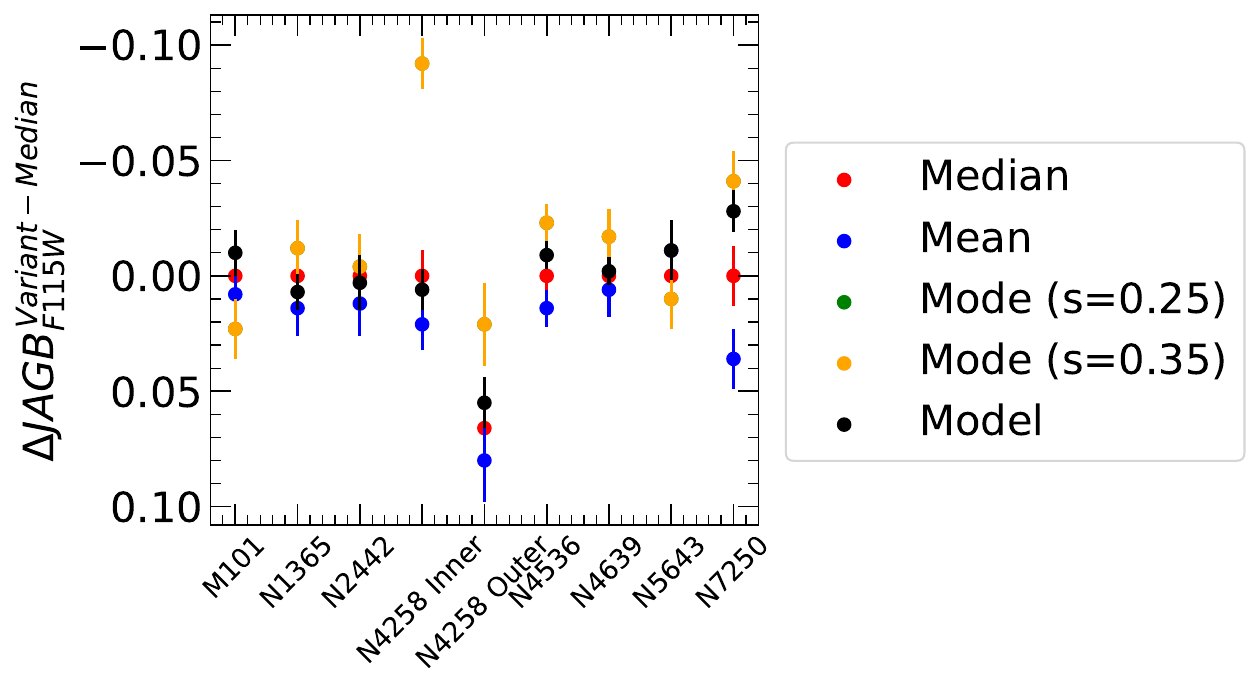}
    }
  \caption{Difference between the measured JAGB (using the measurement variants, i.e. median, clipped mean, mode (s=0.25 and 0.35~mag), and model fit) and median JAGB for the galaxies analyzed in \cite{Lee_JWST_JAGB_H0_2024arXiv240803474L}. The middle values and standard deviations across variants are listed in Table \ref{tab:CCHP_Variants}. For NGC 4258 Inner and Outer, we plot the difference for both with respect to the median measurement for NGC 4258 Inner. For all measurements, the modes with different smoothings overlap.}
  \label{fig:CCHP_variants}
\end{figure*}

For NGC 5468, we find non-negligible incompleteness at the  faint end of the luminosity function (see, for instance, at around 27th~mag in Fig. \ref{fig:CMDs}). We describe the procedure we use to correct for this and slight modifications to the measurements (i.e. binned versions of the median and mean measures) in the Appendix. We do not include NGC 3147 in this analysis due to significant incompleteness that precludes recovering the J-region luminosity function with incompleteness corrections; this is also discussed in the Appendix.

Several different methods have been used to measure the JAGB reference magnitude in the literature. These methods include the mean \citep{Madore_2020ApJ...899...66M} or median \citep{Freedman_2020ApJ...899...67F, Ripoche_2020MNRAS.495.2858R, Parada_2021MNRAS.501..933P, Parada_2023MNRAS.522..195P} magnitude of stars in the J-region luminosity function (see Fig. \ref{fig:CMDs}), mode of a binned and smoothed luminosity function \citep{Lee_M31_2023_ApJ...956...15L}, and model fit \citep{Zgirski_2021ApJ...916...19Z}. \cite{Ripoche_2020MNRAS.495.2858R, Parada_2021MNRAS.501..933P, Parada_2023MNRAS.522..195P, Li_2024ApJ...966...20L} found evidence of non-uniform asymmetry in J-region luminosity functions and demonstrated that the amount of asymmetry can vary by galaxy. This means that JAGB reference magnitudes measured using the mean, median, mode, and model fit will necessarily differ when used to measure the JAGB with an asymmetric luminosity function. In addition, the spread among these differences also varies depending on the level of asymmetry, which differs galaxy to galaxy.

To characterize the effects of this asymmetry on our results, we apply the 25 measurement variants listed in Table \ref{tab:Measurement_Variants} (five color ranges and five measures) to the samples for NGC 4258, NGC 2525, NGC 3370, and NGC 5861, after crowding corrections. These measurement variants were also used and described in \cite{Li_2024ApJ...966...20L}. We plot the measured JAGBs as a function of the 25 measurement variants for NGC 4258 and NGC 2525 in Fig. \ref{fig:Variants_N2525}, as well as those for NGC 3370, NGC 5468, and NGC 5861 in the Appendix. The differences in JAGB reference magnitudes across different measurement approaches seen in Fig. \ref{fig:Variants_N2525} originates from asymmetry of the luminosity function; if a luminosity function is symmetric, the mean will match the median and mode. If a luminosity function is asymmetric, the mean will differ from the median and mode, with the size of this difference a function of the degree of asymmetry. We observe that the spread in measured JAGB reference magnitudes, as estimated using the standard deviation across JAGBs measured using the 25 variants, differs galaxy-to-galaxy, similar to that observed for NGC 1448, NGC 1559, NGC 5584, and NGC 5643 in \cite{Li_2024ApJ...966...20L}. The different spreads in different galaxies also reflects different amounts of asymmetry, similar to that observed in the J-band in \cite{Parada_2021MNRAS.501..933P, Parada_2023MNRAS.522..195P}. We plot the JAGBs across measurement variants for NGC 4258 and NGC 2525 in Fig. \ref{fig:Variants_N2525} and the same for NGC 3370, NGC 3447, NGC 5468, NGC 5861 in the Appendix.


\subsection{Distances} \label{sec:distances}

We measure distances to NGC 2525, NGC 3370, NGC 3447, NGC 5468, and NGC 5861 by tying the JAGB measurements across variants to JAGB measurements in NGC 4258 and the maser distance from \cite{Reid_2019ApJ...886L..27R}. We plot the variants and distances for N2525 in Fig. \ref{fig:Variants_N2525}, respectively and those for NGC 3370, NGC 3447, NGC 5468, and NGC 5861 in the Appendix. We show the middle of all variants with the red lines, \emph{HST} Cepheid distance from fit variant 10 from \cite{Riess_2022ApJ...934L...7R}, also provided in the Appendix in \cite{Riess_JWST_H0_2024arXiv240811770R}, with blue lines and the \emph{JWST} TRGB distances from \cite{Li_TRGB_vs_Ceph_2024arXiv240800065L} with the orange lines. Errors are calculated by adding in quadrature the methodological variations in distance, $15\%$ of the foreground extinction in \emph{F150W} for the host galaxy and NGC 4258 \citep{Schlafly_2011ApJ...737..103S}, JAGB statistical measurement error in the host galaxy and NGC 4258, noting we leave out the 0.032~mag uncertainty from the geometric maser distance measurement to NGC 4258 \citep{Reid_2019ApJ...886L..27R} as it is common to both the Cepheid distances listed in Table \ref{tab:Distance_Comparison}. 

We also show the standard deviation among measurement variants in Fig. \ref{fig:Variants_N2525} and find partial cancellation of methodological variations along the distance ladder. In Fig. \ref{fig:CMDs}, we observe that the mode$-$mean metric all yield the same sign, indicating that the J-region luminosity functions are tilted in the same direction. Methodological variations arise due to J-region luminosity function asymmetry. Due to the relative nature of the distance ladder (i.e. host measurements are made relative to anchor measurements to derive distances), a part of this asymmetry is present in both the anchor and host luminosity function, enabling partial cancellation of the methodological variations and allowing for smaller methodological variations in the distance measurements compared those solely observed in the anchor NGC 4258. 

\section{Complete JWST JAGB Sample}

To maximize number statistics for our $H_0$ measurement, we create the largest sample of JAGB distances currently possible. \cite{Li_2024ApJ...966...20L} analyzed the JAGB in 4 hosts of 6 SNe~Ia from GO-1685 using the same measurement variants as described here. We combine that sample with the new data presented here in Table \ref{tab:Distance_Comparison}.

The Cycle 1 program from the CCHP (GO–1995), provided JAGB distances for 7 hosts of 8 SNe calibrated to NGC 4258 \citep{Lee_JWST_JAGB_H0_2024arXiv240803474L} following a specific algorithm that uses the JAGB LF mode (there is one host of 2 SNe in common, NGC 5643).  \cite{Lee_JWST_JAGB_H0_2024arXiv240803474L}  excluded 3 hosts due to lack of stability in the JAGB. To augment the JAGB sample we include the CCHP sample data but we remeasure the position of the JAGB from the published photometry to ensure consistency with our many variants approach.  We note we find good agreement when we measure the JAGB mode as with \cite{Lee_JWST_JAGB_H0_2024arXiv240803474L}.  However, we also note a significant difference at $>4 \sigma$ ($0.11 \pm 0.022$ mag) in the JAGB measure between two different fields in NGC 4258 which were averaged in \cite{Lee_JWST_JAGB_H0_2024arXiv240803474L} and indicates significant intrinsic dispersion as discussed in \S 5.2.

We show the resulting CMDs and luminosity functions for GO-1995 sample in Fig. \ref{fig:CMDs_J_band_slope}\footnote{We retrieve the data used for the analysis in \cite{Lee_JWST_JAGB_H0_2024arXiv240803474L} from their GitHub repository: \url{https://github.com/abig/lee7/CCHP-JAGB}, and apply the same radial cuts prescribed in their Section 3 as well as the spiral arm masks to NGC 4258, noting the spiral arms in the host galaxies are already masked in the data, where applicable.}. We apply the same measurement variants (i.e. median, clipped mean, mode with two smoothing values, and model fit) listed in Table \ref{tab:Measurement_Variants}. As the data provided for these host galaxies contains stars only in the J-region, we are unable to explore the full range of color cuts listed in Table \ref{tab:Measurement_Variants}, and so we conduct this test using only the baseline color range, noting that typically change in color range has yielded small differences, see for instance Fig. \ref{fig:Variants_N2525}. 

To determine a final JAGB distance we take the difference in JAGB magnitude for the SN Ia hosts and NGC 4258, for each variant (example in Figure 3 and Appendix) and take the {\it the middle value of the variants} (which we will call ``middle'').  To this quantity we then we add the maser distance, 29.398. We estimate the distance uncertainties using 

\begin{equation}
    \sigma_{dist}^2 = \sigma_{var}^2 + \sigma_{A, N4258}^2 + \sigma_{A, host}^2 + \sigma_{stat, N4258}^2 + \sigma_{stat, host}^2
\end{equation}

\noindent where $\sigma_{var}$ is the standard deviation across all measurement variants in distance space, $\sigma_{A, N4258}$ and $\sigma_{A, host}$ are the uncertainties in foreground extinction for NGC 4258 and the host galaxy, and $\sigma_{stat, N4258}$ and $\sigma_{stat, host}$ are the maximum statistical JAGB uncertainties in NGC 4258 and the host galaxy, respectively. The foreground extinction uncertainties are taken to be 15$\%$ of the extinction value in magnitudes, and the statistical uncertainty is calculated using the error on the mean. We do not include the error in the maser distance to NGC 4258 as it is common to all distances and will incorporate that error when calculating $H_0$ later. A full accounting should include the 0.032 mag uncertainty in the maser distance.  We plot the measured JAGBs across these variants in Fig. \ref{fig:CCHP_variants} and list the measured JAGB and standard deviation of the measurement variations in Table \ref{tab:CCHP_Variants}. We will incorporate these values when combining samples to measure $H_0$ in Section \ref{sec:Hubble_constant}. 

Although we find excellent agreement with the JAGB modes in \cite{Lee_JWST_JAGB_H0_2024arXiv240803474L} and those here with a mean difference $<0.01$ mag, see Table 7), the SN Ia host distances differ by a mean of 0.05 mag, a result directly tied to the inhomogeneity of the calibrating fields in NGC 4258.  The JAGB LF's in NGC 4258 are more asymmetric than in the SN hosts, which we also saw in the SH0ES program fields, \cite{Li_2024ApJ...966...20L}, leaving little doubt this is an (unfortunate) feature in the stellar population of NGC 4258.  This may be related to the fact that the JAGB measures of two fields in NGC 4258 differ at the 0.05-0.1 mag level as discussed in \S 5.2.  Because we use the middle of measures (mode,mean,median,model), our JAGB measure in NGC 4258 is $\sim$ 0.05 mag fainter than the mode which explains the mean difference in distance.  While we cannot say which measure is best (middle or mode), in the case of NGC 4258 the non-mode measures cut the field-to-field difference in half and in the case of the LMC, SMC difference where such JAGB inhomogeneity is also present, they reduce the expected difference  \cite{Li_2024ApJ...966...20L}.  The differences seen in NGC 4258 fields and between measures (until resolved) is the leading systematic error in JAGB absolute distances and the Hubble constant.



\section{The Hubble Constant} \label{sec:Hubble_constant}
\subsection{Nominal Result}
We combine the JAGB distances from the full JWST JAGB sample to calibrate the SNe~Ia magnitudes from \cite{Riess_2022ApJ...934L...7R} and measure $H_0$. We demonstrate the consistency of JAGB measurements in the \emph{F115W} (CCHP program) and \emph{F150W} (SH0ES programs) bands, given the same measurement choice, in the Appendix. As NGC 5643 was observed separately by \emph{JWST} programs GO-1685 \citep{Riess_JWST_H0_2024arXiv240811770R} and GO-1995 \citep{Freedman_JWST_H0_2024arXiv240806153F}, we take the weighted mean of JAGB distances from \cite{Li_2024ApJ...966...20L} and \cite{Lee_JWST_JAGB_H0_2024arXiv240803474L} for this measurement.

To measure $H_0$, we follow the formalism outlined in \cite{Scolnic_2023ApJ...954L..31S} and also used in \cite{Li_2024ApJ...966...20L}: we first calibrate the fiducial luminosity of the SNe~Ia in the JAGB hosts (see Table 6) using the JAGB distances and take the weighted mean resulting in $M^0_B = -19.25 \pm 0.03$~mag. We then adopt the intercept of the Hubble diagram from \cite{Riess_2022ApJ...934L...7R} of $a_B = 0.71448 \pm 0.0012$ and the equation 

\begin{equation}
    \textnormal{log} H_0 = 0.2 M^0_B + a_B + 5
\end{equation}

\noindent to find $H_0 = 73.3 \pm 1.4$ km/s/Mpc, where we calculate the error by adding in quadrature the geometric distance uncertainty to NGC 4258 (0.032~mag, \citealt{Reid_2019ApJ...886L..27R}), uncertainty in the SNe~Ia calibration (0.03~mag), and uncertainty in the Hubble diagram intercept, $a_B$ (0.006~mag). This measurement is consistent with the recent measurement from \cite{Riess_JWST_H0_2024arXiv240811770R} that uses Cepheids, TRGB, and JAGB, noting that a majority of the JAGB sample used in \cite{Riess_JWST_H0_2024arXiv240811770R} overlaps with that used here. If we had used only the mode statistic for our $H_0$ measurement, we would have found $H_0 = 71.2 \pm 1.4$ km/s/Mpc. We plot the difference between the JAGB and Cepheid distances as a function of the host SNe~Ia magnitudes in each galaxy and find no statistically significant relation.

\begin{figure*}[ht!]
  \centering
  \fbox{ 
    \begin{tabular}{c}
      \includegraphics[width=0.9\linewidth]{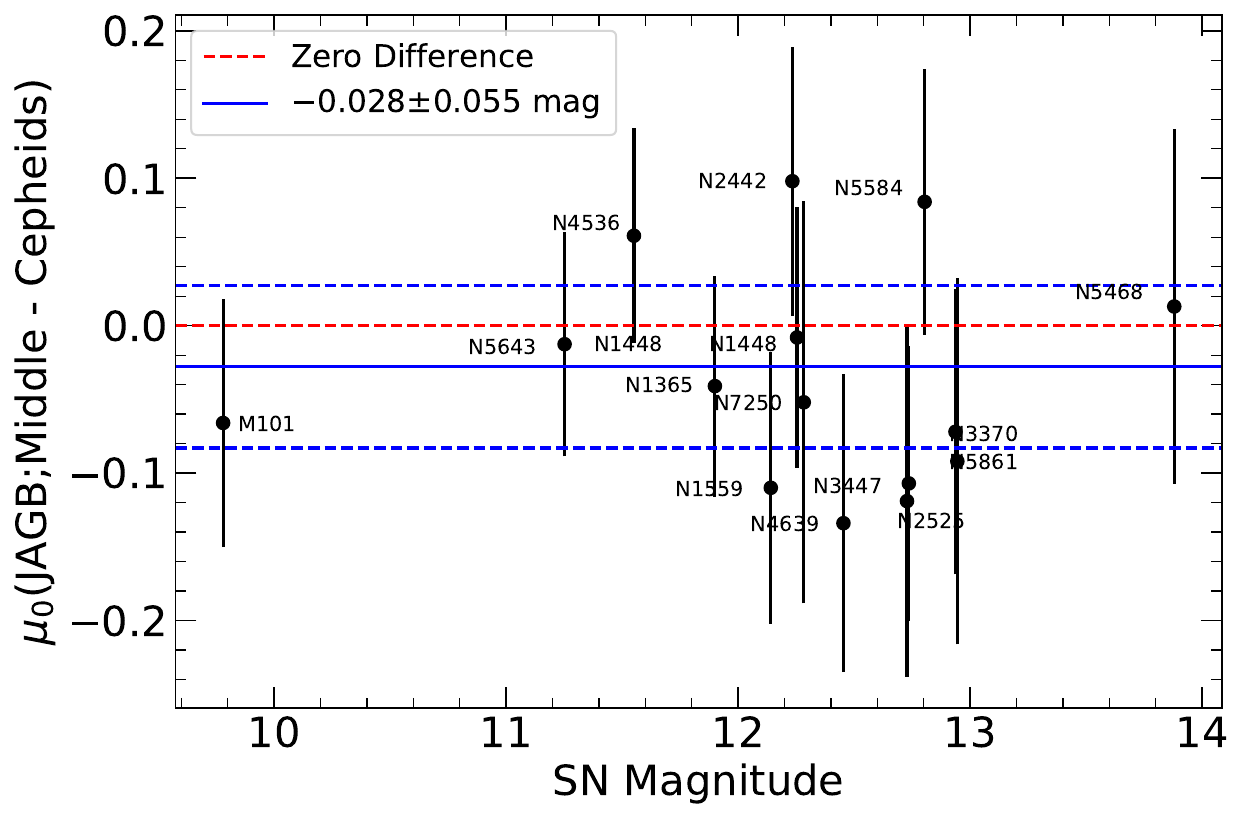} \\
      \includegraphics[width=0.9\linewidth]{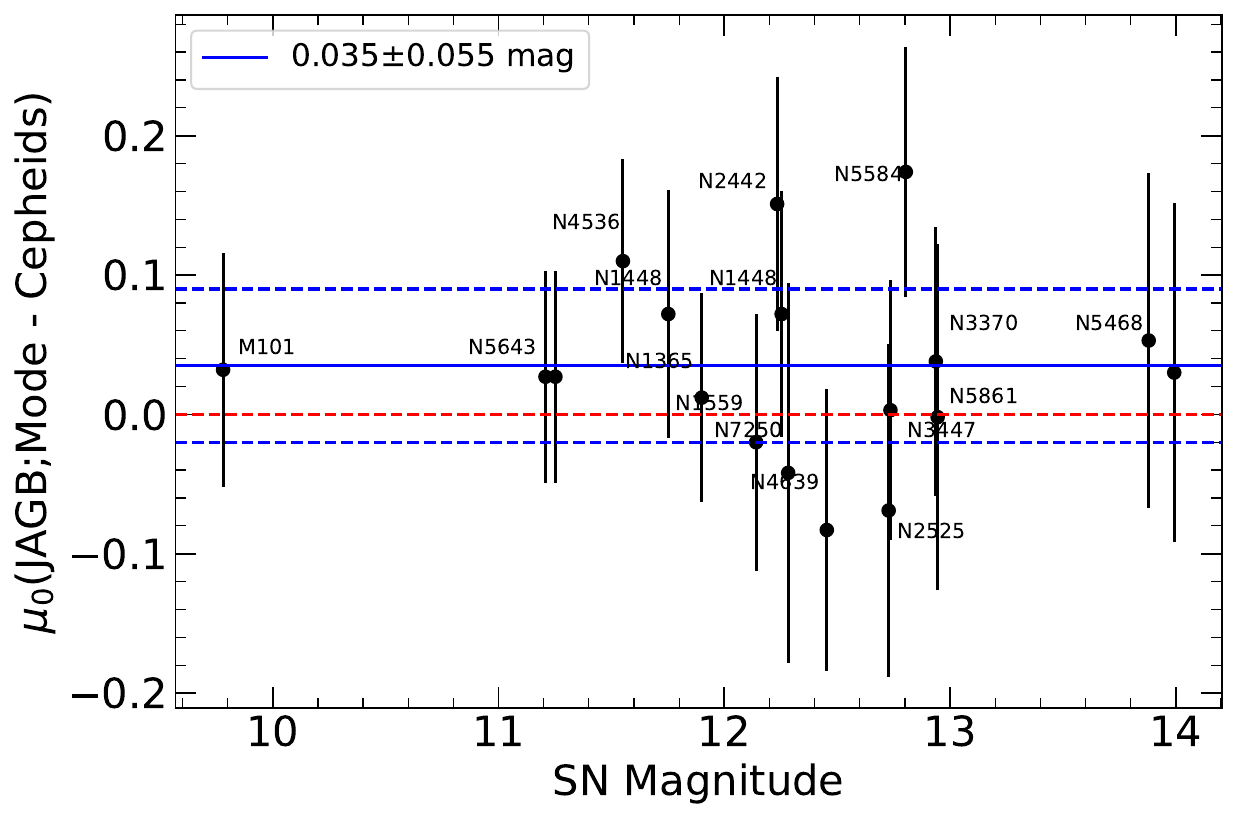} 
    \end{tabular}
  }
  \caption{Difference between the JAGB distance found here and the Cepheids distances from \cite{Riess_JWST_H0_2024arXiv240811770R} plotted as a function of host SNe~Ia magnitude. The red dashed lines are plotted at a JAGB - Cepheid difference of zero for reference, and the blue lines show the weighted mean of the JAGB and Cepheid differences. The top plot shows the differences in distance with Cepheids using the middle JAGB across variants, and the bottom plot shows the same but using the mode ($s=0.35$~mag) JAGB instead. We include the calibration error of 0.05~mag in the error on the weighted mean shown in the plot legends.}
  \label{fig:JAGB_Ceph_SNeIa}
\end{figure*}

\subsection{$H_0$ Uncertainty from Statistical Technique}

\begin{deluxetable*}{ccccc}
\caption{JAGB Measurement for CCHP Galaxies}
\label{tab:CCHP_Variants}
\tablehead{\colhead{Galaxy} & \colhead{Middle JAGB} & \colhead{$\sigma$(JAGB, Variants)} & \colhead{Mode JAGB (Here)} & \colhead{Mode JAGB (Lee+24)}}
\startdata
M101       & 23.22 & 0.01 & 23.23 & 23.24\\
NGC 1365       & 25.44 & 0.01 & 25.42 & 25.42\\
NGC 2442       & 25.81 & 0.01 & 25.80 & 25.79\\
NGC 4258 Inner & 23.47 & 0.03 & 23.41 & 23.43 (Inner+Outer)\\
NGC 4258 Outer & 23.55 & 0.03 & 23.52 & 23.43 (Inner+Outer)\\
NGC 4536       & 25.02 & 0.01 & 25.01 & 25.01\\
NGC 4639       & 25.79 & 0.01 & 25.77 & 25.77 \\
NGC 5643       & 24.74 & 0.01 & 24.75 & 24.74\\
NGC 7250       & 25.76 & 0.03 & 25.74 & 25.74\\
\enddata
\tablecomments{Middle JAGB across five measurement measures and their standard deviations for eight galaxies using the data from \cite{Lee_JWST_JAGB_H0_2024arXiv240803474L, Freedman_JWST_H0_2024arXiv240806153F}.}
\end{deluxetable*}

\begin{deluxetable*}{cccccccccc}
\caption{SNe Ia Host Galaxies and Distances}
\label{tab:Distance_Comparison}
\tablehead{\colhead{Galaxy} & \colhead{$\mu_0$(JAGB)} & \colhead{$\sigma$(JAGB)} & \colhead{$\mu_0$(Cepheids)} & \colhead{$\sigma$(Cepheids)} & \colhead{$m^0_B$(SN)} & \colhead{$\sigma$} & \colhead{SN}}
\startdata
NGC 1448 & 31.29 & 0.07 & 31.298 & 0.051 & 12.254 & 0.136 & 2001el  \\
NGC 1448 & 31.29 & 0.07 & 31.298 & 0.051 & 11.752 & 0.200 & 2021pit  \\
NGC 1559 & 31.39 & 0.06 & 31.500 & 0.071 & 12.141 & 0.086 & 2005df  \\
NGC 5584 & 31.85 & 0.07 & 31.766 & 0.062 & 12.804 & 0.079 & 2007af  \\
NGC 2525 & 31.94 & 0.06 & 32.059 & 0.105 & 12.728 & 0.074 & 2018gv  \\
NGC 3370 & 32.06 & 0.07 & 32.132 & 0.062 & 12.937 & 0.082 & 1994ae  \\
NGC 3447 & 31.84 & 0.08 & 31.947 & 0.049 & 12.736 & 0.089 & 2012ht  \\
NGC 5861 & 32.14 & 0.07 & 32.232 & 0.105 & 12.945 & 0.107 & 2017erp  \\
NGC 5468 & 33.14 & 0.09 & 33.127 & 0.082 & 13.880  & 0.080  & 1999cp   \\
NGC 5468 & 33.14 & 0.09 & 33.127 & 0.082 & 13.993 & 0.072 & 2002cr  \\
M101     & 29.12 & 0.06 & 29.188 & 0.055 & 9.780   & 0.115 & 2011fe \\
NGC 1365 & 31.34 & 0.04 & 31.378 & 0.061 & 11.900   & 0.092 & 2012fr \\
NGC 2442 & 31.56 & 0.05 & 31.459 & 0.073 & 12.234 & 0.082 & 2015F  \\
NGC 4536 & 30.93 & 0.04 & 30.870 & 0.061 & 11.551 & 0.133 & 1981B  \\
NGC 4639 & 31.69 & 0.04 & 31.823 & 0.091 & 12.454 & 0.124 & 1990N \\
NGC 5643 & 30.54 & 0.04 & 30.553 & 0.063 & 11.252 & 0.079 & 2013aa\\
NGC 5643 & 30.54 & 0.04 & 30.553 & 0.063 & 11.208 & 0.074 & 2017cbv\\
NGC 7250 & 31.59 & 0.04 & 31.642 & 0.130 & 12.283 & 0.178 & 2013dy
\enddata
\tablecomments{Middle of variants for JAGB hosts measured here as well as Cepheid and SNe~Ia information for each host galaxy for comparison . We remeasure the distances from \cite{Lee_JWST_JAGB_H0_2024arXiv240803474L, Freedman_JWST_H0_2024arXiv240806153F} using measurement variants, taking the middle JAGB out of all variants, to remain consistent with the measurements made here. The distance errors do not include 0.032~mag from the maser distance to NGC 4258 \citep{Reid_2019ApJ...886L..27R}. Distance uncertainties for the values in the second column are calculated by adding in quadrature 15\% of the uncertainties in extinction in the host galaxy and NGC 4258, standard deviation of the spread in measurement variants in distances, and maximum statistical error in JAGB measurements across variants in the host galaxy and NGC 4258. The uncertainties do not include the error in the maser distance of 0.03~mag.}
\end{deluxetable*}

We find that applying various statistical techniques like mean, median and mode can vary magnitudes by up to 0.2 mag. 
We also notice Fisher-Pearson skews, which are calculated using \texttt{scipy.stats.skew} in Python, are between 0.07 to 0.48. J-region slopes between -0.60 $\pm$ 0.13 mag/mag to 0.36 $\pm$ 0.05 mag/mag \textbf(see also \citealt{Li_2024ApJ...966...20L}). 
Here, following the approach by the Comparative Analysis of TRGBs (CATs) team \citep{Wu_2023ApJ...954...87W, Scolnic_2023ApJ...954L..31S, Li_CATs_2023ApJ...956...32L}, we pursue whether there are internal correlations between various JAGB properties that can be standardized in order to reduce sensitivity to the statistical technique.  If not, these variations will introduce inconsistencies along the distance ladder. We search in Fig. \ref{fig:JAGB_skews} for a potential relationship or metric between the J-region slopes, skew, and mode - mean, and distances using our total sample of galaxies.

\begin{figure*}[htbp]
  \centering
    \fbox{ 
    \includegraphics[width=0.9\linewidth]{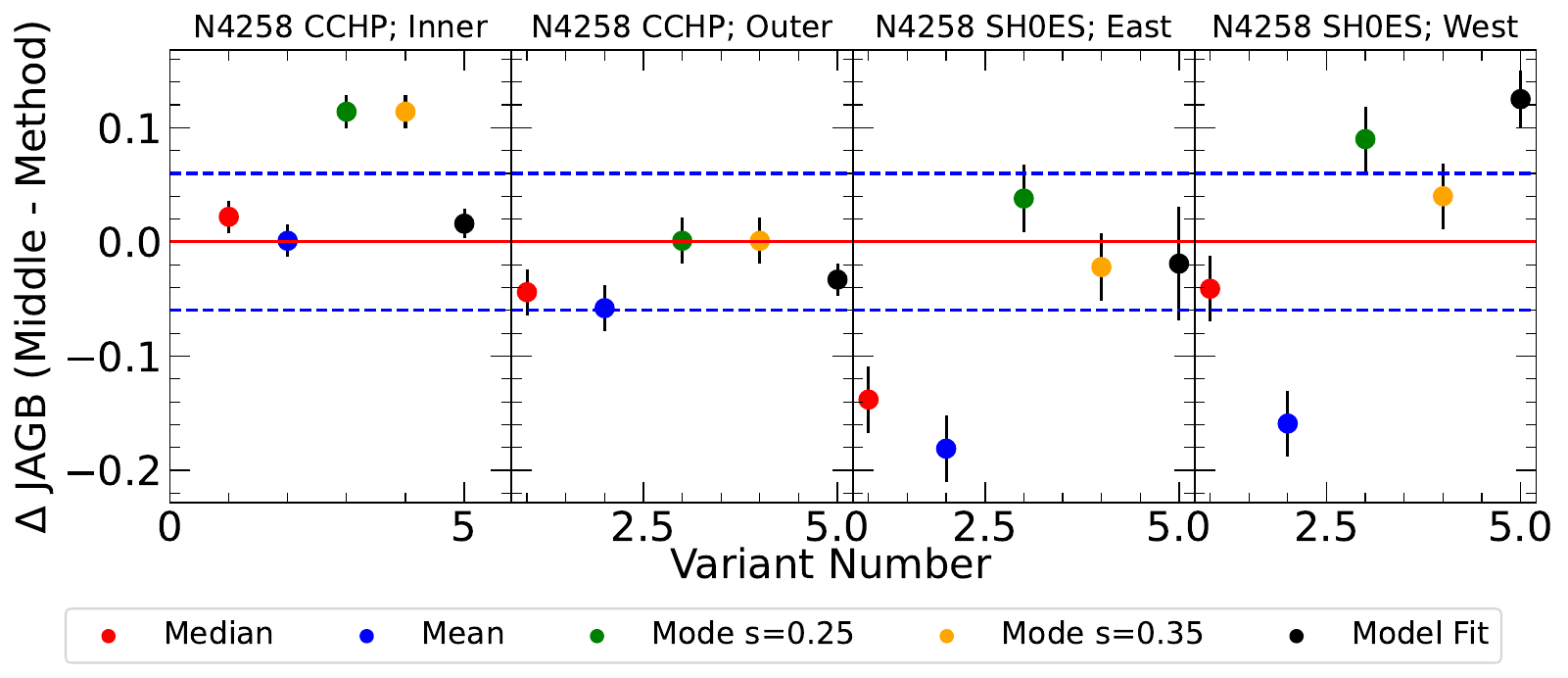}
    }
  \caption{Difference between the middle JAGB across variants for the CCHP and SH0ES observations of NGC 4258 and measurement method in the CCHP Inner and Outer, and SH0ES East and West, separately. The middle measured using after combining the Inner and Outer fields for CCHP and the East and West fields for SH0ES. We show the adopted systematic of 0.06~mag with blue dashed lines and a difference of zero in red for reference.}
  \label{fig:N4258_Variants}
\end{figure*}

\begin{figure}[ht!]
  \centering
    \fbox{ 
    \includegraphics[width=0.9\linewidth]{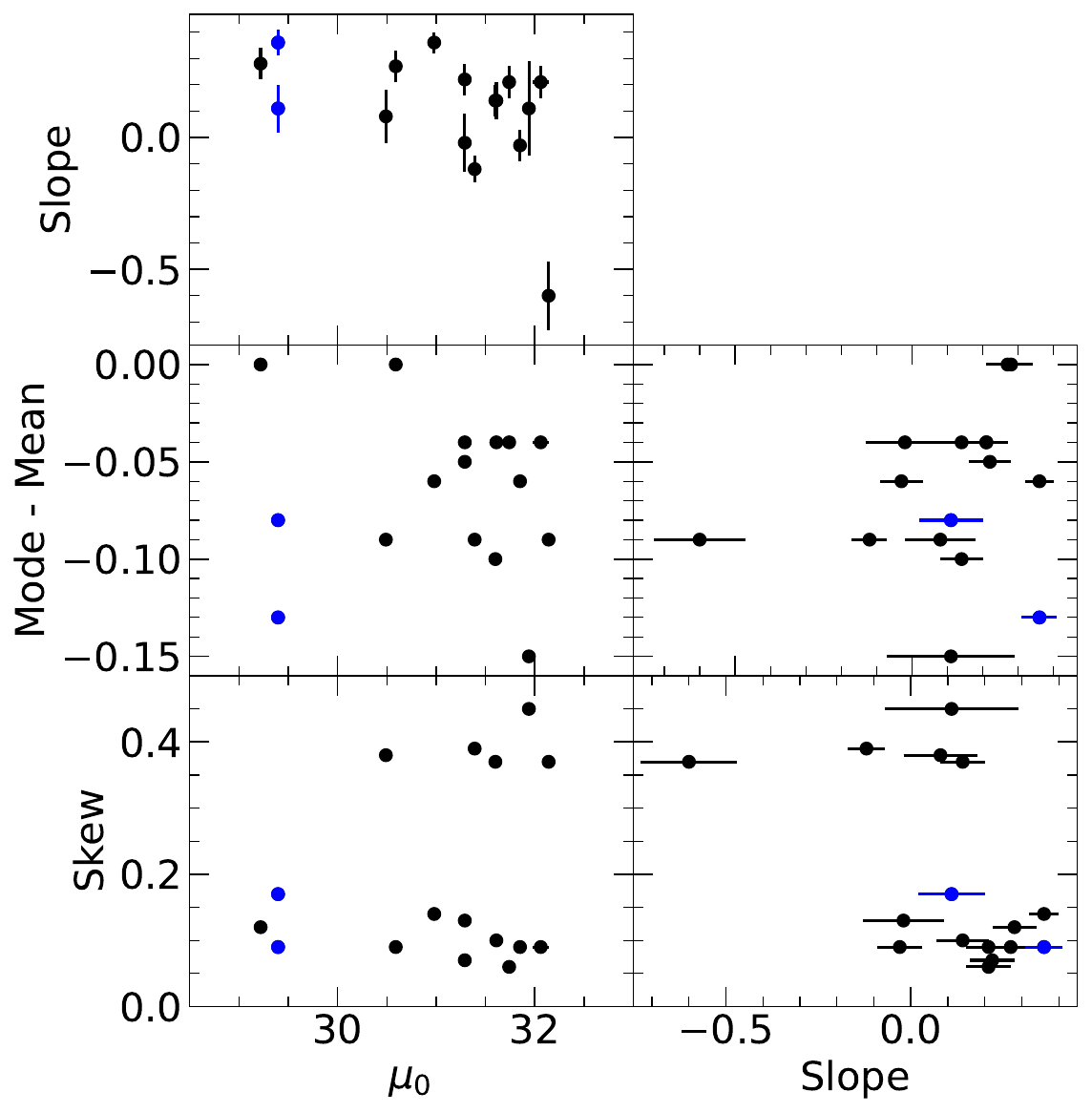} }
  \caption{Skew, mode minus mean magnitudes, and J-region slopes as a function of distances to all galaxies as well as intercomparison . The distances were compiled from \cite{Li_2024ApJ...966...20L, Lee_JWST_JAGB_H0_2024arXiv240803474L, Freedman_JWST_H0_2024arXiv240806153F}, and here.}
  \label{fig:JAGB_skews}
\end{figure}

We are unable to find a metric that can robustly standardize some of the variations seen in the J-region. This does not preclude the possibility of additional standardization, and more information such as metallicity or stellar classifications can provide more insight. The intrinsic variations in the J-region can result in inconsistencies that do not cancel along the distance ladder. We conservatively estimate a minimum uncertainty of 0.06~mag due to intrinsic variation, corresponding to roughly 2.0 km/s/Mpc. We show the variations in JAGB due to selection of different fields and methodological in Fig. \ref{fig:N4258_Variants}.

We note that there is still partial cancellation along the distance ladder.  Although NGC 4258 shows a large amount of measurement variation, possibly due to the smaller fraction of the galaxy observed compared to the host galaxies (NGC 4258 is closer to us, resulting in a larger angular size), the overlapping direction of asymmetry results in partial cancellation of this effect due to the relative nature of the distance ladder. We cannot however, further reduce the level of methodological variation, which is dominated by the asymmetry in NGC 4258, until we have more observations or more anchors. 

\section{Discussion} \label{sec:discussion}

These methodological variations, i.e. differences in JAGB among the measurement variants, are a result of asymmetry in the J-region luminosity functions. The non-uniform asymmetry observed here leads to two considerations. The first are the presence of methodological variations i.e. different measurement statistics will yield different JAGB values. The second is that the non-uniformity of the asymmetry across galaxies (i.e. the luminosity function in one galaxy has a different amount of asymmetry compared to another) prevents these variations from completely canceling across the distance ladder. If the amount of asymmetry was consistent for each galaxy, the relative nature of the distance ladder would facilitate cancellation of these effects. However, because the asymmetry is not consistent galaxy-to-galaxy, this effect does not cancel out; this is an example of a systematic uncertainty in JAGB measurements. In Section \ref{sec:systematic_uncertainties}, we discuss potential contributions to systematics in JAGB measurements.

\subsection{Systematic Uncertainties in the JAGB} \label{sec:systematic_uncertainties}

\subsubsection{Significant Difference in JAGB Between Calibrating Fields in NGC 4258}

The JAGB is most commonly measured in the outer disk where the effects of internal reddening and crowding are lessened but is still close enough to the disk that an adequate number of carbon stars exist for a robust JAGB measurement \citep[see for instance,][]{Lee_M33_2022ApJ...933..201L}. In our analysis of NGC 2525, NGC 3370, and NGC 5861, we selected our spatial cut to define the outer disk (listed in Table \ref{tab:spatial_cuts}) so that the corrected crowding biases, as determined using artificial star tests, are $\leq$0.05~mag and the region contains negligible internal reddening, as estimated using the model from \cite{Menard_2010MNRAS.405.1025M}. \cite{Lee_M31_2023_ApJ...956...15L} had proposed a method to estimate an optimal spatial cut by measuring the JAGB reference magnitude as a function of radial distance, which was improved in \cite{Lee_JWST_JAGB_H0_2024arXiv240803474L}. Their spatial cut is selected to be where the JAGB reference magnitude using stars outside a certain radius does not change significantly (i.e. the rate of change crosses zero; \citealt{Lee_JWST_JAGB_H0_2024arXiv240803474L}). We further investigate this trend. 

\begin{figure*}[ht!]
  \centering
  \begin{tabular}{cc}
    \includegraphics[width=0.55\linewidth]{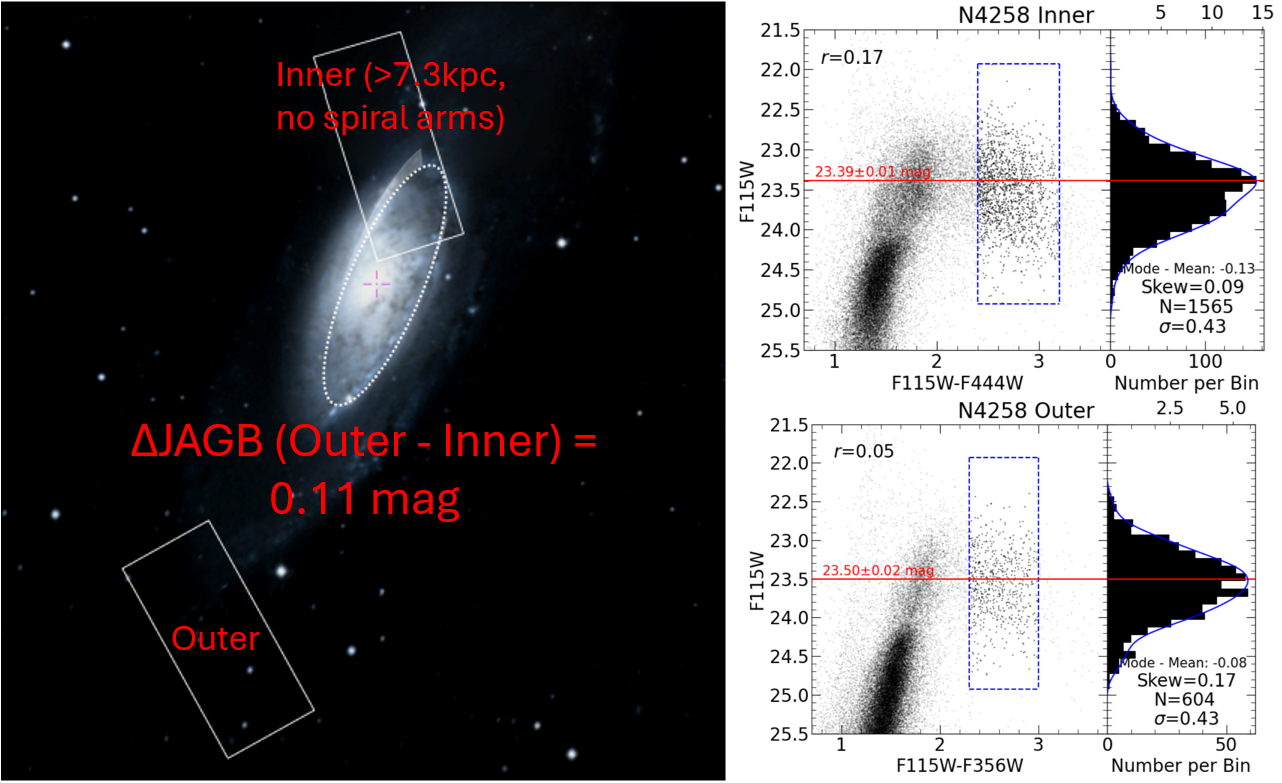} &
    \includegraphics[width=0.45\linewidth]{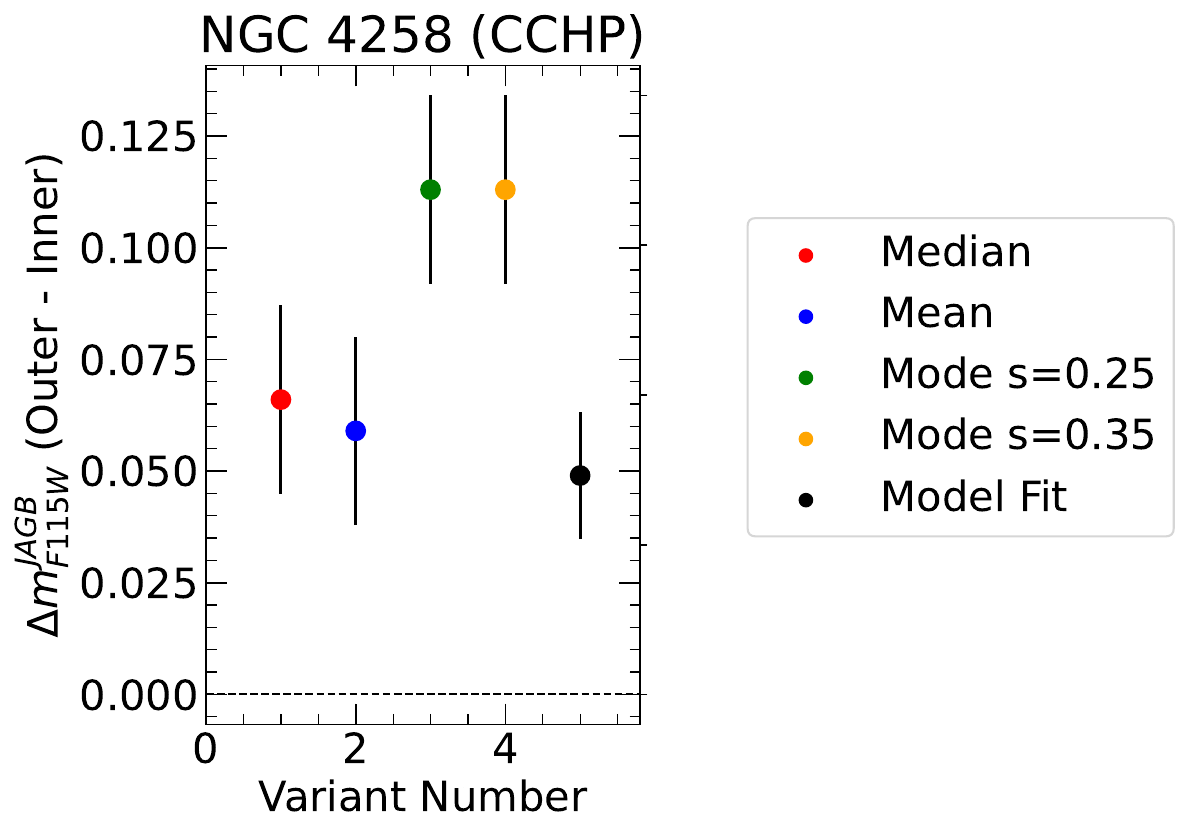} \\
  \end{tabular}
  \caption{In the left and middle subplots, CMDs and luminosity functions for the Inner and Outer fields from \cite{Lee_JWST_JAGB_H0_2024arXiv240803474L} and observations footprints taken from the same paper. For the two subplots on the right, dashed blue lines show the color and magnitude cuts used for the measurement, and the red line and text shows the mode JAGB with a smoothing value of 0.25~mag. To the right shows the luminosity functions of stars in the dashed blue lines and listed difference between the mode (s=0.25~mag), Pearson skew using Python \texttt{scipy.stats}, number of stars in side the dashed blue box, and standard deviation of the stars in the dashed blue box. The number in the upper left hand corner of the CMDs is the Pearson correlation coefficient of stars in side the dashed blue box. in the right subplots, we show the difference between the mode JAGBs in the Inner and Outer fields (using the same spatial cuts as used in \citealt{Lee_JWST_JAGB_H0_2024arXiv240803474L}) using different measurement statistics. Agreement between the two fields would be consistent with 0.00~mag.}
  \label{fig:N4258_Inner_vs_Outer}
\end{figure*}

We can investigate the radial dependence of the JAGB when we separate the Outer and Inner fields of NGC 4258 from the CCHP program. We plot the CMDs for the two fields separately in Fig. \ref{fig:N4258_Inner_vs_Outer}. \cite{Lee_JWST_JAGB_H0_2024arXiv240803474L} measure the JAGB in NGC 4258 after merging the Inner and Outer fields. We observe that the measured mode (s = 0.25~mag) JAGB for each field is brighter and fainter than the JAGB reported in \cite{Lee_JWST_JAGB_H0_2024arXiv240803474L}, suggesting that value results in that study is reflected of the mean of the two populations, weighted by the number of stars. This average from combined fields could be a more representative zero-point to use to anchor distances with, however, it is important to note that the JAGBs in the Inner and Outer fields, separately, differ significantly. The outer field JAGB is also $>4 \sigma$ from the inner field, both of which are beyond the suggested outer disk radius from \cite{Lee_JWST_JAGB_H0_2024arXiv240803474L}, indicating that the JAGB reference magnitude does not level off with radius from the galaxy.  We also show the difference between the measured JAGBs in the Inner and Outer fields for different measurement statistics in Fig. \ref{fig:N4258_Inner_vs_Outer}; we find that the mode yields a larger disagreement compared to the mean, median and model fit. This finding challenges the assumption that the JAGB reaches a consistent value beyond a certain radius.  Rather there appears to be intrinsic scatter of this value.

\subsubsection{Carbon Miras}

\begin{figure}[ht!]
  \centering
    \includegraphics[width=1\linewidth]{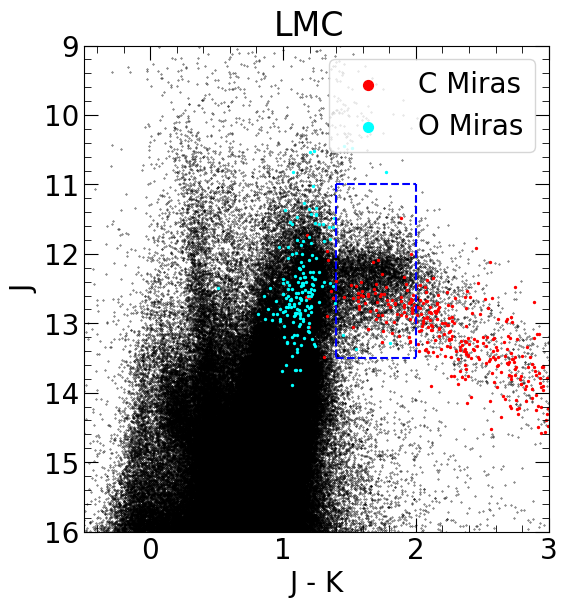}
  \caption{\emph{J} vs. \emph{J} $-$ \emph{K} color magnitude diagram using the LMC catalog from \cite{Macri_2015AJ....149..117M}. We overplot the Oxygen and Carbon Miras from \cite{Yuan_2017AJ....154..149Y} and find that a large number of Carbon Miras, which follow a period luminosity relationship instead of having constant mean magnitudes, are present in the J-region.}
  \label{fig:CMiras}
\end{figure}

J-region variation can be caused by stellar contamination, where stars that do not exhibit constant mean magnitudes are present in the J-region used for the JAGB method. An example of this was provided in \cite{Li_2024ApJ...966...20L}, who showed that extrinsic Carbon stars, which do not have masses constrained by bottom burning and dredge ups, live in the J-region. Carbon Miras are pulsating aysmpototic giant branch stars that follow a period-luminosity relationship and do not have constant mean magnitudes. Fig. 8 in \cite{Yuan_2017AJ....154..149Y} shows the period-luminosity relations of Carbon Miras in the Large Magellanic Cloud. This relationship spans several magnitudes including that overlapping with the J-region \citep[for instance, $M_J$ = $-$6.20~mag from ][]{Madore_2020ApJ...899...66M}. The presence of these Miras in the J-region will depend on their colors; Fig. 4 in \cite{Yuan_2017AJ....154..149Y} shows that Carbon Miras also span the J-region color range, with the J-region in \emph{F150W} $-$ \emph{F277W} from 1 to 1.5~mag \citep{Li_2024ApJ...966...20L}, approximating the \emph{F150W} and \emph{F277W} bands with H and K. We plot the LMC Oxygen and Carbon Miras from \cite{Yuan_2017AJ....154..149Y} together with the LMC catalog from \cite{Macri_2015AJ....149..117M} in Fig. \ref{fig:CMiras} and find that significant number of Carbon Miras are present in the J-region, as marked by the dashed blue box.  These Miras primarily occupy the fainter side of the J-region and can contribute to the J-region luminosity function asymmetry observed here and in past studies \citep{Ripoche_2020MNRAS.495.2858R, Parada_2021MNRAS.501..933P, Parada_2023MNRAS.522..195P, Li_2024ApJ...966...20L}. This plot demonstrates that the J-region, as used for the JAGB standard candle, can be contaminated with stars have variable mean magnitudes. Unless contamination can be removed from the J-region, standardization of the JAGB will be limited. It is difficult to quantify the extent of contamination into a magnitude without further observations or characterizations due to considerations of completeness as well as the possibility of the fraction of Miras varying galaxy to galaxy. However, if we assume Mira contamination contributes to the asymmetry of the LF, we can place an upper estimate of $\sim 0.1$~mag, based on the variations seen in \citep{Li_2024ApJ...966...20L} and here.

\subsubsection{J-region Slopes}

To investigate the possibility of J-region stars varying with color, we retrieve the data used by \cite{Lee_JWST_2024ApJ...961..132L} in J / \emph{F115W}-band and plot  CMDs and luminosity functions for theses data in Fig. \ref{fig:CMDs_J_band_slope}. Immediately, we can qualitatively see that many of the J-regions seem to be tilted with color, contrary to the claim of universally flat J-regions in \cite{Madore_2020ApJ...899...66M, Freedman_2020ApJ...891...57F}. We investigated this quantitatively by applying a linear fit to the stars inside the J-regions, isolated with magnitude cuts $\pm$1.5~mag around the measured JAGB, using \texttt{scipy.optimize.curve\_fit} and estimated uncertainties using the square root of the covariance matrix. We show the fits with the red line and text in Fig. \ref{fig:CMDs_J_band_slope} and find evidence of slopes of \textit{up to 7$\sigma$ significance} (see NGC 4258 Inner field), with all of the galaxies showing slopes with at least 1 $\sigma$ significance. As we are able to closely reproduce the JAGB measurements from \cite{Lee_JWST_JAGB_H0_2024arXiv240803474L}, it is unlikely that there is a significant difference in samples (e.g. a missed spatial cut, etc.) that causes this. This finding demonstrates the existence of variations in J-region stellar populations in different galaxies, as observed in the J-band, and that one cannot assume the J-regions in all galaxies are flat with color in the J-band. This also indicates that the J-band is not necessarily more advantageous over the H-band in terms of J-region tilt for JAGB measurements, noting that the H / \emph{JWST F150W} J-regions show modest variations with color (though to a lesser extent than \emph{F115W} observed here, see Appendix in \citealp{Li_2024ApJ...966...20L}). In addition, nonuniform slopes across galaxies imply that the JAGB measurement will vary due to choice of color cuts, which we account for in our analysis of measurement variants which vary the color ranges (see Table \ref{tab:Measurement_Variants}), as is done in \cite{Li_2024ApJ...966...20L}. 

We also observe that when we expand the magnitude cut from the 0.75~mag used in \cite{Lee_JWST_JAGB_H0_2024arXiv240803474L}, the standard deviation of the luminosity function increases from $\sim$0.3~mag to $\sim$0.4 to 0.5~mag, noting the narrower magnitude cuts cuts off the edges of the luminosity functions and artificially suppresses measurements of the standard deviation.

\subsubsection{Metallicities and Spectral Diversity}

The intrinsic variations in J-region characteristics (e.g. slopes and asymmetry) can also potentially be attributed to differences in stellar metallicities \citep[see for instance, ][]{Karakas_2014MNRAS.445..347K, Ripoche_2020MNRAS.495.2858R, Parada_2021MNRAS.501..933P, Parada_2023MNRAS.522..195P} and/or stellar populations. \cite{Ripoche_2020MNRAS.495.2858R} showed the presence of both N- and J-type carbon stars in the J-region (see their Fig. 10). \cite{Li_2024ApJ...966...20L} plotted the luminosity functions for N- and J-type Carbon stars from \cite{Abia_2022A&A...664A..45A} and found that they have two distinct means that are not consistent with each other, suggesting that superposition of the luminosity functions from these two types of Carbon stars can create a combined asymmetric luminosity function. \cite{Morgan_2003MNRAS.341..534M} had also noted that that different Carbon star types follow different tracks on a \emph{K} vs. \emph{J-K} CMD, with the J-types forming a sequence $\sim$0.6~mag fainter than that of N-types \citep[see Section 5 and Fig. 6 in ][]{Morgan_2003MNRAS.341..534M}. Although this comparison from \cite{Morgan_2003MNRAS.341..534M} was not in the J- or H-bands, it indicates that the different carbon star stellar types can cause photometric differences. 

The origin of this asymmetry could be attributed to differences in stellar metallicities \citep{Karakas_2014MNRAS.445..347K, Parada_2021MNRAS.501..933P, Parada_2023MNRAS.522..195P} and/or contaminating populations of J-type, extrinsic carbon stars \citep{Abia_2022A&A...664A..45A, Li_2024ApJ...966...20L}. To account for these variations in the J-region luminosity function, \cite{Parada_2021MNRAS.501..933P, Parada_2023MNRAS.522..195P} select either the Large Magellanic Cloud (LMC) or Small Magellanic Cloud (SMC) as anchor depending on if the target galaxy had a J-region luminosity function asymmetry that is more LMC- or SMC-like. These studies mark a step forward in recognizing the consequences of variations in stellar populations that may affect JAGB measurements. Here, we adopt a different but similarly conservative approach based on the same underlying principle of accounting for variations among luminosity functions from \cite{Li_2024ApJ...966...20L}. Without knowing \textit{a priori} the full characteristics of the underlying population, degree of contamination, and metallicity dependence, we cannot justify choosing one method over another when they differ applied to asymmetric luminosity functions. To characterize and account for the variation that arises from  methodological choices, we adopted several measurement variants \citep{Li_2024ApJ...966...20L} to characterize their effects on our results.

In sum, we find evidence of tilted J-regions in \emph{F115W}, noting that all of the slopes measured in Fig. \ref{fig:CMDs_J_band_slope} have greater than 1$\sigma$ significance and up 7$\sigma$ significance. Different slopes in the J-region implies that choice of JAGB color cuts can affect measurement results. We incorporate an uncertainty that can arise from this by characterizing the methodological variations, which include difference choices of color cuts. We also examine the J-region slopes and skewnesses and at the moment are unable to find a relation that could account for these variation and encourage the community to better understand and search for such a relation in the future.
\begin{figure}[htbp]
  \centering
    \includegraphics[width=1\linewidth]{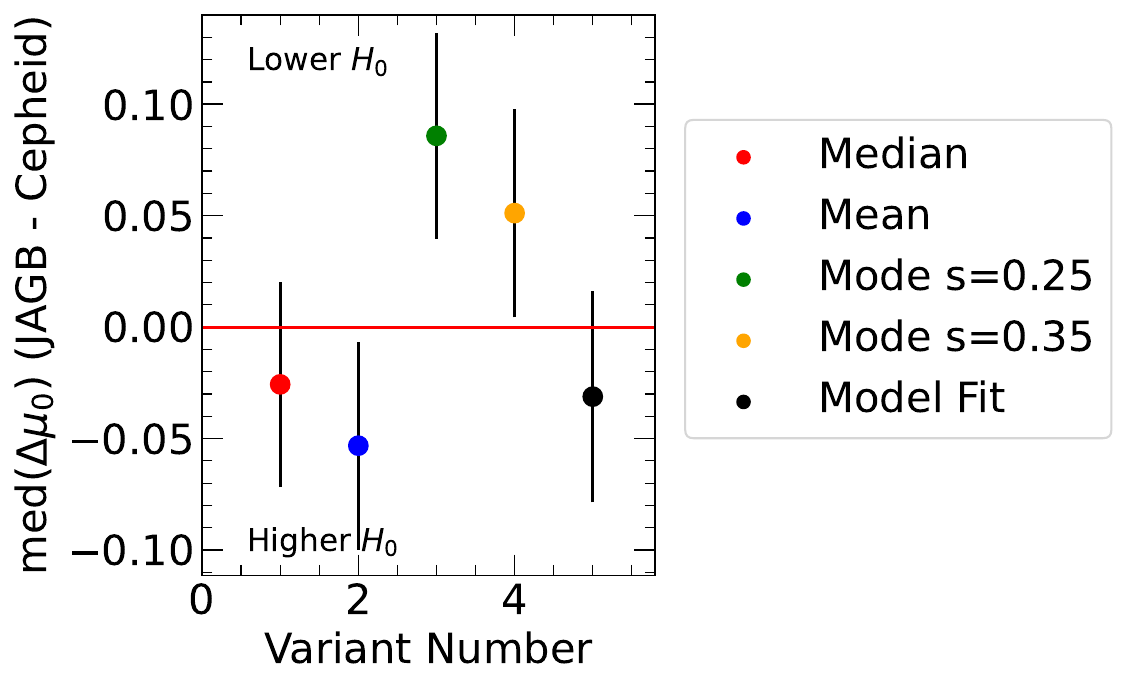}
  \caption{We first measure the differences between the JAGB and Cepheid (\cite{Riess_2022ApJ...934L...7R}; fit variant 10, also provided in the Appendix in \citealt{Riess_JWST_H0_2024arXiv240811770R}) distances for all galaxies using the five measurement measure and baseline color ranges. We then take the median of these differences. The red line is placed at a difference of 0~mag which would indicate exact agreement with Cepheid distances.}
  \label{fig:distances_relative_to_cepheids}
\end{figure}
\subsection{Comparisons with Cepheids}

\begin{figure*}[htbp]
  \centering
    \includegraphics[width=0.8\linewidth]{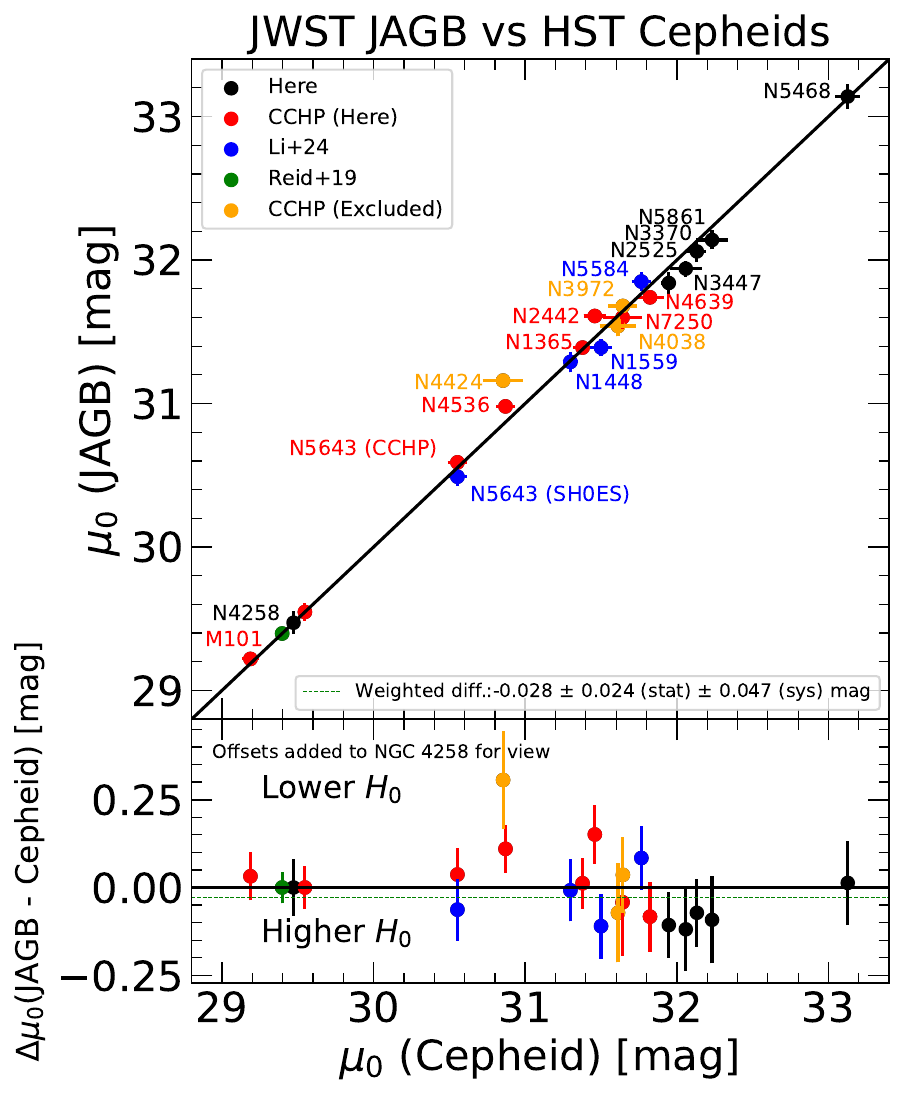}
  \caption{Comparison of \emph{JWST} JAGB distances drawn from the results here, \cite{Li_2024ApJ...966...20L}, and CCHP \citep{Lee_JWST_JAGB_H0_2024arXiv240803474L, Freedman_JWST_H0_2024arXiv240806153F} plotted against \emph{HST} Cepheid distances, fit variant 10, from \cite{Riess_2022ApJ...934L...7R}. We use fit variant 10 for this comparison to remain consistent with having NGC 4258 as the sole anchor. The top panels shows the distance moduli for JAGB and Cepheids, and the bottom plot shows the residuals. We add the maser distance to anchor galaxy NGC 4258 from \cite{Reid_2019ApJ...886L..27R} for reference. The points in orange show galaxies that were removed by \cite{Lee_JWST_JAGB_H0_2024arXiv240803474L, Freedman_JWST_H0_2024arXiv240806153F} in their analysis and so we do not consider them in the mean here. For the NGC 4258, we show three points (green, black, and red) to compare the uncertainties adopted in different studies; these are offset for viewing. For the NGC 4258 measurement uncertainty adopted here, we add include  15$\%$ foreground extinction correction, and standard deviation of variants added in quadrature. We exclude error in the maser distance.}
  \label{fig:JAGB_vs_Ceph}
\end{figure*}

We can combine the JAGB measurements relative to Cepheid measurements to examine which measurement variant yield distances, on average, closest to those of Cepheids. In Fig. \ref{fig:distances_relative_to_cepheids}, we plot the middle difference between the JAGB and Cepheid distances by measurement variant for NGC 4258, NGC 2525, NGC 3370, NGC~5468, and NGC 5861 together with the same version of the plot from \cite{Li_2024ApJ...966...20L}. We show the difference between the JAGB and Cepheid distances of 0.047$\pm$0.066~mag found in \cite{Riess_JWST_H0_2024arXiv240811770R} for comparison.

We find that for these galaxies, the mode methods yields the distance closest with those from Cepheids in the \emph{F150W}, which is similar to the comparison made in Fig. 11 from \cite{Li_2024ApJ...966...20L}. This may indicate that the mode measurement provides the more `standard' measurement, if we assume Cepheid variables are the standard reference point and are not using the JAGB to crosscheck Cepheids. It is unclear if this persists for the J-band, as shown in Fig. 13 in \cite{Li_2024ApJ...966...20L} where the mean yields the JAGB measurements between the LMC and SMC that most consistent with the geometric distances from \cite{Pietrzynski_2019Natur.567..200P, Graczyk_2020ApJ...904...13G}.

The analysis presented here increases the number of JAGB-based distance measurements to SN~Ia host galaxies to 13 \citep{Li_2024ApJ...966...20L, Lee_JWST_JAGB_H0_2024arXiv240803474L, Freedman_JWST_H0_2024arXiv240806153F}, roughly a third of the sample used to construct the \emph{HST} Cepheid-based distance ladder in \cite{Riess_2022ApJ...934L...7R}. We can compare the \emph{JWST} JAGB distances to these host galaxies with the same set having \emph{HST} Cepheid distances to test the possibility of unaccounted for systematics in the Cepheid-based measurements. We plot this comparison in Fig. \ref{fig:JAGB_vs_Ceph} and find a weighted mean difference that is consistent with zero \footnote{The claim of an 0.086 mag difference between JAGB and Cepheids from \cite{Freedman_JWST_H0_2024arXiv240806153F} resulted from a comparison of a small sample of JAGB hosts and as-of-yet unpublished Cepheid measurements, not with those we compared with from \citep{Riess_2022ApJ...934L...7R}.  Comparing the CCHP JAGB measures with the Cepheids from R22 yields an insignificant difference of 0.047 $\pm 0.066$ mag as given in \citep{Riess_JWST_H0_2024arXiv240811770R}.}
  With agreement between the primary distance indicators, it is now crucial that the largest available, complete SNe~Ia calibrator samples are used to measure $H_0$ to avoid errors due to sample variance, see \cite{Riess_JWST_H0_2024arXiv240811770R}.

\section{Conclusions}

The JAGB remains a promising method that can be used to measure extragalactic distances and crosscheck systematics in Cepheid-based distance measurements used to derive $H_0$. In this study, we combine analysis of JAGB observations across three \emph{JWST} programs to find a Hubble constant of $H_0$ = 73.3 $\pm$ 1.4 (stat) $\pm 2.0$ (sys) km/s/Mpc.  Differences in the method used to measure the JAGB reference magnitude change the value of $H_0$ by $\sim4$ km/s/Mpc. We find evidence of variations within J-regions across hosts, including a wide-range of skews,  dependencies with color, and differences between the mode and mean. We estimate an appropriate systematic is 0.06 mag, or roughly 2.0 km/s/Mpc. At the moment, combining the available JAGB distance to host galaxies provides no evidence in favor of a systematic in \emph{HST} Cepheid measurements that would resolve the Hubble Tension but also exhibits intrinsic variations that should either be accounted for in an error budget or ideally standardized.  Future observations with \emph{HST}, e.g. GO-17520 \citep{Breuval_2023hst..prop17520B}, and \emph{JWST} will improve the statistical leverage of this comparison and $H_0$ measurements.

\bibliography{sample631}{}

\begin{thebibliography}{}
\expandafter\ifx\csname natexlab\endcsname\relax\def\natexlab#1{#1}\fi
\providecommand{\url}[1]{\href{#1}{#1}}
\providecommand{\dodoi}[1]{doi:~\href{http://doi.org/#1}{\nolinkurl{#1}}}
\providecommand{\doeprint}[1]{\href{http://ascl.net/#1}{\nolinkurl{http://ascl.net/#1}}}
\providecommand{\doarXiv}[1]{\href{https://arxiv.org/abs/#1}{\nolinkurl{https://arxiv.org/abs/#1}}}

\bibitem[{{Abia} {et~al.}(2022){Abia}, {de Laverny}, {Romero-G{\'o}mez}, \& {Figueras}}]{Abia_2022A&A...664A..45A}
{Abia}, C., {de Laverny}, P., {Romero-G{\'o}mez}, M., \& {Figueras}, F. 2022, \aap, 664, A45, \dodoi{10.1051/0004-6361/202243595}

\bibitem[{{Anand} {et~al.}(2024){Anand}, {Riess}, {Yuan}, {Beaton}, {Casertano}, {Li}, {Makarov}, {Makarova}, {Tully}, {Anderson}, {Breuval}, {Dolphin}, {Karachentsev}, {Macri}, \& {Scolnic}}]{Anand_2024ApJ...966...89A}
{Anand}, G.~S., {Riess}, A.~G., {Yuan}, W., {et~al.} 2024, \apj, 966, 89, \dodoi{10.3847/1538-4357/ad2e0a}

\bibitem[{{Astropy Collaboration} {et~al.}(2013){Astropy Collaboration}, {Robitaille}, {Tollerud}, {Greenfield}, {Droettboom}, {Bray}, {Aldcroft}, {Davis}, {Ginsburg}, {Price-Whelan}, {Kerzendorf}, {Conley}, {Crighton}, {Barbary}, {Muna}, {Ferguson}, {Grollier}, {Parikh}, {Nair}, {Unther}, {Deil}, {Woillez}, {Conseil}, {Kramer}, {Turner}, {Singer}, {Fox}, {Weaver}, {Zabalza}, {Edwards}, {Azalee Bostroem}, {Burke}, {Casey}, {Crawford}, {Dencheva}, {Ely}, {Jenness}, {Labrie}, {Lim}, {Pierfederici}, {Pontzen}, {Ptak}, {Refsdal}, {Servillat}, \& {Streicher}}]{Astropy_2013A&A...558A..33A}
{Astropy Collaboration}, {Robitaille}, T.~P., {Tollerud}, E.~J., {et~al.} 2013, \aap, 558, A33, \dodoi{10.1051/0004-6361/201322068}

\bibitem[{{Astropy Collaboration} {et~al.}(2018){Astropy Collaboration}, {Price-Whelan}, {Sip{\H{o}}cz}, {G{\"u}nther}, {Lim}, {Crawford}, {Conseil}, {Shupe}, {Craig}, {Dencheva}, {Ginsburg}, {VanderPlas}, {Bradley}, {P{\'e}rez-Su{\'a}rez}, {de Val-Borro}, {Aldcroft}, {Cruz}, {Robitaille}, {Tollerud}, {Ardelean}, {Babej}, {Bach}, {Bachetti}, {Bakanov}, {Bamford}, {Barentsen}, {Barmby}, {Baumbach}, {Berry}, {Biscani}, {Boquien}, {Bostroem}, {Bouma}, {Brammer}, {Bray}, {Breytenbach}, {Buddelmeijer}, {Burke}, {Calderone}, {Cano Rodr{\'\i}guez}, {Cara}, {Cardoso}, {Cheedella}, {Copin}, {Corrales}, {Crichton}, {D'Avella}, {Deil}, {Depagne}, {Dietrich}, {Donath}, {Droettboom}, {Earl}, {Erben}, {Fabbro}, {Ferreira}, {Finethy}, {Fox}, {Garrison}, {Gibbons}, {Goldstein}, {Gommers}, {Greco}, {Greenfield}, {Groener}, {Grollier}, {Hagen}, {Hirst}, {Homeier}, {Horton}, {Hosseinzadeh}, {Hu}, {Hunkeler}, {Ivezi{\'c}}, {Jain}, {Jenness}, {Kanarek}, {Kendrew}, {Kern}, {Kerzendorf}, {Khvalko}, {King}, {Kirkby}, {Kulkarni},
  {Kumar}, {Lee}, {Lenz}, {Littlefair}, {Ma}, {Macleod}, {Mastropietro}, {McCully}, {Montagnac}, {Morris}, {Mueller}, {Mumford}, {Muna}, {Murphy}, {Nelson}, {Nguyen}, {Ninan}, {N{\"o}the}, {Ogaz}, {Oh}, {Parejko}, {Parley}, {Pascual}, {Patil}, {Patil}, {Plunkett}, {Prochaska}, {Rastogi}, {Reddy Janga}, {Sabater}, {Sakurikar}, {Seifert}, {Sherbert}, {Sherwood-Taylor}, {Shih}, {Sick}, {Silbiger}, {Singanamalla}, {Singer}, {Sladen}, {Sooley}, {Sornarajah}, {Streicher}, {Teuben}, {Thomas}, {Tremblay}, {Turner}, {Terr{\'o}n}, {van Kerkwijk}, {de la Vega}, {Watkins}, {Weaver}, {Whitmore}, {Woillez}, {Zabalza}, \& {Astropy Contributors}}]{Astropy_2018AJ....156..123A}
{Astropy Collaboration}, {Price-Whelan}, A.~M., {Sip{\H{o}}cz}, B.~M., {et~al.} 2018, \aj, 156, 123, \dodoi{10.3847/1538-3881/aabc4f}

\bibitem[{{Astropy Collaboration} {et~al.}(2022){Astropy Collaboration}, {Price-Whelan}, {Lim}, {Earl}, {Starkman}, {Bradley}, {Shupe}, {Patil}, {Corrales}, {Brasseur}, {N{\"o}the}, {Donath}, {Tollerud}, {Morris}, {Ginsburg}, {Vaher}, {Weaver}, {Tocknell}, {Jamieson}, {van Kerkwijk}, {Robitaille}, {Merry}, {Bachetti}, {G{\"u}nther}, {Aldcroft}, {Alvarado-Montes}, {Archibald}, {B{\'o}di}, {Bapat}, {Barentsen}, {Baz{\'a}n}, {Biswas}, {Boquien}, {Burke}, {Cara}, {Cara}, {Conroy}, {Conseil}, {Craig}, {Cross}, {Cruz}, {D'Eugenio}, {Dencheva}, {Devillepoix}, {Dietrich}, {Eigenbrot}, {Erben}, {Ferreira}, {Foreman-Mackey}, {Fox}, {Freij}, {Garg}, {Geda}, {Glattly}, {Gondhalekar}, {Gordon}, {Grant}, {Greenfield}, {Groener}, {Guest}, {Gurovich}, {Handberg}, {Hart}, {Hatfield-Dodds}, {Homeier}, {Hosseinzadeh}, {Jenness}, {Jones}, {Joseph}, {Kalmbach}, {Karamehmetoglu}, {Ka{\l}uszy{\'n}ski}, {Kelley}, {Kern}, {Kerzendorf}, {Koch}, {Kulumani}, {Lee}, {Ly}, {Ma}, {MacBride}, {Maljaars}, {Muna}, {Murphy}, {Norman},
  {O'Steen}, {Oman}, {Pacifici}, {Pascual}, {Pascual-Granado}, {Patil}, {Perren}, {Pickering}, {Rastogi}, {Roulston}, {Ryan}, {Rykoff}, {Sabater}, {Sakurikar}, {Salgado}, {Sanghi}, {Saunders}, {Savchenko}, {Schwardt}, {Seifert-Eckert}, {Shih}, {Jain}, {Shukla}, {Sick}, {Simpson}, {Singanamalla}, {Singer}, {Singhal}, {Sinha}, {Sip{\H{o}}cz}, {Spitler}, {Stansby}, {Streicher}, {{\v{S}}umak}, {Swinbank}, {Taranu}, {Tewary}, {Tremblay}, {de Val-Borro}, {Van Kooten}, {Vasovi{\'c}}, {Verma}, {de Miranda Cardoso}, {Williams}, {Wilson}, {Winkel}, {Wood-Vasey}, {Xue}, {Yoachim}, {Zhang}, {Zonca}, \& {Astropy Project Contributors}}]{Astropy_2022ApJ...935..167A}
{Astropy Collaboration}, {Price-Whelan}, A.~M., {Lim}, P.~L., {et~al.} 2022, \apj, 935, 167, \dodoi{10.3847/1538-4357/ac7c74}

\bibitem[{{Battinelli} \& {Demers}(2005)}]{Battinelli_2005AA...442..159B}
{Battinelli}, P., \& {Demers}, S. 2005, \aap, 442, 159, \dodoi{10.1051/0004-6361:20053357}

\bibitem[{{Breuval} {et~al.}(2023){Breuval}, {Riess}, {Anand}, {Anderson}, {Casertano}, {Huang}, {Li}, {Macri}, {Trahin}, \& {Yuan}}]{Breuval_2023hst..prop17520B}
{Breuval}, L., {Riess}, A., {Anand}, G.~S., {et~al.} 2023, {A 1\% cross-calibration of Cepheids, TRGB, and JAGB in five nearby galaxies with HST}, HST Proposal. Cycle 31, ID. \#17520

\bibitem[{{Breuval} {et~al.}(2024){Breuval}, {Riess}, {Casertano}, {Yuan}, {Macri}, {Romaniello}, {Murakami}, {Scolnic}, {Anand}, \& {Soszy{\'n}ski}}]{Breuval_2024ApJ...973...30B}
{Breuval}, L., {Riess}, A.~G., {Casertano}, S., {et~al.} 2024, \apj, 973, 30, \dodoi{10.3847/1538-4357/ad630e}

\bibitem[{{Cook} {et~al.}(1986){Cook}, {Aaronson}, \& {Norris}}]{Cook_1986ApJ...305..634C}
{Cook}, K.~H., {Aaronson}, M., \& {Norris}, J. 1986, \apj, 305, 634, \dodoi{10.1086/164277}

\bibitem[{{Di Valentino} {et~al.}(2021){Di Valentino}, {Mena}, {Pan}, {Visinelli}, {Yang}, {Melchiorri}, {Mota}, {Riess}, \& {Silk}}]{Di_valentino_2021CQGra..38o3001D}
{Di Valentino}, E., {Mena}, O., {Pan}, S., {et~al.} 2021, Classical and Quantum Gravity, 38, 153001, \dodoi{10.1088/1361-6382/ac086d}

\bibitem[{{Dolphin}(2016)}]{Dolphin_2016ascl.soft08013D}
{Dolphin}, A. 2016, {DOLPHOT: Stellar photometry}, Astrophysics Source Code Library, record ascl:1608.013.
\newblock \doeprint{1608.013}

\bibitem[{{Dolphin}(2000)}]{Dolphin_2000PASP..112.1383D}
{Dolphin}, A.~E. 2000, \pasp, 112, 1383, \dodoi{10.1086/316630}

\bibitem[{{Dolphin}(2002)}]{Dolphin_2002MNRAS.332...91D}
---. 2002, \mnras, 332, 91, \dodoi{10.1046/j.1365-8711.2002.05271.x}

\bibitem[{{Fitzpatrick}(1999)}]{Fitzpatrick_1999PASP..111...63F}
{Fitzpatrick}, E.~L. 1999, \pasp, 111, 63, \dodoi{10.1086/316293}

\bibitem[{{Freedman} \& {Madore}(2020)}]{Freedman_2020ApJ...899...67F}
{Freedman}, W.~L., \& {Madore}, B.~F. 2020, \apj, 899, 67, \dodoi{10.3847/1538-4357/aba9d8}

\bibitem[{{Freedman} {et~al.}(2021){Freedman}, {Madore}, {Hoyt}, {Jang}, {Lee}, \& {Owens}}]{Freedman_2021jwst.prop.1995F}
{Freedman}, W.~L., {Madore}, B.~F., {Hoyt}, T., {et~al.} 2021, {Answering the Most Important Problem in Cosmology Today: Is the Tension in the Hubble Constant Real?}, JWST Proposal. Cycle 1, ID. \#1995

\bibitem[{{Freedman} {et~al.}(2024){Freedman}, {Madore}, {Jang}, {Hoyt}, {Lee}, \& {Owens}}]{Freedman_JWST_H0_2024arXiv240806153F}
{Freedman}, W.~L., {Madore}, B.~F., {Jang}, I.~S., {et~al.} 2024, arXiv e-prints, arXiv:2408.06153.
\newblock \doarXiv{2408.06153}

\bibitem[{{Freedman} {et~al.}(2020){Freedman}, {Madore}, {Hoyt}, {Jang}, {Beaton}, {Lee}, {Monson}, {Neeley}, \& {Rich}}]{Freedman_2020ApJ...891...57F}
{Freedman}, W.~L., {Madore}, B.~F., {Hoyt}, T., {et~al.} 2020, \apj, 891, 57, \dodoi{10.3847/1538-4357/ab7339}

\bibitem[{{Graczyk} {et~al.}(2020){Graczyk}, {Pietrzy{\'n}ski}, {Thompson}, {Gieren}, {Zgirski}, {Villanova}, {G{\'o}rski}, {Wielg{\'o}rski}, {Karczmarek}, {Narloch}, {Pilecki}, {Taormina}, {Smolec}, {Suchomska}, {Gallenne}, {Nardetto}, {Storm}, {Kudritzki}, {Ka{\l}uszy{\'n}ski}, \& {Pych}}]{Graczyk_2020ApJ...904...13G}
{Graczyk}, D., {Pietrzy{\'n}ski}, G., {Thompson}, I.~B., {et~al.} 2020, \apj, 904, 13, \dodoi{10.3847/1538-4357/abbb2b}

\bibitem[{{Karakas}(2014)}]{Karakas_2014MNRAS.445..347K}
{Karakas}, A.~I. 2014, \mnras, 445, 347, \dodoi{10.1093/mnras/stu1727}

\bibitem[{{Lee}(2023)}]{Lee_M31_2023_ApJ...956...15L}
{Lee}, A.~J. 2023, \apj, 956, 15, \dodoi{10.3847/1538-4357/acee69}

\bibitem[{{Lee} {et~al.}(2024{\natexlab{a}}){Lee}, {Freedman}, {Jang}, {Madore}, \& {Owens}}]{Lee_JWST_2024ApJ...961..132L}
{Lee}, A.~J., {Freedman}, W.~L., {Jang}, I.~S., {Madore}, B.~F., \& {Owens}, K.~A. 2024{\natexlab{a}}, \apj, 961, 132, \dodoi{10.3847/1538-4357/ad12c7}

\bibitem[{{Lee} {et~al.}(2024{\natexlab{b}}){Lee}, {Freedman}, {Madore}, {Jang}, {Owens}, \& {Hoyt}}]{Lee_JWST_JAGB_H0_2024arXiv240803474L}
{Lee}, A.~J., {Freedman}, W.~L., {Madore}, B.~F., {et~al.} 2024{\natexlab{b}}, arXiv e-prints, arXiv:2408.03474, \dodoi{10.48550/arXiv.2408.03474}

\bibitem[{{Lee} {et~al.}(2021{\natexlab{a}}){Lee}, {Freedman}, {Madore}, {Owens}, {Monson}, \& {Hoyt}}]{Lee_WLM_2021ApJ...907..112L}
---. 2021{\natexlab{a}}, \apj, 907, 112, \dodoi{10.3847/1538-4357/abd253}

\bibitem[{{Lee} {et~al.}(2021{\natexlab{b}}){Lee}, {Freedman}, {Madore}, {Owens}, \& {Sung Jang}}]{Lee_MW_2021ApJ...923..157L}
{Lee}, A.~J., {Freedman}, W.~L., {Madore}, B.~F., {Owens}, K.~A., \& {Sung Jang}, I. 2021{\natexlab{b}}, \apj, 923, 157, \dodoi{10.3847/1538-4357/ac2f4c}

\bibitem[{{Lee} {et~al.}(2022){Lee}, {Rousseau-Nepton}, {Freedman}, {Madore}, {Cioni}, {Hoyt}, {Jang}, {Javadi}, \& {Owens}}]{Lee_M33_2022ApJ...933..201L}
{Lee}, A.~J., {Rousseau-Nepton}, L., {Freedman}, W.~L., {et~al.} 2022, \apj, 933, 201, \dodoi{10.3847/1538-4357/ac7321}

\bibitem[{{Lee} {et~al.}(2024{\natexlab{c}}){Lee}, {Monson}, {Freedman}, {Madore}, {Owens}, {Beaton}, {Espinoza}, {Ren}, \& {Ren}}]{Lee_Magellan_2024ApJ...967...22L}
{Lee}, A.~J., {Monson}, A.~J., {Freedman}, W.~L., {et~al.} 2024{\natexlab{c}}, \apj, 967, 22, \dodoi{10.3847/1538-4357/ad32c7}

\bibitem[{{Li} {et~al.}(2024{\natexlab{a}}){Li}, {Riess}, {Casertano}, {Anand}, {Scolnic}, {Yuan}, {Breuval}, \& {Huang}}]{Li_2024ApJ...966...20L}
{Li}, S., {Riess}, A.~G., {Casertano}, S., {et~al.} 2024{\natexlab{a}}, \apj, 966, 20, \dodoi{10.3847/1538-4357/ad2f2b}

\bibitem[{{Li} {et~al.}(2023){Li}, {Riess}, {Scolnic}, {Anand}, {Wu}, {Casertano}, {Yuan}, {Beaton}, \& {Anderson}}]{Li_CATs_2023ApJ...956...32L}
{Li}, S., {Riess}, A.~G., {Scolnic}, D., {et~al.} 2023, \apj, 956, 32, \dodoi{10.3847/1538-4357/acf4fb}

\bibitem[{{Li} {et~al.}(2024{\natexlab{b}}){Li}, {Anand}, {Riess}, {Casertano}, {Yuan}, {Breuval}, {Macri}, {Scolnic}, {Beaton}, \& {Anderson}}]{Li_TRGB_vs_Ceph_2024arXiv240800065L}
{Li}, S., {Anand}, G.~S., {Riess}, A.~G., {et~al.} 2024{\natexlab{b}}, arXiv e-prints, arXiv:2408.00065, \dodoi{10.48550/arXiv.2408.00065}

\bibitem[{{Macri} {et~al.}(2015){Macri}, {Ngeow}, {Kanbur}, {Mahzooni}, \& {Smitka}}]{Macri_2015AJ....149..117M}
{Macri}, L.~M., {Ngeow}, C.-C., {Kanbur}, S.~M., {Mahzooni}, S., \& {Smitka}, M.~T. 2015, \aj, 149, 117, \dodoi{10.1088/0004-6256/149/4/117}

\bibitem[{{Madore} \& {Freedman}(2020)}]{Madore_2020ApJ...899...66M}
{Madore}, B.~F., \& {Freedman}, W.~L. 2020, \apj, 899, 66, \dodoi{10.3847/1538-4357/aba045}

\bibitem[{{Madore} {et~al.}(2022){Madore}, {Freedman}, {Lee}, \& {Owens}}]{Madore_MW_2022ApJ...938..125M}
{Madore}, B.~F., {Freedman}, W.~L., {Lee}, A.~J., \& {Owens}, K. 2022, \apj, 938, 125, \dodoi{10.3847/1538-4357/ac92fd}

\bibitem[{{M{\'e}nard} {et~al.}(2010){M{\'e}nard}, {Scranton}, {Fukugita}, \& {Richards}}]{Menard_2010MNRAS.405.1025M}
{M{\'e}nard}, B., {Scranton}, R., {Fukugita}, M., \& {Richards}, G. 2010, \mnras, 405, 1025, \dodoi{10.1111/j.1365-2966.2010.16486.x}

\bibitem[{{Morgan} {et~al.}(2003){Morgan}, {Cannon}, {Hatzidimitriou}, \& {Croke}}]{Morgan_2003MNRAS.341..534M}
{Morgan}, D.~H., {Cannon}, R.~D., {Hatzidimitriou}, D., \& {Croke}, B.~F.~W. 2003, \mnras, 341, 534, \dodoi{10.1046/j.1365-8711.2003.06424.x}

\bibitem[{{Murakami} {et~al.}(2023){Murakami}, {Riess}, {Stahl}, {D'Arcy Kenworthy}, {Pluck}, {Macoretta}, {Brout}, {Jones}, {Scolnic}, \& {Filippenko}}]{Murakami_2023JCAP...11..046M}
{Murakami}, Y.~S., {Riess}, A.~G., {Stahl}, B.~E., {et~al.} 2023, \jcap, 2023, 046, \dodoi{10.1088/1475-7516/2023/11/046}

\bibitem[{{Parada} {et~al.}(2021){Parada}, {Heyl}, {Richer}, {Ripoche}, \& {Rousseau-Nepton}}]{Parada_2021MNRAS.501..933P}
{Parada}, J., {Heyl}, J., {Richer}, H., {Ripoche}, P., \& {Rousseau-Nepton}, L. 2021, \mnras, 501, 933, \dodoi{10.1093/mnras/staa3750}

\bibitem[{{Parada} {et~al.}(2023){Parada}, {Heyl}, {Richer}, {Ripoche}, \& {Rousseau-Nepton}}]{Parada_2023MNRAS.522..195P}
---. 2023, \mnras, 522, 195, \dodoi{10.1093/mnras/stad965}

\bibitem[{{Pietrzy{\'n}ski} {et~al.}(2019){Pietrzy{\'n}ski}, {Graczyk}, {Gallenne}, {Gieren}, {Thompson}, {Pilecki}, {Karczmarek}, {G{\'o}rski}, {Suchomska}, {Taormina}, {Zgirski}, {Wielg{\'o}rski}, {Ko{\l}aczkowski}, {Konorski}, {Villanova}, {Nardetto}, {Kervella}, {Bresolin}, {Kudritzki}, {Storm}, {Smolec}, \& {Narloch}}]{Pietrzynski_2019Natur.567..200P}
{Pietrzy{\'n}ski}, G., {Graczyk}, D., {Gallenne}, A., {et~al.} 2019, \nat, 567, 200, \dodoi{10.1038/s41586-019-0999-4}

\bibitem[{{Planck Collaboration} {et~al.}(2020){Planck Collaboration}, {Aghanim}, {Akrami}, {Ashdown}, {Aumont}, {Baccigalupi}, {Ballardini}, {Banday}, {Barreiro}, {Bartolo}, {Basak}, {Battye}, {Benabed}, {Bernard}, {Bersanelli}, {Bielewicz}, {Bock}, {Bond}, {Borrill}, {Bouchet}, {Boulanger}, {Bucher}, {Burigana}, {Butler}, {Calabrese}, {Cardoso}, {Carron}, {Challinor}, {Chiang}, {Chluba}, {Colombo}, {Combet}, {Contreras}, {Crill}, {Cuttaia}, {de Bernardis}, {de Zotti}, {Delabrouille}, {Delouis}, {Di Valentino}, {Diego}, {Dor{\'e}}, {Douspis}, {Ducout}, {Dupac}, {Dusini}, {Efstathiou}, {Elsner}, {En{\ss}lin}, {Eriksen}, {Fantaye}, {Farhang}, {Fergusson}, {Fernandez-Cobos}, {Finelli}, {Forastieri}, {Frailis}, {Fraisse}, {Franceschi}, {Frolov}, {Galeotta}, {Galli}, {Ganga}, {G{\'e}nova-Santos}, {Gerbino}, {Ghosh}, {Gonz{\'a}lez-Nuevo}, {G{\'o}rski}, {Gratton}, {Gruppuso}, {Gudmundsson}, {Hamann}, {Handley}, {Hansen}, {Herranz}, {Hildebrandt}, {Hivon}, {Huang}, {Jaffe}, {Jones}, {Karakci}, {Keih{\"a}nen},
  {Keskitalo}, {Kiiveri}, {Kim}, {Kisner}, {Knox}, {Krachmalnicoff}, {Kunz}, {Kurki-Suonio}, {Lagache}, {Lamarre}, {Lasenby}, {Lattanzi}, {Lawrence}, {Le Jeune}, {Lemos}, {Lesgourgues}, {Levrier}, {Lewis}, {Liguori}, {Lilje}, {Lilley}, {Lindholm}, {L{\'o}pez-Caniego}, {Lubin}, {Ma}, {Mac{\'\i}as-P{\'e}rez}, {Maggio}, {Maino}, {Mandolesi}, {Mangilli}, {Marcos-Caballero}, {Maris}, {Martin}, {Martinelli}, {Mart{\'\i}nez-Gonz{\'a}lez}, {Matarrese}, {Mauri}, {McEwen}, {Meinhold}, {Melchiorri}, {Mennella}, {Migliaccio}, {Millea}, {Mitra}, {Miville-Desch{\^e}nes}, {Molinari}, {Montier}, {Morgante}, {Moss}, {Natoli}, {N{\o}rgaard-Nielsen}, {Pagano}, {Paoletti}, {Partridge}, {Patanchon}, {Peiris}, {Perrotta}, {Pettorino}, {Piacen tini}, {Polastri}, {Polenta}, {Puget}, {Rachen}, {Reinecke}, {Remazeilles}, {Renzi}, {Rocha}, {Rosset}, {Roudier}, {Rubi{\~n}o-Mart{\'\i}n}, {Ruiz-Granados}, {Salvati}, {Sandri}, {Savelainen}, {Scott}, {Shellard}, {Sirignano}, {Sirri}, {Spencer}, {Sunyaev}, {Suur-Uski}, {Tauber},
  {Tavagnacco}, {Tenti}, {Toffolatti}, {Tomasi}, {Trombetti}, {Valenziano}, {Valiviita}, {Van Tent}, {Vibert}, {Vielva}, {Villa}, {Vittorio}, {Wandelt}, {Wehus}, {White}, {White}, {Zacchei}, \& {Zonca}}]{Planck_2020A&A...641A...6P}
{Planck Collaboration}, {Aghanim}, N., {Akrami}, Y., {et~al.} 2020, \aap, 641, A6, \dodoi{10.1051/0004-6361/201833910}

\bibitem[{{Pritchet} {et~al.}(1987){Pritchet}, {Richer}, {Schade}, {Crabtree}, \& {Yee}}]{Pritchet_1987ApJ...323...79P}
{Pritchet}, C.~J., {Richer}, H.~B., {Schade}, D., {Crabtree}, D., \& {Yee}, H.~K.~C. 1987, \apj, 323, 79, \dodoi{10.1086/165808}

\bibitem[{{Reid} {et~al.}(2019){Reid}, {Pesce}, \& {Riess}}]{Reid_2019ApJ...886L..27R}
{Reid}, M.~J., {Pesce}, D.~W., \& {Riess}, A.~G. 2019, \apjl, 886, L27, \dodoi{10.3847/2041-8213/ab552d}

\bibitem[{{Richer}(1981)}]{Richer_1981ApJ...243..744R}
{Richer}, H.~B. 1981, \apj, 243, 744, \dodoi{10.1086/158642}

\bibitem[{{Richer} {et~al.}(1984){Richer}, {Crabtree}, \& {Pritchet}}]{Richer_1984ApJ...287..138R}
{Richer}, H.~B., {Crabtree}, D.~R., \& {Pritchet}, C.~J. 1984, \apj, 287, 138, \dodoi{10.1086/162671}

\bibitem[{{Richer} {et~al.}(1985){Richer}, {Pritchet}, \& {Crabtree}}]{Richer_1985ApJ...298..240R}
{Richer}, H.~B., {Pritchet}, C.~J., \& {Crabtree}, D.~R. 1985, \apj, 298, 240, \dodoi{10.1086/163602}

\bibitem[{{Riess} {et~al.}(2021){Riess}, {Anderson}, {Breuval}, {Casertano}, {Macri}, {Scolnic}, \& {Yuan}}]{Riess_2021jwst.prop.1685R}
{Riess}, A., {Anderson}, R.~I., {Breuval}, L., {et~al.} 2021, {Uncrowding the Cepheids for an Improved Determination of the Hubble Constant}, JWST Proposal. Cycle 1, ID. \#1685

\bibitem[{{Riess} {et~al.}(2023){Riess}, {Anderson}, {Breuval}, {Casertano}, {Macri}, {Scolnic}, \& {Yuan}}]{Riess_2023jwst.prop.2875R}
---. 2023, {Scrutinizing the Dirtiest Cepheids, a Test of the Hubble Tension}, JWST Proposal. Cycle 2, ID. \#2875

\bibitem[{{Riess} {et~al.}(2022){Riess}, {Yuan}, {Macri}, {Scolnic}, {Brout}, {Casertano}, {Jones}, {Murakami}, {Anand}, {Breuval}, {Brink}, {Filippenko}, {Hoffmann}, {Jha}, {D'arcy Kenworthy}, {Mackenty}, {Stahl}, \& {Zheng}}]{Riess_2022ApJ...934L...7R}
{Riess}, A.~G., {Yuan}, W., {Macri}, L.~M., {et~al.} 2022, \apjl, 934, L7, \dodoi{10.3847/2041-8213/ac5c5b}

\bibitem[{{Riess} {et~al.}(2024{\natexlab{a}}){Riess}, {Scolnic}, {Anand}, {Breuval}, {Casertano}, {Macri}, {Li}, {Yuan}, {Huang}, {Jha}, {Murakami}, {Beaton}, {Brout}, {Wu}, {Addison}, {Bennett}, {Anderson}, {Filippenko}, \& {Carr}}]{Riess_JWST_H0_2024arXiv240811770R}
{Riess}, A.~G., {Scolnic}, D., {Anand}, G.~S., {et~al.} 2024{\natexlab{a}}, arXiv e-prints, arXiv:2408.11770, \dodoi{10.48550/arXiv.2408.11770}

\bibitem[{{Riess} {et~al.}(2024{\natexlab{b}}){Riess}, {Anand}, {Yuan}, {Casertano}, {Dolphin}, {Macri}, {Breuval}, {Scolnic}, {Perrin}, \& {Anderson}}]{Riess_2024ApJ...962L..17R}
{Riess}, A.~G., {Anand}, G.~S., {Yuan}, W., {et~al.} 2024{\natexlab{b}}, \apjl, 962, L17, \dodoi{10.3847/2041-8213/ad1ddd}

\bibitem[{{Ripoche} {et~al.}(2020){Ripoche}, {Heyl}, {Parada}, \& {Richer}}]{Ripoche_2020MNRAS.495.2858R}
{Ripoche}, P., {Heyl}, J., {Parada}, J., \& {Richer}, H. 2020, \mnras, 495, 2858, \dodoi{10.1093/mnras/staa1346}

\bibitem[{{Schlafly} \& {Finkbeiner}(2011)}]{Schlafly_2011ApJ...737..103S}
{Schlafly}, E.~F., \& {Finkbeiner}, D.~P. 2011, \apj, 737, 103, \dodoi{10.1088/0004-637X/737/2/103}

\bibitem[{{Scolnic} {et~al.}(2022){Scolnic}, {Brout}, {Carr}, {Riess}, {Davis}, {Dwomoh}, {Jones}, {Ali}, {Charvu}, {Chen}, {Peterson}, {Popovic}, {Rose}, {Wood}, {Brown}, {Chambers}, {Coulter}, {Dettman}, {Dimitriadis}, {Filippenko}, {Foley}, {Jha}, {Kilpatrick}, {Kirshner}, {Pan}, {Rest}, {Rojas-Bravo}, {Siebert}, {Stahl}, \& {Zheng}}]{Scolnic_Pantheon_2022ApJ...938..113S}
{Scolnic}, D., {Brout}, D., {Carr}, A., {et~al.} 2022, \apj, 938, 113, \dodoi{10.3847/1538-4357/ac8b7a}

\bibitem[{{Scolnic} {et~al.}(2023){Scolnic}, {Riess}, {Wu}, {Li}, {Anand}, {Beaton}, {Casertano}, {Anderson}, {Dhawan}, \& {Ke}}]{Scolnic_2023ApJ...954L..31S}
{Scolnic}, D., {Riess}, A.~G., {Wu}, J., {et~al.} 2023, \apjl, 954, L31, \dodoi{10.3847/2041-8213/ace978}

\bibitem[{Tange(2011)}]{Tange_2011a}
Tange, O. 2011, ;login: The USENIX Magazine, 36, 42.
\newblock \url{http://www.gnu.org/s/parallel}

\bibitem[{{Uddin} {et~al.}(2024){Uddin}, {Burns}, {Phillips}, {Suntzeff}, {Freedman}, {Brown}, {Morrell}, {Hamuy}, {Krisciunas}, {Wang}, {Hsiao}, {Goobar}, {Perlmutter}, {Lu}, {Stritzinger}, {Anderson}, {Ashall}, {Hoeflich}, {Shappee}, {Persson}, {Piro}, {Baron}, {Contreras}, {Galbany}, {Kumar}, {Shahbandeh}, {Davis}, {Anais}, {Busta}, {Campillay}, {Castell{\'o}n}, {Corco}, {Diamond}, {Gall}, {Gonzalez}, {Holmbo}, {Roth}, {Ser{\'o}n}, {Taddia}, {Torres}, {Baltay}, {Folatelli}, {Hadjiyska}, {Kasliwal}, {Nugent}, {Rabinowitz}, \& {Ryder}}]{Uddin_CSP_2024ApJ...970...72U}
{Uddin}, S.~A., {Burns}, C.~R., {Phillips}, M.~M., {et~al.} 2024, \apj, 970, 72, \dodoi{10.3847/1538-4357/ad3e63}

\bibitem[{{Verde} {et~al.}(2023){Verde}, {Sch{\"o}neberg}, \& {Gil-Mar{\'\i}n}}]{Verde_2023arXiv231113305V}
{Verde}, L., {Sch{\"o}neberg}, N., \& {Gil-Mar{\'\i}n}, H. 2023, arXiv e-prints, arXiv:2311.13305, \dodoi{10.48550/arXiv.2311.13305}

\bibitem[{{Weinberg} \& {Nikolaev}(2001)}]{Weinberg_2001ApJ...548..712W}
{Weinberg}, M.~D., \& {Nikolaev}, S. 2001, \apj, 548, 712, \dodoi{10.1086/319001}

\bibitem[{{Weisz} {et~al.}(2024){Weisz}, {Dolphin}, {Savino}, {McQuinn}, {Newman}, {Williams}, {Kallivayalil}, {Anderson}, {Boyer}, {Correnti}, {Geha}, {Sandstrom}, {Cole}, {Warfield}, {Skillman}, {Cohen}, {Beaton}, {Bressan}, {Bolatto}, {Boylan-Kolchin}, {Brooks}, {Bullock}, {Conroy}, {Cooper}, {Dalcanton}, {Dotter}, {Fritz}, {Garling}, {Gennaro}, {Gilbert}, {Girardi}, {Johnson}, {Johnson}, {Kalirai}, {Kirby}, {Lang}, {Marigo}, {Richstein}, {Schlafly}, {Tollerud}, \& {Wetzel}}]{Weisz_2024ApJS..271...47W}
{Weisz}, D.~R., {Dolphin}, A.~E., {Savino}, A., {et~al.} 2024, \apjs, 271, 47, \dodoi{10.3847/1538-4365/ad2600}

\bibitem[{{Wu} {et~al.}(2023){Wu}, {Scolnic}, {Riess}, {Anand}, {Beaton}, {Casertano}, {Ke}, \& {Li}}]{Wu_2023ApJ...954...87W}
{Wu}, J., {Scolnic}, D., {Riess}, A.~G., {et~al.} 2023, \apj, 954, 87, \dodoi{10.3847/1538-4357/acdd7b}

\bibitem[{{Yuan} {et~al.}(2017){Yuan}, {Macri}, {He}, {Huang}, {Kanbur}, \& {Ngeow}}]{Yuan_2017AJ....154..149Y}
{Yuan}, W., {Macri}, L.~M., {He}, S., {et~al.} 2017, \aj, 154, 149, \dodoi{10.3847/1538-3881/aa86f1}

\bibitem[{{Zgirski} {et~al.}(2021){Zgirski}, {Pietrzy{\'n}ski}, {Gieren}, {G{\'o}rski}, {Wielg{\'o}rski}, {Karczmarek}, {Bresolin}, {Kervella}, {Kudritzki}, {Storm}, {Graczyk}, {Hajdu}, {Narloch}, {Pilecki}, {Suchomska}, \& {Taormina}}]{Zgirski_2021ApJ...916...19Z}
{Zgirski}, B., {Pietrzy{\'n}ski}, G., {Gieren}, W., {et~al.} 2021, \apj, 916, 19, \dodoi{10.3847/1538-4357/ac04b2}

\end{thebibliography}
\bibliographystyle{aasjournal}

\begin{acknowledgments}

We are indebted to all of those who spent years and even decades bringing {\it JWST} to fruition. We thank Louise Breuval, Yukei Murakami and Abigail Lee for helpful conversations and Wenlong Yuan for providing photometry of Miras in the Large Magellanic Cloud. SL is supported by the National Science Foundation Graduate Research Fellowship Program under grant number DGE2139757. GSA acknowledges financial support from JWST GO-1685 and JWST GO-2875. DS is supported by Department of Energy grant DE-SC0010007, the David and Lucile Packard Foundation, the Templeton Foundation and Sloan Foundation. 

This research made use of the NASA Astrophysics Data System. This work is based on observations made with the NASA/ESA/CSA JWST. The data were obtained from the Mikulski Archive for Space Telescopes at the Space Telescope Science Institute, which is operated by the Association of Universities for Research in Astronomy, Inc., under NASA contract NAS 5-03127 for JWST. These observations are associated with programs $\#$1685  and $\#$2875. The JWST data presented in this article were obtained from the Mikulski Archive for Space Telescopes (MAST) at the Space Telescope Science Institute. The specific observations analyzed can be accessed via \dataset[DOI: ].

This research has made use of the NASA/IPAC Infrared Science Archive, which is funded by the National Aeronautics and Space Administration and operated by the California Institute of Technology.

The Digitized Sky Surveys were produced at the Space Telescope Science Institute under U.S. Government grant NAG W-2166. The images of these surveys are based on photographic data obtained using the Oschin Schmidt Telescope on Palomar Mountain and the UK Schmidt Telescope. The plates were processed into the present compressed digital form with the permission of these institutions.

\end{acknowledgments}

\vspace{5mm}
\facilities{JWST(NIRCAM)}

\software{astropy \citep{Astropy_2013A&A...558A..33A, Astropy_2018AJ....156..123A, Astropy_2022ApJ...935..167A},  
          DOLPHOT \citep{Dolphin_2000PASP..112.1383D, Dolphin_2016ascl.soft08013D, Weisz_2024ApJS..271...47W}, 
          GNU Parallel \citep{Tange_2011a}
          }

\appendix

\section{Comparing JAGB Measurements in \emph{F115W} and \emph{F150W}}

The CCHP observations from \emph{JWST} program GO-1995 are conducted in the \emph{F115W} band, while the observations we use for NGC 2525, N3370, and NGC 5861 as well as those from \cite{Li_2024ApJ...966...20L} are in the \emph{F150W} band. We can test the consistency between \emph{F115W} and \emph{F150W} measurements of the JAGB using the outer disk of NGC 5643, which was observed in both \emph{JWST} program GO-1685 \citep{Riess_2021jwst.prop.1685R} and GO-1995 \citep{Freedman_2021jwst.prop.1995F} and contain overlapping fields. This allows us to select stars common to both observations and compare the measured JAGBs in each sample.

To perform this test, we first retrieve the photometry used in \cite{Lee_JWST_JAGB_H0_2024arXiv240803474L} from their GitHub repository and select stars outside a radius of 39.3 kpc as done by \cite{Lee_JWST_JAGB_H0_2024arXiv240803474L}. While $A_{F115W}$ can be directly calculated, to ensure we are using the same foreground extinction in \emph{F115W} and extinction law, we reconstruct their distance measurement. The JAGB measurement for NGC 5643 is $m = 24.74$~mag in their Fig. 5, which is tied to the JAGB measurement in NGC 4258 of 23.41~mag (after subtracting foreground extinction of 0.02~mag) and the geometric distance of 29.40~mag \citep{Reid_2019ApJ...886L..27R}. This yields a distance before foreground extinction of $\mu$ = 30.73~mag. The final JAGB distance listed in their Table 5 is $\mu$ = 30.59~mag, implying a 0.14~mag foreground extinction correction in \emph{F115W}. As a foreground extinction correction for \emph{F356W} is not described, we assume it is negligible and for consistency do not apply it in this comparison. 

Next, we perform photometry on the NGC 5643 observations from GO-1685 and apply the same quality cuts as used in \cite{Lee_JWST_JAGB_H0_2024arXiv240803474L}. We apply foreground extinction corrections using E(B-V) values from the NASA Extragalactic Database (NED) and the same coefficients from \cite{Anand_2024ApJ...966...89A, Riess_2024ApJ...962L..17R}, then crossmatch the SH0ES sample with the clipped CCHP sample using their coordinates. We find 857 matched stars using a 1 arcsec crossmatch radius.

\begin{figure*}[ht!]
  \centering
  \begin{tabular}{cc}
    \includegraphics[width=0.45\linewidth]{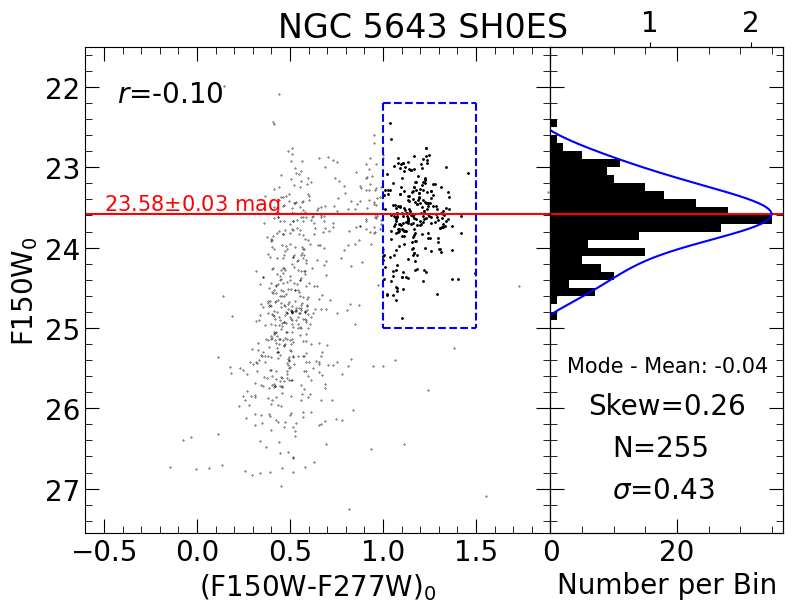} &
    \includegraphics[width=0.45\linewidth]{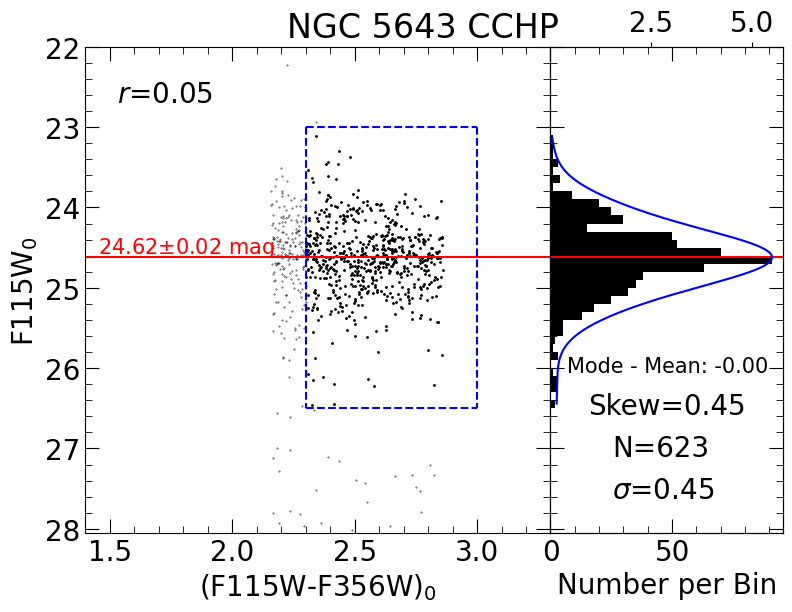} \\
  \end{tabular}
  \caption{Color magnitude diagrams and luminosity functions for NGC 5643 using crossmatched stars between observations taken in \emph{JWST} program GO-1685 \citep{Riess_2021jwst.prop.1685R} and GO-1995 \citep{Freedman_2021jwst.prop.1995F}. The left panels of each plot show the color magnitude diagrams after foreground extinction and crowding corrections. The vertical blue dashed lines use the color cuts of 1.0 $<$ \emph{F150W $-$ F277W} $<$ 1.5~mag and 2.3 $<$ \emph{F115W $-$ F356W} $<$ 3.0~mag \citep{Lee_JWST_JAGB_H0_2024arXiv240803474L} with magnitude cuts $\sim$5 $\sigma$ from the anticipated JAGB, selected to fully encapsulate the J-region. The value in the upper left corner of each plot shows the Pearson correlation coefficient for the stars inside the dashed blue box. The right panel shows the luminosity function for stars inside the dashed blue box, with a GLOESS smoothed luminosity function (s=0.25~mag) shown in blue. We also list the difference between the mode (s=0.25~mag) and mean reference magnitudes, Fisher-Pearson coefficient of skewness using the Python \texttt{scipy.stats.skew}, the number (N) of stars in the dashed-blue box, and the standard deviation of the luminosity function.}
  \label{fig:CMDs_N5643}
\end{figure*}

Next, we measure the mode JAGB using a smoothing value of 0.25~mag for both samples. We show the color magnitude diagrams in Fig. \ref{fig:CMDs_N5643}. For the SH0ES sample, we find $m = 23.58$~mag. When tied to the NGC 4258 JAGB measurement of the corresponding variant in \cite{Li_2024ApJ...966...20L} of 22.36~mag and the maser distance from \cite{Reid_2019ApJ...886L..27R}, this yields $\mu_0 = 30.62$~mag. For the CCHP sample, using the same algorithm we find $m = 24.64$~mag. Tying this measurement to 23.41~mag from Table 5 in \cite{Lee_JWST_JAGB_H0_2024arXiv240803474L} and the maser distance yields $\mu_0 = 30.61$~mag, in excellent agreement with the JAGB measured with the SH0ES sample as well as the measurement with the full CCHP sample of $\mu_0 = $31.60~mag. This agreement suggests consistency in photometry and GLOESS algorithm (here, see \cite{Wu_2023ApJ...954...87W} and for CCHP, see \cite{Lee_Magellan_2024ApJ...967...22L}, both publicly available) between the two samples as well suggests that using the \emph{F115W} or \emph{F150W} to measure distances yields no statistically significant difference, though more samples to perform similar comparisons are needed to avoid the fallacy of hasty generalization. We also note that the mode-based JAGB measurement in \emph{F150W} from \cite{Li_2024ApJ...966...20L}, using the baseline color range and a smoothing value of 0.25~mag, is 31.58~mag using the full field (see their Fig. 10). The distance to NGC 5643 in \emph{F115W} is 30.59~mag from \cite{Lee_JWST_JAGB_H0_2024arXiv240803474L} using their full sample. Both measurements using the same measurement method, either with the full fields or with crossmatched fields, are in excellent agreement.

\section{Completeness Characterizations and Comparisons}

In this section, we describe the completeness characterization procedure and provide a table listing the differences in measured JAGB for each galaxy before and after applying completeness correction.

\subsection{Completeness} \label{subsec:Completeness}

NGC 2525, NGC 3147, NGC 3370, NGC 5468, and NGC 5861 all lie at distances beyond $\mu_0 = 32$~mag, or 25 Mpc \citep{Riess_2022ApJ...934L...7R}, and are farther than the JAGB distances measured for SN Ia host galaxies in the literature \citep{Li_2024ApJ...966...20L, Lee_JWST_JAGB_H0_2024arXiv240803474L}.
Limited integration times (set by simultaneous observations of Cepheids, where the exposure times were dictated by the requirements set for Cepheids) for observations of stars at these distances can potentially result in nontrivial observational incompleteness, where a decreasing fraction of stars are successfully observed at increasingly faint magnitudes. If the incompleteness of an observation begins to drop significantly near the JAGB, incompleteness corrections would need to be applied  to the J-region luminosity function to avoid biasing a JAGB measurement. In principle, one could correct for incompleteness by dividing the number of stars in each magnitude and color bin in the J-region luminosity function by the completeness fraction.

To investigate potential effects of incompleteness on our JAGB measurements and the possibility of applying corrections, we first measure completeness fractions in the outer disks of our target galaxies using artificial stars tests from the DOLPHOT software suite. 

To characterize the completeness for each galaxy, we place 100 $-$ 200 stars in each 0.1 mag bin in a magnitude range $\pm$ 1~mag wider than the magnitudes cuts shown in Fig. \ref{fig:CMDs}. Because the J-region is a 2D region (i.e. it is defined within both a magnitude and color range), and completeness in \emph{F277W} can affect completeness in \emph{F150W} in the color direction, we repeat the artificial star tests across the \emph{F150W} - \emph{F277W} color range of 1.0~mag and 1.5~mag in 0.1~mag bins. For each galaxy, we compute the completeness by dividing the number of stars recovered after applying the quality cuts described in Section \ref{sec:photometry} by the total number of stars injected into the images per bin. We show the completeness fractions for each galaxy in Fig. \ref{fig:Completeness_Fractions}. We observe that the four galaxies at $\mu_0 = \sim$32~mag \citep{Riess_2022ApJ...934L...7R} have completeness fractions that drop to $\sim$0.8 at 1~mag fainter than the anticipated JAGB magnitude ($\sim26$~mag). For NGC 5458 and NGC 3147, which lie at $\mu_0 \sim$33~mag this drops closer to $\sim$0.6 and  $\sim$0.2, respectively. Qualitatively, we can also observe a slight color dependence of the completeness at a given \emph{F150W} magnitude demonstrated by the diagonal color gradients in Fig. \ref{fig:Completeness_Fractions}.

\begin{figure*}[ht!]
  \centering
  \begin{tabular}{cc}
    \includegraphics[width=0.45\linewidth]{JAGB_2D_Completeness_Grid_NGC_4258.pdf} &
    \includegraphics[width=0.45\linewidth]{JAGB_2D_Completeness_Grid_NGC_2525.pdf} \\
    \includegraphics[width=0.45\linewidth]{JAGB_2D_Completeness_Grid_NGC_3370.pdf} &
    \includegraphics[width=0.45\linewidth]{JAGB_2D_Completeness_Grid_NGC_5861.pdf} \\
    \includegraphics[width=0.45\linewidth]
    {JAGB_2D_Completeness_Grid_NGC_3147.pdf} &
    \includegraphics[width=0.45\linewidth]{JAGB_2D_Completeness_Grid_NGC_5468.pdf} \\
    \multicolumn{2}{c}{\includegraphics[width=0.45\linewidth]{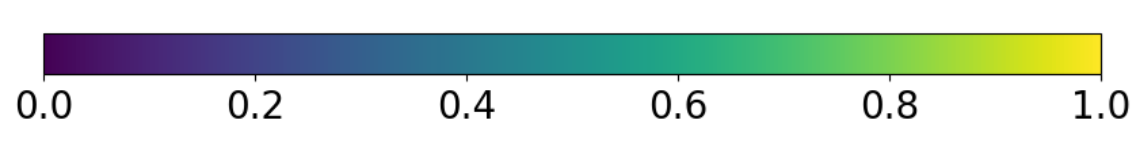}}
  \end{tabular}
  \caption{Interpolated completeness fractions (color scale set by the color bar at the bottom) across magnitude and color for the galaxies analyzed in this study. J-region luminosity functions for each galaxy in 0.01~mag bins are plotted in the right panels of each subplot.}
  \label{fig:Completeness_Fractions}
\end{figure*}

\begin{figure*}[ht!]
  \centering
  \begin{tabular}{cc}
    \includegraphics[width=0.45\linewidth]{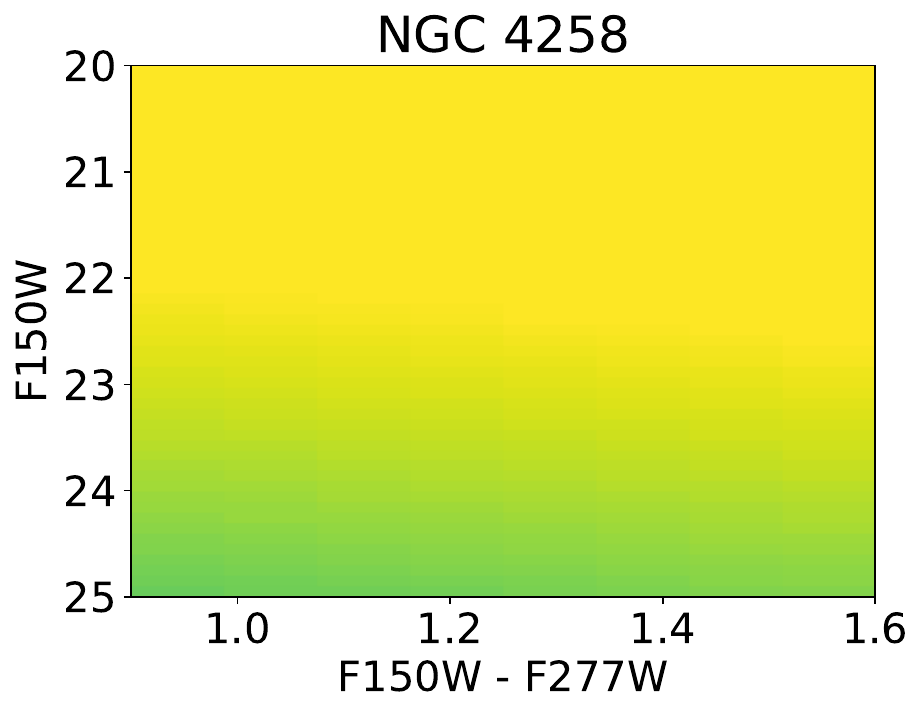} &
    \includegraphics[width=0.45\linewidth]{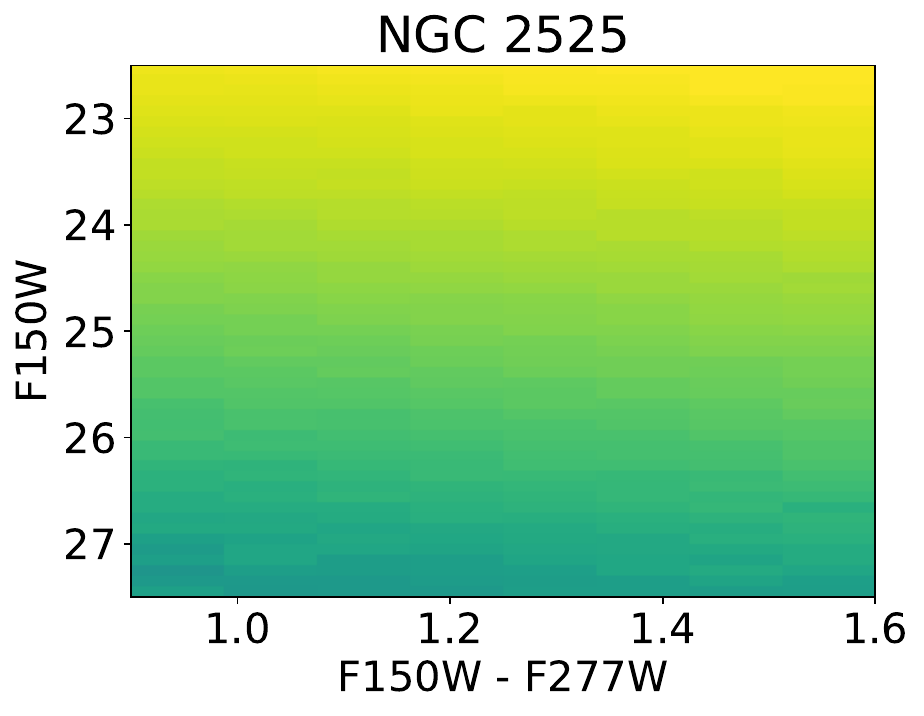} \\
    \includegraphics[width=0.45\linewidth]{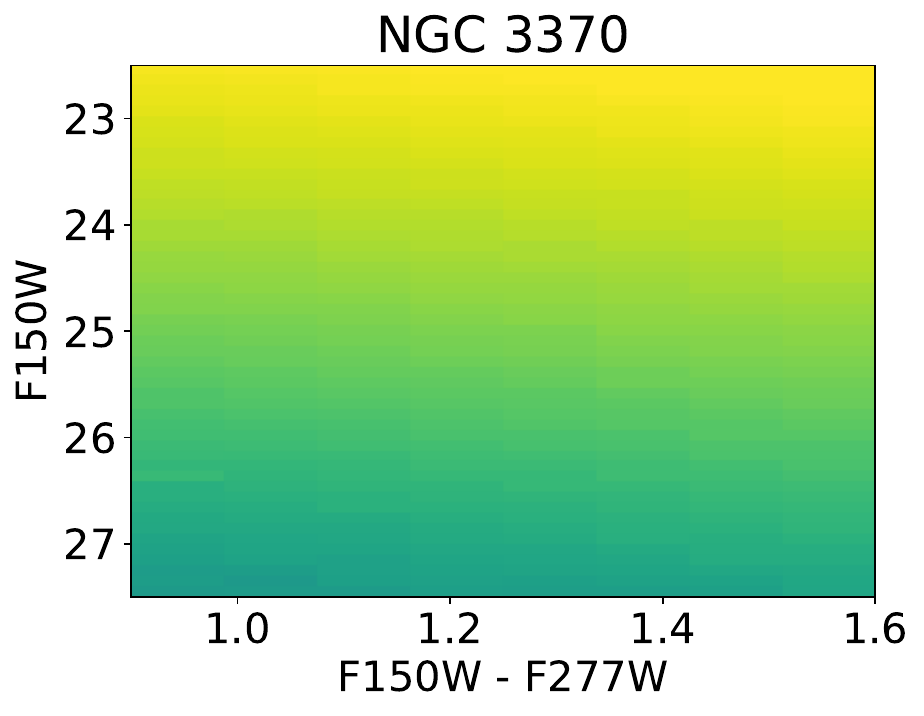} &
    \includegraphics[width=0.45\linewidth]{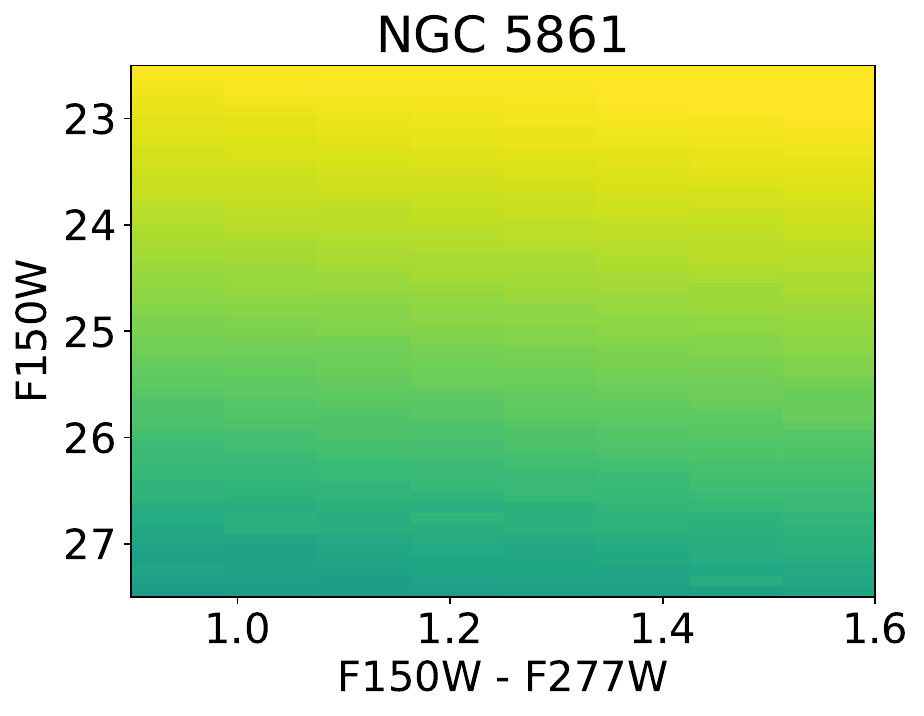} \\
    \includegraphics[width=0.42\linewidth]
    {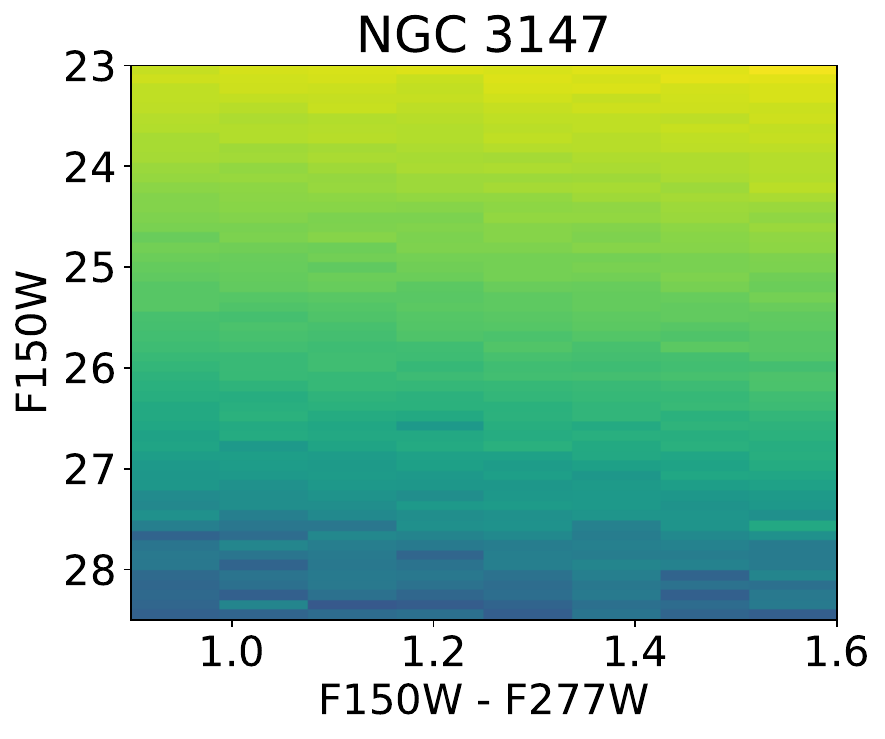} &
    \includegraphics[width=0.45\linewidth]{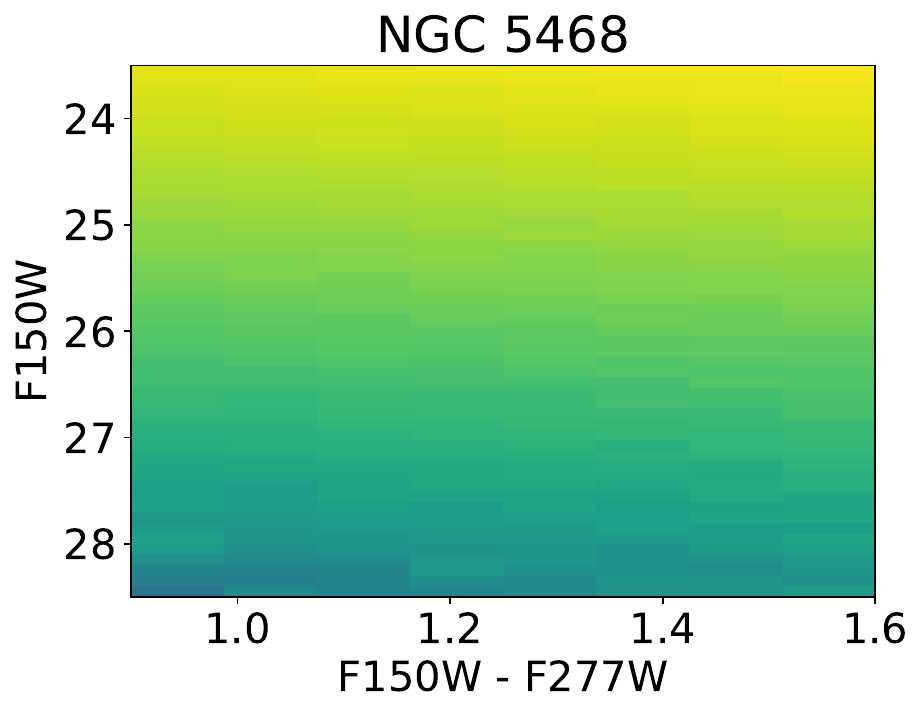} \\
    \multicolumn{2}{c}{\includegraphics[width=0.45\linewidth]{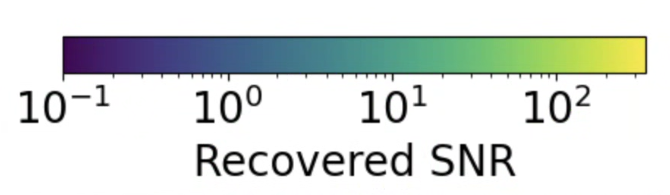}}
  \end{tabular}
  \caption{Recovered SNR from artificial star tests across magnitude and color for the galaxies analyzed in this study.}
  \label{fig:Completeness_Fractions}
\end{figure*}

Next, we interpolate the completeness fractions using \texttt{RectBivariateSpline} from the Python \texttt{scipy} package and apply completeness corrections to the J-region luminosity functions by first binning the luminosity function in 0.01~mag bins and dividing the number of stars in each bin by the completeness fraction. 

We find that we are able to recover the shape of the LF for all galaxies except for NGC 3147.  NGC 3147 shows especially low ($\leq $50$\%$) completeness fractions around the anticipated JAGB reference magnitude of $\sim$26~mag. While incompleteness can be corrected for, the uncertainties for these completeness corrections for high incompletenesses can be large due to the mismatch between how stars are recovered in the actual photometry and artificial star test. For instance, at low incompleteness, stars typically have a low signal-to-noise and the photometric procedure can pick up non-stellar objects (which are later cut out with quality cuts). Artificial star tests, however, assume that each input star is real. The low completeness fractions for NGC 3147 can be largely attributed to a higher density of stars in the outer disk when compared to NGC 5468, for instance. This feature would cause the DOLPHOT quality cuts to cut out a larger fraction of stars due to the \texttt{crowd} cut. The low signal-to-noise of the NGC 3147 observation further exacerbates this; we show the \emph{F150W} DOLPHOT SNR in Table \ref{tab:SNR}. In addition, at low SNR, RGB and AGB stars are more likely to contaminate the J-region compared to at high SNR due to larger photometric uncertainties; this can be seen, for instance, qualitatively at the fainter side of the J-region for NGC 5468 in Fig. \ref{fig:CMDs}.

For these reasons, we decide that we are unable to recover a J-region luminosity function that can yield a robust JAGB measurement for NGC 3147, with the criteria that the recovered DOLPHOT SNR should be greater than $\sim$10 1~mag fainter than the anticipated JAGB. For comparison, we also list the SNR at the anticipated or measured JAGB and 1~mag fainter than that magnitude for the five galaxies analyzed in Table \ref{tab:SNR}.

\begin{deluxetable}{ccc}
\caption{SNR for the New Host Galaxies Analyzed}
\label{tab:SNR}
\tablehead{\colhead{Galaxy} & \colhead{SNR(JAGB)} & \colhead{SNR(JAGB$+$1 mag)}}
\startdata
NGC 2525 & 31 & 15 \\
NGC 3370 & 32 & 15 \\
NGC 5861 & 38 & 17 \\
NGC 3147 & 14 & 7 \\
NGC 5468 & 29 & 13 \\
\enddata
\tablecomments{Mean \emph{F150W} signal-to-noise ratios of stars $\pm$0.01~mag from the measured JAGB and 1 mag fainter than the measured or anticipated JAGB.}
\end{deluxetable}

We explore the effects of incompleteness corrections on the final JAGB measurement by first measuring the JAGB across 25 measurement variants similar to that described in Section \ref{subsec:measurement_variants}, but using binned histograms for the mean and median measurements. For the model fit method, we increase the bin width to 0.05~mag to remain consistent with \cite{Li_2024ApJ...966...20L}. We repeat the measurements without incompleteness corrections and the modified measurement variants. We list differences between the measured JAGBs in the Table \ref{tab:JAGB_comparison_completeness}, which are largely a couple hundredths of a magnitude noting that differences in the same direction as differences for NGC 4258 will largely cancel. Due to this cancellation and because we take the middle of all variants as our distance measurement, the change in distance for NGC 2525, NGC 3370, and NGC 5861 are 0.01~mag; for this reason, we do not use completeness corrections in our baseline results for these galaxies. However, due to the larger corrections the change in distance for NGC 5468 are greater than 0.01~mag, and so we elect to incorporate that correction into our baseline result.

\begin{deluxetable}{cccccccccc}
\tablecaption{Comparison of JAGB Measurement Variants}
\label{tab:JAGB_comparison_completeness}
\tablehead{
    \colhead{Variant Number} & 
    \colhead{Method} & 
    \colhead{Color Range Lower} & 
    \colhead{Color Range Upper} & 
    \colhead{Smoothing} & 
    \colhead{NGC2525} & 
    \colhead{NGC3370} & 
    \colhead{NGC4258} & 
    \colhead{NGC5468} & 
    \colhead{NGC5861}
}
\startdata
1 & Median & 1.05 & 1.42 & nan & -0.013 & -0.022 & -0.008 & -0.025 & 0.003 \\
2 & Mean & 1.05 & 1.42 & nan & -0.021 & -0.024 & -0.002 & -0.028 & 0.009 \\
3 & Mode & 1.05 & 1.42 & 0.25 & -0.01 & -0.022 & 0.001 & -0.025 & -0.01 \\
4 & Mode & 1.05 & 1.42 & 0.35 & -0.02 & -0.022 & -0.009 & -0.035 & -0.02 \\
5 & Model & 1.05 & 1.42 & nan & -0.009 & -0.01 & -0.012 & -0.004 & 0.005 \\
6 & Median & 1.0 & 1.5 & nan & -0.015 & -0.027 & 0.001 & -0.016 & 0.001 \\
7 & Mean & 1.0 & 1.5 & nan & -0.025 & -0.033 & -0.001 & -0.027 & 0.005 \\
8 & Mode & 1.0 & 1.5 & 0.25 & -0.02 & -0.022 & -0.009 & -0.015 & -0.01 \\
9 & Mode & 1.0 & 1.5 & 0.35 & -0.03 & -0.032 & -0.009 & -0.035 & -0.02 \\
10 & Model & 1.0 & 1.5 & nan & -0.027 & -0.015 & -0.005 & -0.005 & 0.011 \\
11 & Median & 0.95 & 1.58 & nan & -0.011 & -0.03 & 0.001 & -0.02 & -0.004 \\
12 & Mean & 0.95 & 1.58 & nan & -0.026 & -0.035 & 0.003 & -0.015 & 0.001 \\
13 & Mode & 0.95 & 1.58 & 0.25 & -0.01 & -0.032 & -0.009 & -0.029 & -0.01 \\
14 & Mode & 0.95 & 1.58 & 0.35 & -0.03 & -0.032 & -0.009 & -0.039 & -0.02 \\
15 & Model & 0.95 & 1.58 & nan & -0.013 & -0.012 & -0.003 & -0.006 & 0.018 \\
16 & Median & 1.05 & 1.5 & nan & -0.012 & -0.024 & -0.001 & -0.024 & -0.001 \\
17 & Mean & 1.05 & 1.5 & nan & -0.021 & -0.026 & -0.001 & -0.029 & 0.007 \\
18 & Mode & 1.05 & 1.5 & 0.25 & -0.02 & -0.022 & 0.001 & -0.025 & 0.0 \\
19 & Mode & 1.05 & 1.5 & 0.35 & -0.03 & -0.022 & -0.009 & -0.035 & -0.01 \\
20 & Model & 1.05 & 1.5 & nan & -0.024 & -0.013 & -0.01 & -0.004 & 0.007 \\
21 & Median & 1.0 & 1.42 & nan & -0.025 & -0.026 & 0.002 & -0.016 & 0.003 \\
22 & Mean & 1.0 & 1.42 & nan & -0.03 & -0.03 & -0.002 & -0.024 & 0.008 \\
23 & Mode & 1.0 & 1.42 & 0.25 & -0.02 & -0.032 & -0.009 & -0.025 & -0.01 \\
24 & Mode & 1.0 & 1.42 & 0.35 & -0.02 & -0.032 & -0.009 & -0.045 & -0.02 \\
25 & Model & 1.0 & 1.42 & nan & -0.009 & -0.012 & -0.007 & -0.007 & 0.01 \\
\enddata
\tablecomments{Difference between the JAGB measurements across measurement variants with the baseline result for host galaxies.}
\end{deluxetable}

We plot the JAGB as function of the same 25 variants described above for NGC 4258 and NGC 5468, as well as distance tied to NGC 4258 \citep{Reid_2019ApJ...886L..27R}, in the next section, Fig. \ref{fig:Variants}. For these variants, the mode and model fit procedures remain the same (though with bin sizes of 0.1 mag), and for the mean and median measurements we take those measures applied to discretized magnitudes from the binned luminosity functions. We apply the same binning-based measurement procedures for NGC 4258 as done with NGC 5468 to remain consistent along the distance ladder. We find agreement with the Cepheid distance of $\mu_0 = 33.127 \pm 0.082$ from \citep{Riess_2022ApJ...934L...7R}, fit variant 10. We observe that although the dispersion across variants decreases with the binning-based methods due to loss of information in magnitudes, there is additional uncertainty of 0.05~mag due to the binning. In addition, we use bootstrap resampling to estimate uncertainties in the variant measurements except for the model fit; due to binning, these uncertainties are also larger than those used in the non-binning versions.

\begin{figure*}[ht!]
  \centering
  \begin{tabular}{cccc}
    \includegraphics[width=0.2\linewidth]{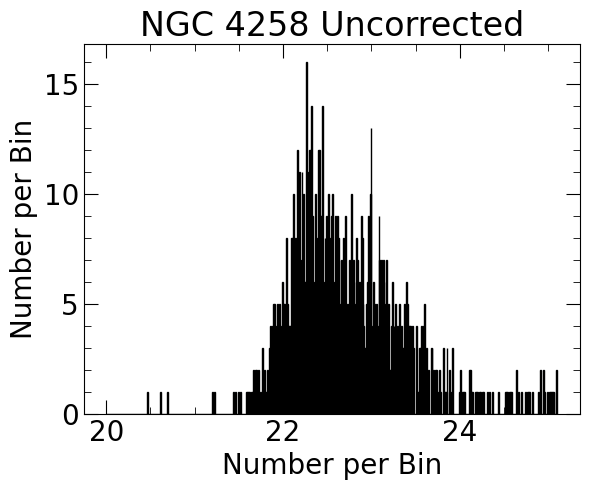} &
    \includegraphics[width=0.2\linewidth]{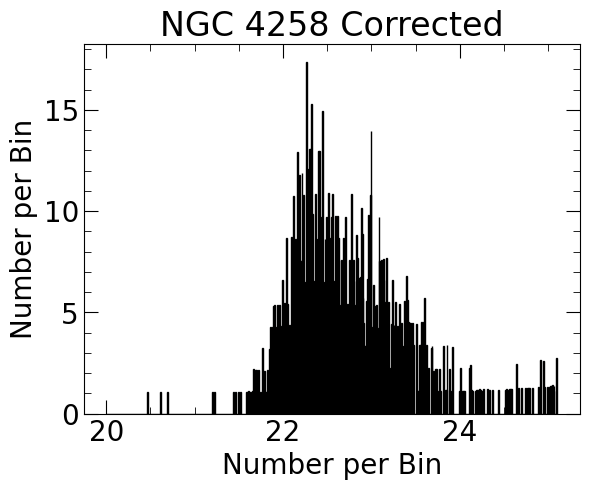} &
    \includegraphics[width=0.2\linewidth]{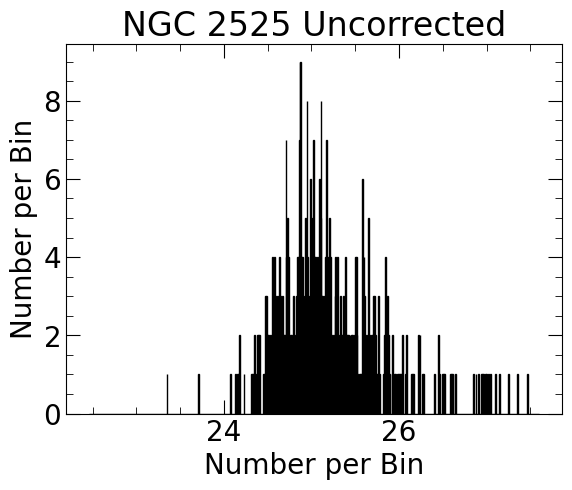} &
    \includegraphics[width=0.2\linewidth]{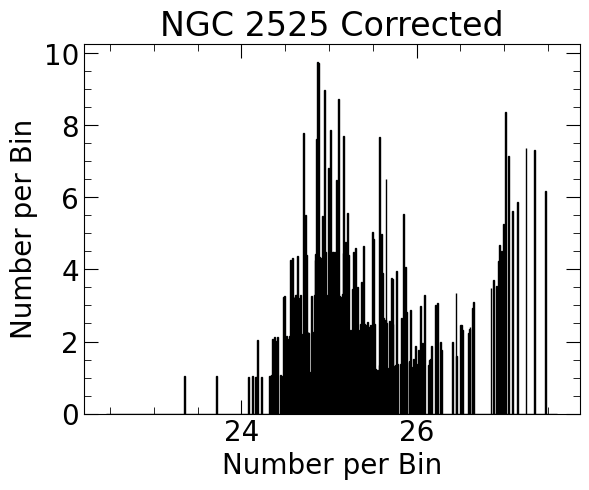} \\
    \includegraphics[width=0.2\linewidth]
    {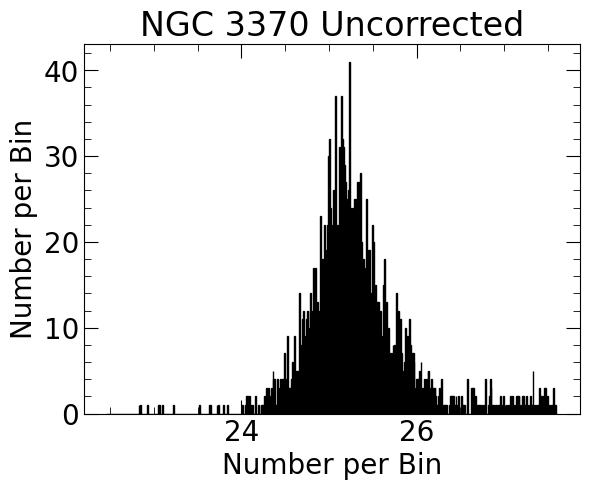} &
    \includegraphics[width=0.2\linewidth]{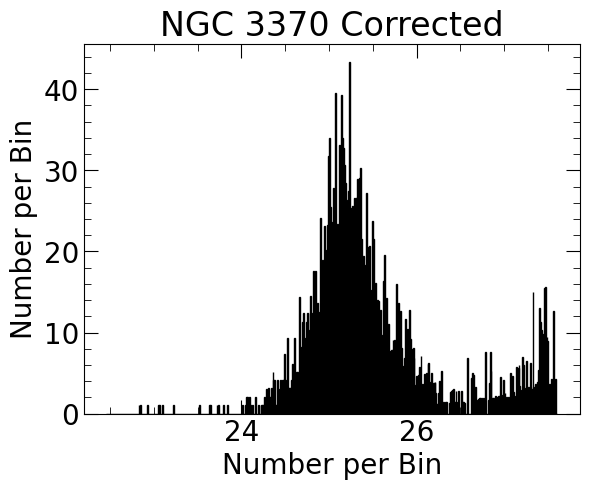} &
    \includegraphics[width=0.2\linewidth]
    {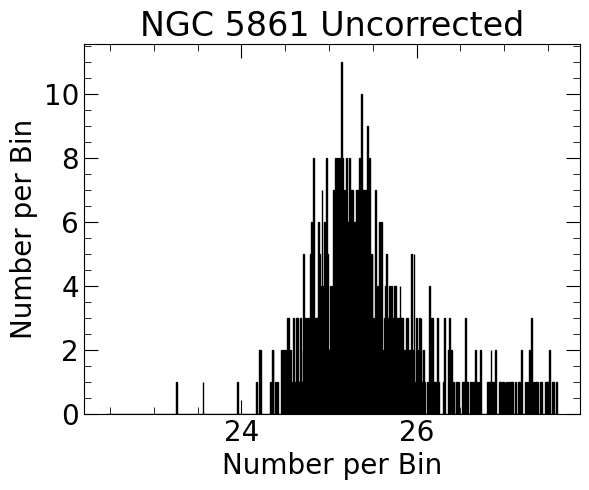} &
    \includegraphics[width=0.2\linewidth]{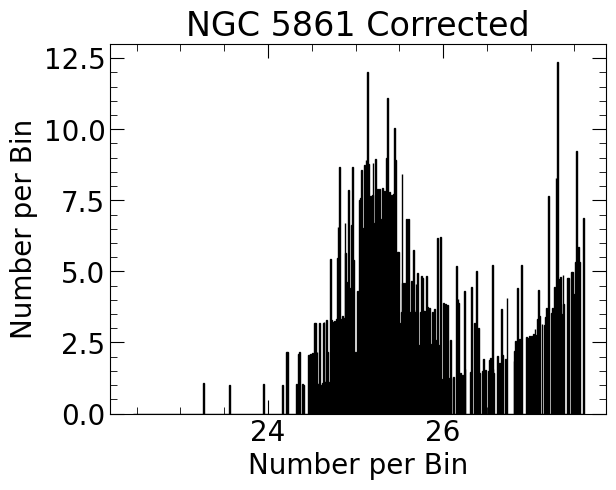} \\
    \includegraphics[width=0.2\linewidth]
    {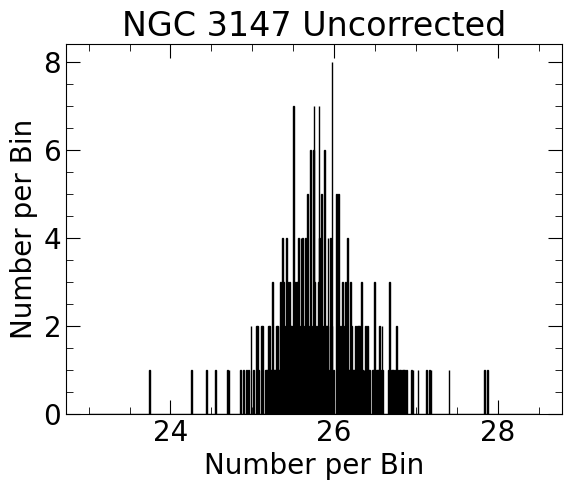} &
    \includegraphics[width=0.2\linewidth]{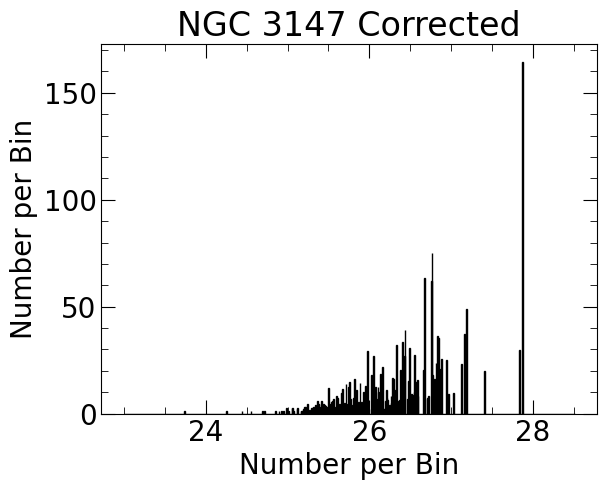} &
    \includegraphics[width=0.2\linewidth]
    {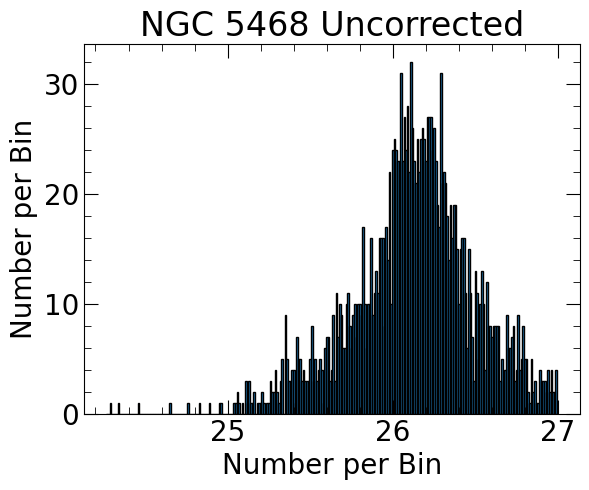} &
    \includegraphics[width=0.2\linewidth]{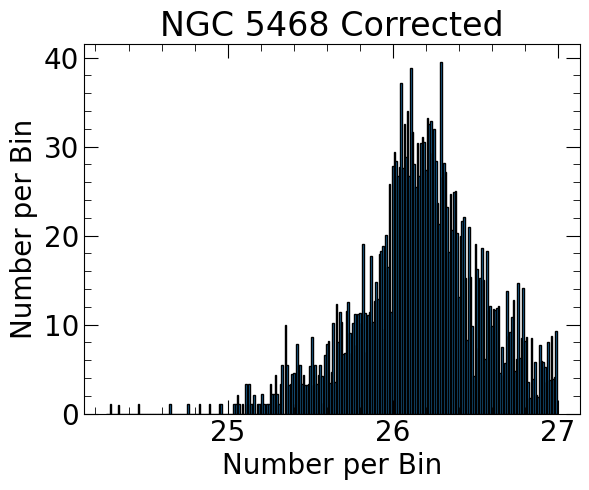} \\
  \end{tabular}
  \caption{Luminosity functions for NGC 4258, NGC 2525, NGC 3370, NGC 5861, NGC 3147, and NGC 5468 before and after incompleteness corrections.}
  \label{fig:LF_after_completeness_corr}
\end{figure*}

\section{Measurement Variants Plots}

In this section, we provide the measured JAGB reference magnitudes and distances for NGC 3370, NGC 3447, NGC 5861, and NGC 5468 across all measurement variants.

\begin{figure*}[ht!]
  \centering
  \fbox{
  \begin{tabular}{cc}
    \includegraphics[width=0.4\linewidth]{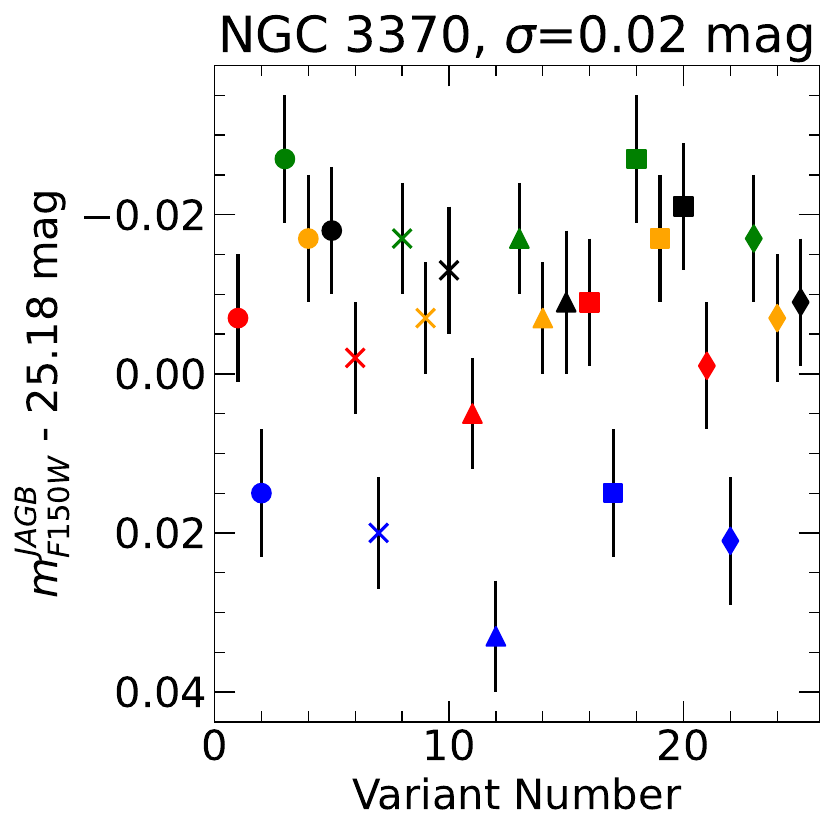} &
    \includegraphics[width=0.4\linewidth]{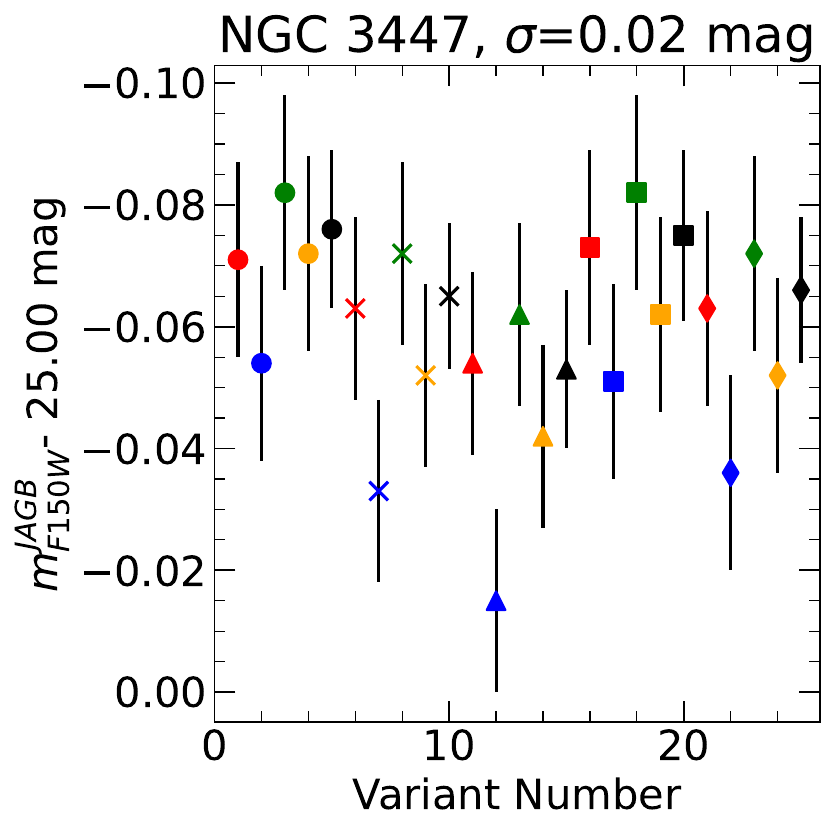} \\
    \includegraphics[width=0.4\linewidth]{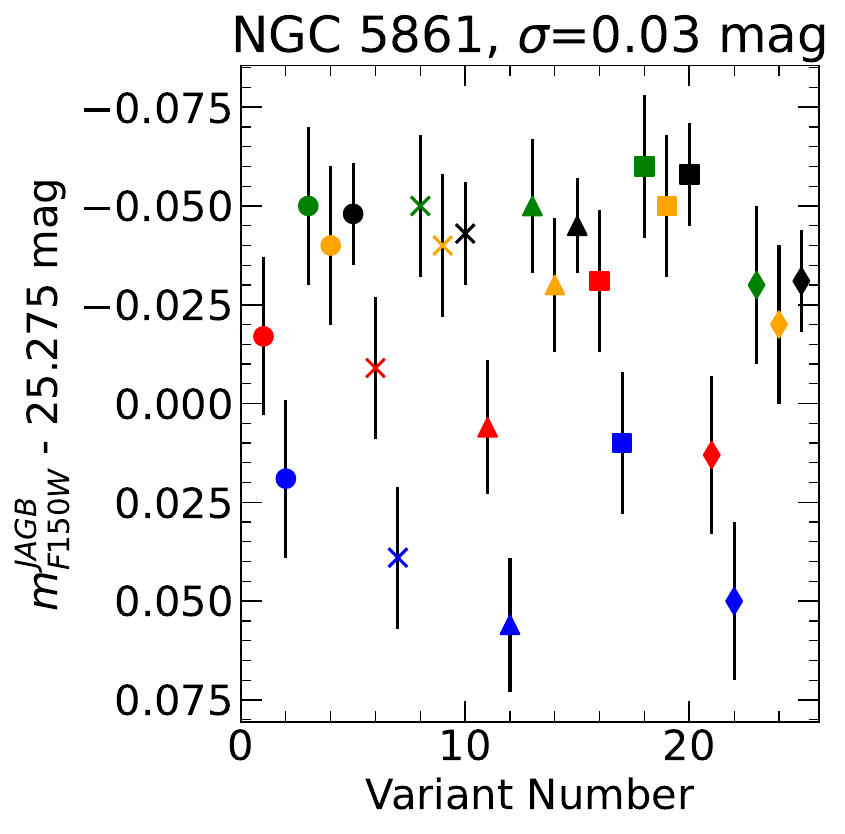} &
    \includegraphics[width=0.4\linewidth]{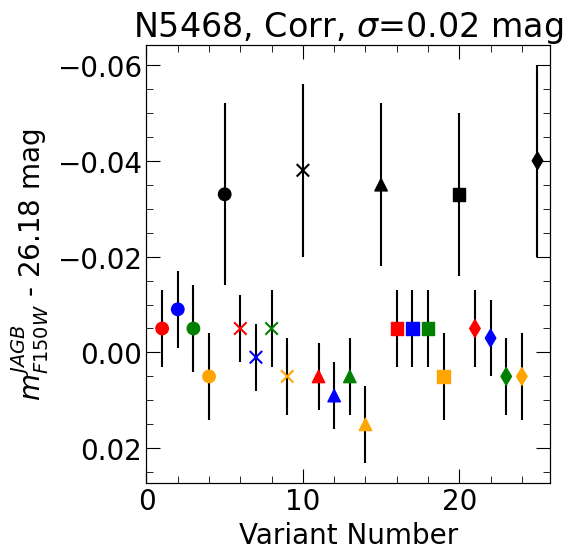} \\
   \multicolumn{2}{c}{\includegraphics[width=0.5\linewidth]{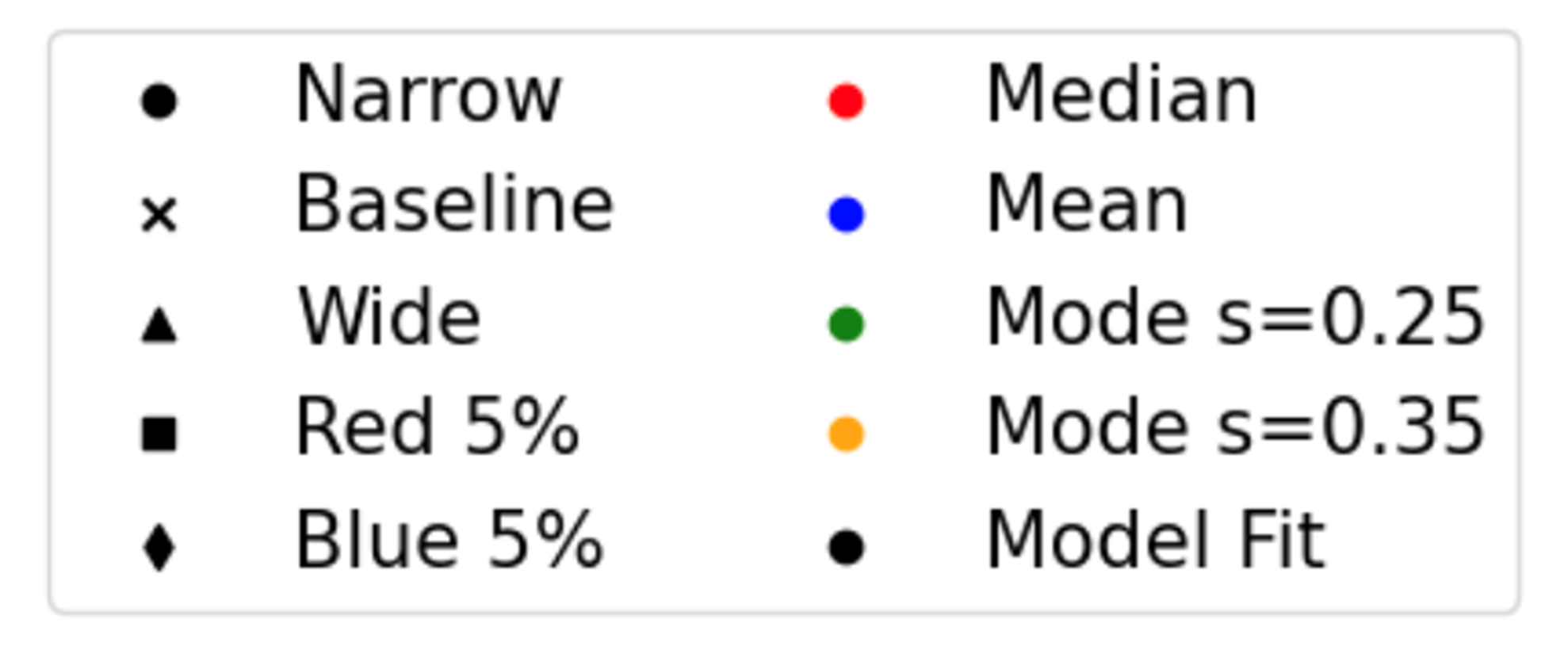}}
  \end{tabular}}
  \caption{JAGB reference magnitudes as a function of the 25 measurement variants listed in Table \ref{tab:Measurement_Variants} for NGC 3370, NGC 5861, and NGC 5468. The measurement variants are detailed in Table \ref{tab:Measurement_Variants}. The variants for NGC 5468 are performed using the binned versions of the variants and after completeness corrections.}
  \label{fig:Variants}
\end{figure*}


\begin{figure*}[ht!]
  \centering
  \fbox{
  \begin{tabular}{cc}
    \includegraphics[width=0.4\linewidth]{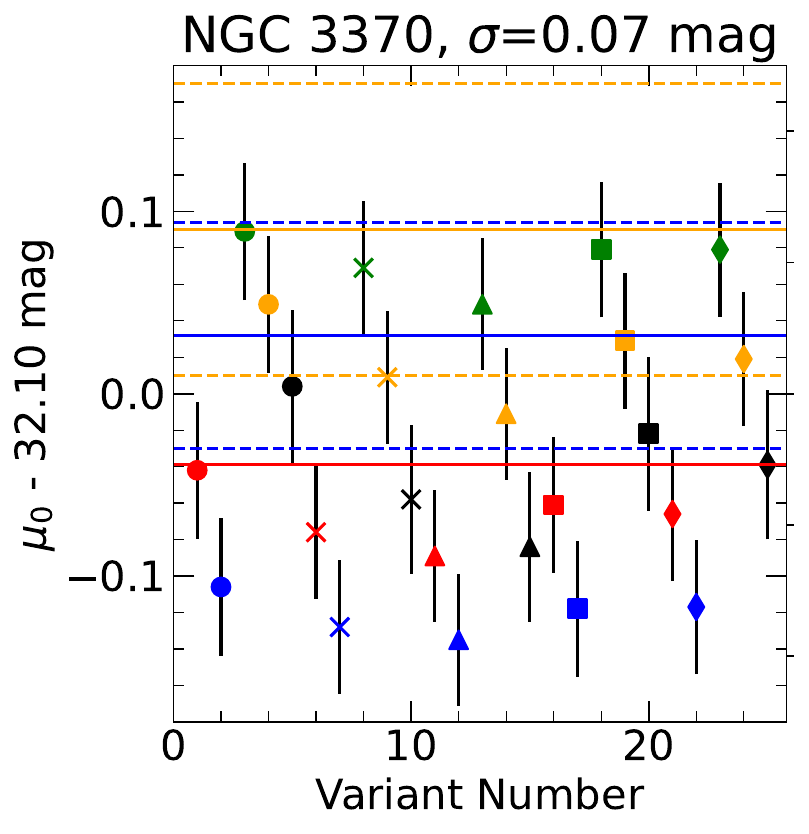} &
    \includegraphics[width=0.4\linewidth]{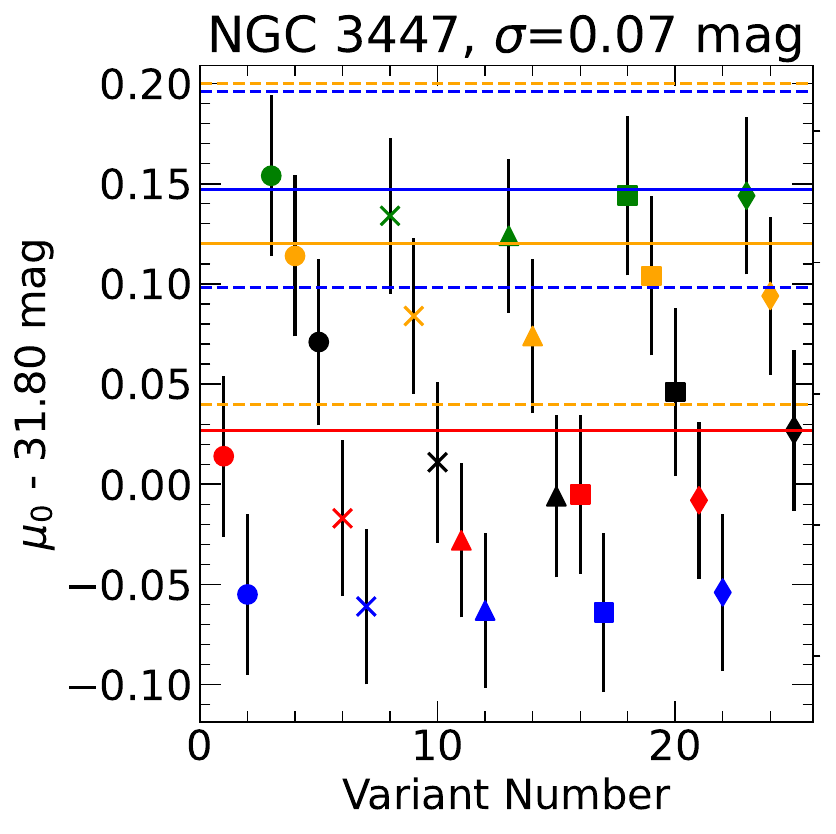} \\
    \includegraphics[width=0.4\linewidth]{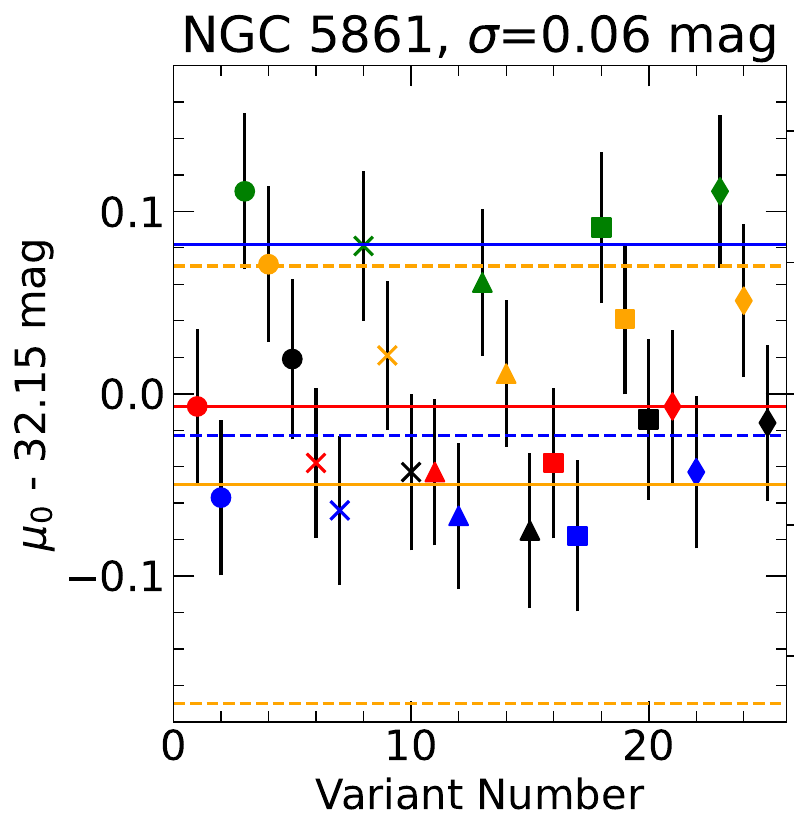} &
    \includegraphics[width=0.4\linewidth]{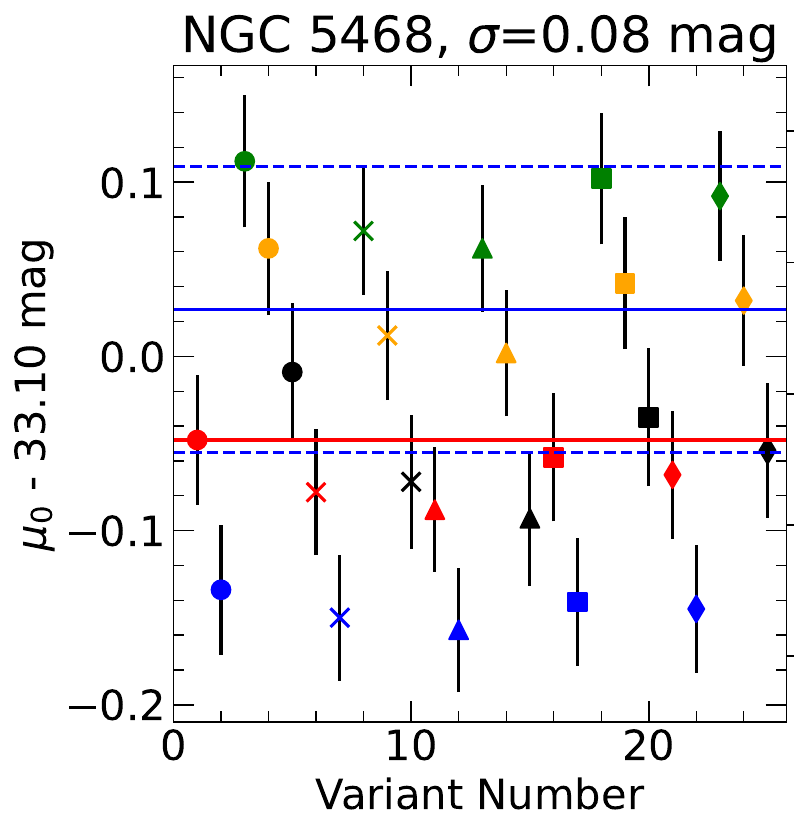} \\
   \multicolumn{2}{c}{\includegraphics[width=0.5\linewidth]{Distance_Legend_w_Median.png}}
  \end{tabular}}
  \caption{JAGB-based distances to NGC 3370, NGC 3447, NGC 5861, and NGC 5468 plotted against the 25 measurement variants listed in Table \ref{tab:Measurement_Variants}. The red line shows the middle JAGB distance, which we use as our baseline result. The blue lines shows the \emph{HST} Cepheid-based distance, fit variant 10, from \cite{Riess_2022ApJ...934L...7R}, and the orange line shows the \emph{JWST} TRGB distance from \cite{Li_TRGB_vs_Ceph_2024arXiv240800065L}.  Uncertainties for the Cepheid and TRGB distances are shown with the dashed lines.}
  \label{fig:distances}
\end{figure*}

\end{document}